\documentclass[aps,prd,twocolumn,preprintnumbers,nofootinbib,longbibliography,floatfix]{revtex4-1}

\usepackage{multirow,csvsimple,verbatim,graphicx,bm,color,float,amsmath,amssymb,siunitx,hyperref,bold-extra,afterpage,xspace}
\usepackage[caption=false]{subfig}
\usepackage[utf8]{inputenc}
\usepackage[all]{hypcap} 

\definecolor{darkblue}{rgb}{0,0,0.7}
\definecolor{darkred}{rgb}{0.7,0,0}
\definecolor{darkgreen}{rgb}{0,0.7,0}
\definecolor{orange}{rgb}{0.9, 0.25, 0}
\definecolor{purple}{rgb}{0.4, 0.04, 0.2}

\DeclareRobustCommand{\Sec}[1]{Sec.~\ref{#1}}
\DeclareRobustCommand{\Secs}[2]{Secs.~\ref{#1} and \ref{#2}}
\DeclareRobustCommand{\App}[1]{App.~\ref{#1}}
\DeclareRobustCommand{\Tab}[1]{Table~\ref{#1}}

\DeclareRobustCommand{\Fig}[1]{Fig.~\ref{#1}}
\DeclareRobustCommand{\Figs}[2]{Figs.~\ref{#1} and \ref{#2}}

\DeclareRobustCommand{\Eq}[1]{Eq.~(\ref{#1})}
\DeclareRobustCommand{\Eqs}[2]{Eqs.~(\ref{#1}) and (\ref{#2})}
\DeclareRobustCommand{\Ref}[1]{Ref.~\cite{#1}}
\DeclareRobustCommand{\Refs}[1]{Refs.~\cite{#1}}

\newcommand{\figO}{\mathcal{O}}
\newcommand{\figL}{\mathcal{L}}
\newcommand{\ka}{\kappa}
\newcommand{\GeV}{\text{GeV}}

\newcommand{\Zenodo}{{\sc Zenodo}\xspace}
\newcommand{\Pythia}{{\sc Pythia}\xspace}
\newcommand{\OmniFold}{{\sc OmniFold}\xspace}
\newcommand{\Jet}{{\tt Jet}\xspace}

\newcommand{\vev}[1]{\langle #1 \rangle}

\newcommand{\cor}[1]{#1}
\newcommand{\cortwo}[1]{#1}

\begin{document}

\title{Disentangling Quarks and Gluons with CMS Open Data}

\author{Patrick T. Komiske}
\email{pkomiske@mit.edu}
\affiliation{Center for Theoretical Physics, Massachusetts Institute of Technology, Cambridge, MA 02139, USA}
\affiliation{The NSF AI Institute for Artificial Intelligence and Fundamental Interactions}

\author{Serhii Kryhin}
\email{serhin@mit.edu}
\affiliation{Center for Theoretical Physics, Massachusetts Institute of Technology, Cambridge, MA 02139, USA}
\affiliation{The NSF AI Institute for Artificial Intelligence and Fundamental Interactions}

%\author{Eric M. Metodiev}
%\email{metodiev@mit.edu}
%\affiliation{Center for Theoretical Physics, Massachusetts Institute of Technology, Cambridge, MA 02139, USA}

\author{Jesse Thaler}
\email{jthaler@mit.edu}
\affiliation{Center for Theoretical Physics, Massachusetts Institute of Technology, Cambridge, MA 02139, USA}
\affiliation{The NSF AI Institute for Artificial Intelligence and Fundamental Interactions}

\begin{abstract}
We study quark and gluon jets separately using public collider data from the CMS experiment.
Our analysis is based on \SI{2.3}{fb^{-1}} of proton-proton collisions at $\sqrt{s}=\SI{7}{TeV}$, collected at the Large Hadron Collider in 2011.
We define two non-overlapping samples via a pseudorapidity cut---central jets with $|\eta|\le0.65$ and forward jets with $|\eta|>0.65$---and employ jet topic modeling to extract individual distributions for the maximally separable categories.
Under certain assumptions, such as sample independence and mutual irreducibility, these categories correspond to ``quark'' and ``gluon'' jets, as given by a recently proposed operational definition.
We consider a number of different methods for extracting reducibility factors from the central and forward datasets, from which the fractions of quark jets in each sample can be determined.
The greatest stability and robustness to statistical uncertainties is achieved by a novel method based on parametrizing the endpoints of a receiver operating characteristic (ROC) curve.
To mitigate detector effects, which would otherwise induce unphysical differences between central and forward jets, we use the \OmniFold method to perform central value unfolding.
As a demonstration of the power of this method, we extract the intrinsic dimensionality of the quark and gluon jet samples, which exhibit Casimir scaling, as expected from the strongly-ordered limit. 
To our knowledge, this work is the first application of full phase space unfolding to real collider data \cor{(albeit without a full systematics analysis)}, and 
%one of the first applications of topic modeling 
\cor{the first application of topic modeling using a machine-learned classifier} to extract separate quark and gluon distributions at the LHC.
\end{abstract}

\preprint{MIT-CTP 5422}
\maketitle
{\small
\tableofcontents
}

%%%%%%%%%%%%%%%%%%%%%%%%%%%%%%%%%%%%%%%%%%%%%%%%%%%%%%%%%%%%%%%%%%%%%%%%%%

\section{Introduction}

Despite being fundamental ingredients in quantum chromodynamics (QCD), quarks and gluons are ambiguous categories experimentally.
While one can cluster sprays of hadrons into jets and use them as proxies for short-distance partons, one cannot definitively say whether a jet was initiated by a quark or gluon due to color confinement.
Nevertheless, one can use an \emph{operational definition} of ``quarks'' and ``gluons'' to extract interesting jet information from experimental data, as advocated in \Ref{Komiske:2018vkc}.
This approach, which is based on the statistical framework of \emph{jet topic modeling}~\cite{Metodiev:2018ftz}, has been applied by the ATLAS experiment to measured jet samples \cite{ATLAS:2019rqw} and explored in phenomenological studies~\cite{radha_phdthesis,Brewer:2020och,Stewart:2022ari,Ying:2022jvy}.
See \Refs{Dillon:2019cqt,Alvarez:2019knh,Dillon:2020quc} for related topic modeling studies in particle physics.

Because this quark/gluon definition is operational, it depends on the detailed choice of topic modeling algorithm.
This is especially true when dealing with finite data samples and their associated statistical uncertainties.
The approach described in \Ref{Metodiev:2018ftz} was based on histogramming observables and therefore sensitive to the choice of binning, as studied in \Ref{radha_phdthesis}.
This in turn impacts the uncertainties of the results, since the statistical noise scales roughly with the inverse square root of the number of events in the most constraining bin.
To deal with low-statistics datasets, \Ref{Brewer:2020och} introduced a more robust method based on functional fitting, though this approach is sensitive to the choice of priors.
In addition to these statistical issues, one always has to confront systematic uncertainties in any experimental context.

In this paper, we develop two new approaches to topic modeling that exhibit more stable behavior and apply them to jets extracted from the 2011 CMS Open Data~\cite{CMS:JetPrimary2011A,Komiske:2019jim,MOD:ZenodoCMS}.
Jets are divided by pseudorapidity into forward and central datasets, which can be disentangled into ``quark'' and ``gluon'' categories using various topic modeling procedures.
We compare the quark and gluon fractions obtained from different jet observables, including those based on machine learning~\cite{Komiske:2018cqr}, and emphasize the relative statistical uncertainties of the different approaches.
Using our most robust topic modeling method---based on analyzing the endpoints of the receiver operator characteristic (ROC) curve of a machine-learned classifier---we then present various quark and gluon observables, including their intrinsic dimensionality~\cite{Grassberger:1983zz,CAMASTRA20032945,NIPS2002_1177967c,Komiske:2019fks}.
As a cross check, we compare results from the CMS Open Data to those obtained from the parton shower generator \Pythia 6.4.25 \cite{Sjostrand:2006za}. 
\cor{This \Pythia~6 sample was provided as part of the CMS Open Data release.}

Because the jet substructure response of the CMS detector~\cite{Chatrchyan:2008aa} depends on the jet pseudorapidity, we find that detector effects have a big impact on topic modeling.
Specifically, a necessary condition for topic modeling is \emph{sample independence}, which means that the underlying quark and gluon distributions must be the same in each mixed jet sample.
Though a complete study of experimental systematics is beyond the scope of this work, we do implement \emph{central value unfolding} using \OmniFold~\cite{Andreassen:2019cjw,Andreassen:2021zzk}, which mitigates these detector effects.
Weights from this unfolding are publicly available at \Ref{MOD:ZenodoCMSOmniFold}.
%
%When applied to unfolded data, the results of topic modeling on the CMS Open Data are comparable to those from \Pythia, after accounting for differences in the quark/gluon fractions.
%
\cor{The results of topic modeling on unfolded particle-level CMS Open Data are comparable to those from Pythia, after accounting for differences in the quark/gluon fractions.}
%
%\sk{After reading the last sentence several times I see why it is confusing for the referee. I don't think we explicitely compare the results produced by the presented method for Unfolded data and for plain Pythia. So now I am not sure what this sentence supposed to mean too.}

Our two new topic modeling algorithms show better robustness to statistical noise and training imperfections than the \emph{anchor bin} method used in \Refs{Metodiev:2018ftz,Komiske:2018vkc,ATLAS:2019rqw,radha_phdthesis}.
Anchor bins correspond to maximally pure regions of phase space with very high background rejection but correspondingly low signal efficiency.
These anchor bins therefore exhibit large statistical uncertainties due to low count rates.
To mitigate these uncertainties, we first introduce a \emph{log-likelihood ratio fit} procedure based on probability distribution functions and quantile binning.
We then show how to sidestep binning through a \emph{ROC curve fit} defined by cumulative distribution functions.
Using a sample of 2 million jets, we find consistent results from all three methods, but better stability from the two fitting procedures.

The remainder of this paper is organized as follows. 
In \Sec{sec:theory}, we discuss the motivation behind our new topic modeling methods from a statistical perspective and introduce the machine learning models we use.
We discuss the features of the CMS Open Data dataset in \Sec{sec:cmsod}, and we justify the application of central value unfolding to mitigate sample dependence.
In \Sec{section:extracting_k}, we apply our topic modeling procedures to the CMS Open Data dataset and compare the extracted reducibility factors.
We then present separate quark and gluon distributions in \Sec{sec:quark_and_gluon_dist}, including ROC curves, rapidity spectra, and the intrinsic dimensionality of quark and gluon jets, which are our main results.
We conclude in \Sec{sec:conc} with a discussion of possible extensions of these methods.

%%%%%%%%%%%%%%%%%%%%%%%%%%%%%%%%%%%%%%%%%%%%%%%%%%%%%%%%%%%%%%%%%%%%%%%%%%

\section{Jet Topic Modeling} \label{sec:theory}

In this section, we review the framework of jet topic modeling as well as the anchor bin method used in previous studies.
We then introduce our log-likelihood ratio fit (L-fit) and ROC curve fit (R-fit) methods, which exhibit better stability.
We then discuss machine-learned classification and regression as inputs to the topic modeling procedure.

\subsection{Review of Mixtures and Reducibility Factors}
\label{subsection:reduce_coef}

The following discussion uses the notation and terminology from the original jet topics paper \cite{Metodiev:2018ftz}, which is based on the \textsc{Demix} algorithm of \Ref{katz-samuels_decontamination_2019}.
Assume you have two mixed samples of jets, denoted as $M_1$ and $M_2$, which can be decomposed into their quark ($q$) and gluon ($g$) components.
This means that for some jet observable $\figO$, the observable distributions for the two mixtures can be written as:
\begin{align}
\label{eq:mixture_M1}
p_{M_1} (\figO) &= f_1 \, p_{q} (\figO) + (1 - f_1) \, p_{g} (\figO), \\
\label{eq:mixture_M2}
p_{M_2} (\figO) &= f_2 \, p_{q} (\figO) + (1 - f_2) \, p_{g} (\figO),
\end{align}
where $p_{q} (\figO)$ and $p_{g} (\figO)$ are the quark and gluon distributions for $\figO$, and $f_1 \in [0,1]$ and $f_2 \in [0,1]$ are the quark fractions in $M_1$ and $M_2$, respectively.
We further assume that the mixtures are different and take $f_1 > f_2$, such that $M_1$ is more quark enriched than $M_2$.

The above equations are valid under the assumption of \emph{sample independence}.
This means that the observable distributions for quark and gluon jets are the same in the phase space regions defined by $M_1$ and $M_2$, such that the two mixtures differ only by their quark fractions.
In principle, soft color correlations and non-perturbative effects can introduce sample dependence for quarks and gluons, though the analysis in \Ref{Komiske:2018vkc} found this to be a small effect (see also \cor{\Refs{ATLAS:2014vax,Bright-Thonney:2018mxq}}).
As we will see in \Sec{sec:sample_dependence}, detector effects will introduce large sample dependence, which we must mitigate for our analysis.

Making the further assumption that quarks and gluons are \emph{mutually irreducible} in $\figO$ (see \Eq{eq:quark_gluon_redicibility} below), we can invert \Eqs{eq:mixture_M1}{eq:mixture_M2} and solve for the quark and gluon distributions.
Defining the likelihood ratio
\begin{equation}
\label{eq:lratio}
L_{A/B}(\figO)  = \left(L_{B/A}(\figO) \right)^{-1} = \frac{p_{A}(\figO)}{p_{B}(\figO)},
\end{equation}
the \emph{reducibility factors} are the maximum amount of one mixture that can be subtracted from the other and still leave a valid probability density:
\begin{equation}
\label{eq:kappas}
\ka_{21} \equiv \min_{\figO}\, L_{M_2/M_1}\left(\figO\right), \quad \ka_{12} \equiv \min_{\figO}\,L_{M_1/M_2}\left(\figO\right),
\end{equation}
In particular, the quark and gluon distributions are given by
\begin{align}
\label{eq:p_q}
p_q(\figO) & =\frac{p_{M_1}(\figO)-\ka_{12}\, p_{M_2}(\figO)}{1-\ka_{12}},\\
\label{eq:p_g}
p_g(\figO) &=\frac{p_{M_2}(\figO)-\ka_{21}\, p_{M_1}(\figO)}{1-\ka_{21}},
\end{align}
which are non-negative distributions by construction.

We emphasize that the mutual irreducibility of quarks and gluons is an assumption, but a powerful one since it allows us to operationally define what we mean by ``quarks'' and ``gluons'' \cite{Komiske:2018vkc}.
To state the assumption more explicitly, we can define the reducibility factors for the quark and gluon distributions themselves as:
\begin{equation}
\label{eq:kappasqg}
\ka_{gq} \equiv \min_{\figO} \left(L_{g/q}\left(\figO\right)\right), \quad \ka_{qg} \equiv \min_{\figO}{\left(L_{q/g}\left(\figO\right)\right)}.
\end{equation}
With this notation, mutual irreducibility corresponds to the assumption that
\begin{equation}
\label{eq:quark_gluon_redicibility}
\ka_{gq} = \ka_{qg}  = 0.
\end{equation}
In various limits of quantum chromodynamics, one can prove that this relation holds \cite{Metodiev:2018ftz,Larkoski:2019nwj,Stewart:2022ari}, which is why we are motivated to use it for our studies, despite its potential limitations \cite{Gras:2017jty}.

A useful consequence of \Eq{eq:quark_gluon_redicibility} is that we can derive the quark fractions for $M_1$ and $M_2$ by plugging \Eqs{eq:p_q}{eq:p_g} into \Eqs{eq:mixture_M1}{eq:mixture_M2}, yielding:
\begin{equation}
\label{eq:fracs}
f_1=\frac{1-\ka_{12}}{1-\ka_{12}\,\ka_{21}},\quad f_2=\frac{\ka_{21}\left(1-\ka_{12}\right)}{1-\ka_{12}\,\ka_{21}}.
\end{equation}
Via \Eq{eq:fracs}, we can extract the quark fractions in a given dataset only knowing $\ka_{21}$ and $\ka_{12}$.
Via \Eqs{eq:p_q}{eq:p_g}, we can also extract the distributions of quark and gluon jets for any observable, even one that was not used to extract the reducibility factors in the first place.
In this way, jet topic modeling boils down to determining the reducibility factors in \Eq{eq:kappas}, and we present three different strategies for their extraction in the next three subsections.

%%%%%%%%%%%%%%%%%%%%%%%%%%%%%

\subsection{Anchor Bin Method}
\label{subsection:bin_methods}

The anchor bin method is the most direct interpretation of \Eq{eq:kappas} and was used in \Refs{Metodiev:2018ftz,Komiske:2018vkc,ATLAS:2019rqw,radha_phdthesis}.
If one only has access to finite samples drawn from the probability densities $p_{M_1}(\figO)$ and $p_{M_2}(\figO)$, then one needs some way to estimate the likelihood ratio.
One way to do so is to bin the data into histograms, yielding the probabilities $p_{M_1}(\figO_i)$ and $p_{M_2}(\figO_i)$, where $i$ is the bin number.
The anchor bin method therefore depends on the choice of binning, as discussed further in \Sec{sec:anchor}.

Because it is more convenient to work with log-likelihood ratios, we define the \emph{anchor values} as:
\begin{equation}
\label{eq:anchors}
a_{21} \equiv \min_{i}\, \ln L_{{M_2}/{M_1}}(\figO_i), \quad a_{12} \equiv \max_{i}\, \ln L_{{M_2}/{M_1}}(\figO_i).
\end{equation}
We refer to the corresponding arguments of the minimum/maximum as the \emph{anchor bins}. 
The reducibility factors are then given by
\begin{equation}
\label{eq:kappasanchor}
\ka_{21} = \exp[a_{21}], \quad \ka_{12} = \exp[-a_{12}].
\end{equation}
In practice, \Eq{eq:anchors} is modified to account for statistical uncertainties as described in \Sec{sec:anchor}.
With the reducibility factors in hand, we can use \Eq{eq:fracs} to extract the quark fractions and \Eqs{eq:p_q}{eq:p_g} to extract the quark and gluon distributions for any observable.

The main advantage of the anchor bin method is its conceptual simplicity, since it is just a binned version of \Eq{eq:kappas}.
The main drawback is that it depends sensitively on the end points of the probability densities, where bin counts can be low.
The next two methods aim to mitigate this issue by taking into account information from the whole distribution.

%%%%%%%%%%%%%%%%%%%%%%%%%%%%%%%%

\subsection{Log-Likelihood Ratio Fit Method}
\label{subsection:l-fit}

The log-likelihood ratio fit (L-fit) method is a more statistically robust way to extract reducibility factors.
Like the anchor bin method, it depends on the choice of histogram binning.
Unlike the anchor bin method, it takes advantage of data from the whole observable domain, not only from the neighborhoods around the maximum and minimum values.

After estimating the log-likelihood ratio $\ln L_{M_2/M_1}$, we fit it to a polynomial of degree $K$:
\begin{equation} \label{eq:lfit_fit_form}
	f(\figO; b) = \sum_{k = 0}^{K} b_k \, f_k(\figO),
\end{equation}
where $b_k$ are fit coefficients, and $f_k(\figO)$ is a polynomial of degree $k$ from some polynomial basis.
To avoid sensitivity to possible changes of variables, we find it convenient in \Sec{sec:lfit} to characterize $\figO$ in terms of its quantiles (strictly speaking, fractiles), such that the argument of $f$ ranges from 0 to 1.
Quantile binning makes it such that different observables have similar shapes to their log-likelihood ratios, simplifying the fitting procedure.

Assuming that the function $f(\figO; b)$ faithfully represents $L_{M_2/M_1}(\figO)$ when $b = b^\text{opt}$, the anchor values can be extracted via:
\begin{equation}
\label{eq:lfitanchors}
a_{21} = \min_\figO f(\figO; b^\text{opt}), \quad a_{12} = \max_\figO f(\figO; b^\text{opt}), 
\end{equation}
and the reducibility factors are given by \Eq{eq:kappasanchor} as before.
In the special case that the quark/gluon likelihood ratio is a monotonic function of the observable $\figO$, with smaller values corresponding to more quark-like, \Eq{eq:lfitanchors} simplifies to
\begin{equation}
\label{eq:lfitanchors2}
a_{21} = f(\figO_\text{min}; b^\text{opt}), \quad a_{12} = f(\figO_\text{max}; b^\text{opt}).
\end{equation}
where $\figO_\text{min} = \min (\figO)$ and $\figO_\text{max} = \max (\figO)$.
When characterizing $\figO$ in terms of its quantiles, $\figO^\text{quant}_\text{min} = 0$ and $\figO^\text{quant}_\text{max} = 1$.

The main advantage of the L-fit method is that it allows for a better accounting of statistical noise and is less dependent on binning than the anchor bin method.
The main drawback is that it has residual binning dependence, especially if one does not use the quantile parametrization of the observable.
It also depends on the precise choice of functional fit form, though this dependence can be assessed as an uncertainty in the method.

%%%%%%%%%%%%%%%%%%%%%%%%%%%%%%%

\subsection{ROC Curve Fit Method}
\label{subsection:roc}

The ROC curve fit (R-fit) method allows us to extract the reducibility factors without needing to bin the observable.
The method assumes that the quark/gluon likelihood ratio is a monotonic function of the observable, which will be the case for the observables studied in this paper.

The R-fit method starts from the cumulative probabilities $P(\figO_{\rm cut})$, defined as
\begin{equation}
P(\figO_{\rm cut}) = \int_{\figO_{\rm min}}^{\figO_{\rm cut}} d\figO \, p(\figO),
\end{equation}
where $\figO_{\rm min}$ is the minimum value of the observable.
For the mixtures $M_1$ and $M_2$, the ROC curve is the parametric curve traced out by the value of the upper cut $\figO_{\rm cut}$:
\begin{equation}
\label{eq:mix_ROC}
\{P_{M_1}(\figO_{\rm cut}), P_{M_2}(\figO_{\rm cut})\}.
\end{equation}
Assuming that the ROC curve is single valued, it tells us the efficiency for selecting the ``background'' ($M_2$, more gluon-like) as a function of the efficiency for selecting the ``signal'' ($M_1$, more quark-like):
\begin{equation}
P_{M_2} (P_{M_1}).
\end{equation}
In the statistics language, this is the false positive rate as a function of the true positive rate.%
\footnote{This definition of the ROC curve is standard in the HEP literature.
In the statistics literature, it is more common to plot the true positive rate as a function of the false positive rate.}

Since the cumulative probabilities $P_{M_1}(\figO_{\rm cut})$ and $P_{M_2}(\figO_{\rm cut})$ contain the same information as the probability densities, we can use them to extract the reducibility factors $\kappa_{21}$ and $\kappa_{12}$.
Perhaps less obvious is that the ROC curve built from $P_{M_1}$ and $P_{M_2}$ contains sufficient information for this extraction.
Assuming the quark/gluon likelihood ratio is a monotonic function of the observable, then as discussed around \Eq{eq:lfitanchors2}, the anchor bins are located at the end points of the observable.
One endpoint has $P_{M_1} \to 0$ and $P_{M_2} \to 0$, and the slope of the ROC curve is:
\begin{align}
\nonumber
\lim_{P_{M_1} \to 0} \frac{d P_{M_2}}{d P_{M_1} }& = \lim_{\figO_{\rm cut} \to \figO_{\rm min}} \frac{d P_{M_2}}{d \figO_{\rm cut}} \frac{d \figO_{\rm cut}}{d P_{M_1} }\\
&= \frac{p_{M_2}(\figO_{\rm min})}{p_{M_1}(\figO_{\rm min})} = \kappa_{21},
\label{eq:Rfit_kappa21}
\end{align}
where we have used the fundamental theorem of calculus in the second line.
Similarly, the other endpoint has $P_{M_1} \to 1$ and $P_{M_2} \to 1$, and the slope of the ROC curve is:
\begin{equation}
\lim_{P_{M_1} \to 1} \frac{d (1-P_{M_2})}{d (1- P_{M_1})} = \frac{p_{M_2}(\figO_{\rm max})}{p_{M_1}(\figO_{\rm max})} = \frac{1}{\kappa_{12}}.
\label{eq:Rfit_kappa12}
\end{equation}
We emphasize that these relations assume the monotonicity of the (log-)likelihood ratio as a function of the observable.
In this way, the slopes of the ROC curve at the endpoints directly yield the reducibility factors.

With finite statistics datasets, we can fit the ROC curve to a polynomial and extract the slopes from the obtained functional fit.
To ensure the fit intersects the points $\{0, 0\}$ and $\{1, 1\}$, we use the form:
\begin{equation}
	f(r)=r+r\left(1-r\right)\left(\sum_{k=0}^{K}{b_k \,f_k(r)}\right),
	\label{eq:fit_form_roc}
\end{equation}
where $b_k$ are the fit coefficients and $f_k\left(x\right)$ are polynomials of degree $k$ defined on the interval $[0, 1]$.

Like the L-fit method, the R-fit method is robust to statistical noise since it uses information from the whole observable range.
That said, the procedure for accounting for statistical uncertainties with the R-fit method is more involved, as discussed in \Sec{sec:roc}.
The main caution when using this method is that it relies on the quark/gluon likelihood ratio being a monotonic function of the observable.
%
%Fortunately, this turns out to be the case for the machine-learned classifiers that we use for our default reducibility factor extraction.
\cor{Fortunately, this is a quite common property (for example, constituent multiplicity and mass), and this also turns out to be the case for the machine-learned classifiers that we use for our default reducibility factor extraction.}

%%%%%%%%%%%%%%%%%%%%

\subsection{Machine-Learned Classification}
\label{subsection:classification}

Jet topic modeling can be applied to any observable, but as advocated in \Ref{Komiske:2018vkc}, it is particularly powerful when applied to a machine-learned classifier.
As reviewed below, such classifiers are monotonically related to the likelihood ratio $L_{g/q}(\figO)$ in the asymptotic limit, and therefore well suited for extracting reducibility factors.

Using the classification without labels (CWoLa) technique~\cite{Metodiev:2017vrx}, we can train a classifier to distinguish $M_1$ and $M_2$, and this yields a function that defines optimal decision boundaries between quark and gluon jets.
For concreteness, we use the binary cross entropy loss functional to train a classifier $c(x)$ over the full phase space $x$:
\begin{multline}
L[c] = \int dx \Big(p(M_1) \, p_{M_1}(x) \,  \ln c(x) \\
~+  p(M_2) \, p_{M_2}(x) \, \ln [1 - c(x)] \Big), 
\end{multline}
where $p(M_a)$ is the proportion of the events drawn from sample $a$.
This functional is minimized when
\begin{equation}
\label{eq:BinCrossEnt}
\frac{1 - c(x)}{c(x)} = \frac{p(M_2)}{p(M_1)} \, L_{M_2 / M_1}(x),
\end{equation}
so $c(x)$ is indeed a monotonic function of $L_{M_2 / M_1}$ in the limit of infinite training data.
Following \Ref{Metodiev:2017vrx}, it is straightforward to show that it is also a monotonic function of $L_{g/q}$.
Therefore, the classifier $c(x)$ is a suitable observable for extracting reducibility factors using the L-fit method with \Eq{eq:lfitanchors2} or using the R-fit method with \Eqs{eq:Rfit_kappa21}{eq:Rfit_kappa12}.

The specific classifiers we use for our study are energy flow networks (EFNs) and particle flow network (PFNs) \cite{Komiske:2018cqr}, based on the Deep Sets formalism~\cite{zaheer2017deep}.
The distinctive feature of EFNs is that they have an infrared-and-collinear-safe (IRC-safe) latent space:
\begin{equation}
	c(x) = F \left( \sum_{i} z_i \Phi(\hat{p}_i) \right),
\end{equation}
where the sum is taken over all the particles in the jet, $z_i$ is the energy of particle $i$, and $\hat{p}_i$ is the direction of particle $i$.
For a latent space of dimension $\ell$, $\Phi$ is a per-particle trainable function with $\ell$ outputs, and $F$ is a per-jet trainable function with $\ell$ inputs.
Relaxing the requirement of IRC safety yields PFNs:
\begin{equation}
	\label{eq:PFN_def}
	c(x) = F \left( \sum_{i} \Phi(p_i) \right),
\end{equation}
where $p_i$ is the full information about particle $i$, including possible charge/flavor labels.
Both EFNs and PFNs can handle variable sized inputs and respect particle permutations, both of which are desirable properties for analyzing jets.
For the observables in \Sec{subsec:observables}, we omit charge/flavor information, since otherwise we would be sensitive to the relative fraction of up-type versus down-type quarks as a function of jet rapidity.

%%%%%%%%%%%%%%%%%%%%%%%%%%%%%%%%%%%%%%%%%%%%%%%%%%%%%%%%%%%%%%%%%%%%%%%%%%%

\subsection{Machine-Learned Regression}
\label{subsection:regressions}

CWoLa is actually a special case of a more general strategy to extract discriminants from mixed training data.
Instead of defining two jet mixtures $M_1$ and $M_2$ and performing binary classification, we can instead label jets via a continuous parameter $\eta$ and regress $\eta$ on the jet properties.
This strategy requires that the fraction of quark and gluon jets change as a function of $\eta$, but the properties of the jets themselves are unaffected.
\cor{(While $\eta$ can be any observable that satisfies the requirement above, in the scope of this paper we will constrain ourself to $\eta$ be a jet pseudorapidity.)}
This implies the factorized probability structure:
\begin{equation}
\label{eq:factorized_eta}
p(\eta, x) = f \, p_q(\eta) \, p_q(x) + (1-f) \, p_g(\eta) \, p_g(x),
\end{equation}
where $0< f < 1$ is the total quark fraction of the sample, and the probability densities in the above expression each integrate to 1.

We perform regression with the mean squared error loss functional:
\begin{equation} \label{eq:loss}
	L[h] = \int d\eta \, dx \, p(\eta,x) \, \big(\eta - h(x)\big)^2,
\end{equation}
where $h(x)$ is the regression function to be learned.
This loss functional is minimized for
\begin{equation}
\label{eq:opt_r}
	h(x) = \frac{\int d\eta \, \eta  \, p(\eta,x)}
		{\int d\eta \, p(\eta,x)},
\end{equation}
such that $h(x)$ learns the expectation value of $\eta$ as a function of $x$.

To see that $h(x)$ is monotonically related to $L_{g/q}$, we can plug \Eq{eq:factorized_eta} into \Eq{eq:opt_r}:
\begin{equation}
r(x) = \frac{f \vev{\eta}_q + (1-f) \vev{\eta}_g \, L_{g/q} (x)}{f + (1-f) \, L_{g/q} (x)},
\end{equation}
where $\vev{\eta}_a = \int d\eta \,\eta \, p_a(\eta)$.
As long as $\vev{\eta}_q$ and $\vev{\eta}_g$ differ, this is a monotonic function of $L_{g/q}$ and therefore defines optimal decision boundaries.

In the studies below, we show jet topics results using both classification and regression observables, where we find consistent results within uncertainties.
Note that if a change of variables is applied to $\eta$, then the extracted $h(x)$ will change by a Jacobian factor.
With finite training data, there is an optimal change of variables that minimizes the statistical uncertainty on the inferred $L_{g/q}(x)$, though we leave an investigation of that to future work.

%%%%%%%%%%%%%%%%%%%%%%%%%%%%%%%%%%%%%%%%%%%%%%%%%%%%%%%%%%%%%

\begin{table*}
\renewcommand\arraystretch{1.1}
\setlength{\tabcolsep}{12pt}
\begin{tabular}{r c c l}
\hline
\hline
Sample & Forward ($f_1$) & Central ($f_2$) & Equation\\
 \hline
 \hline
\Pythia Parton & 0.689 & 0.510 & \Eq{eq:frac_exp_pythia}\\
\Pythia Parton with CVU & 0.711 & 0.543 & \Eq{eq:quark_fractions_CVU} \\
CMS 2011 with \OmniFold: \textbf{CVU Jet Topics} & $\mathbf{0.708 \pm 0.025}$ & $\mathbf{0.561 \pm 0.033}$ &\Eq{eq:quark_fractions_PFN}\\
\hline
\Pythia Jet Topics & $0.649 \pm 0.012$ & $0.466\pm 0.011$ & \Eq{eq:quark_fractions_PFN_pythia}\\
\hline
\hline
\end{tabular}
\caption{Various quark fractions used in this analysis.
The first row correspond to parton labels, as determined by \Pythia.
The second row is also from \Pythia but includes weights from central value unfolding; this is a parton-level reference that can be compared to our jet topics result.
The third row (in bold) is the main result from our study, using jet topics to extract an operational definition of ``quark'' fractions in the central-value-unfolded CMS 2011 Open Data.
The fourth row, studied in \App{sec:pythia_only_analysis}, comes from performing the jet topics procedure on the \Pythia Monte Carlo samples.
}
\label{tab:fractions}
\end{table*}

\section{Jets in the CMS Open Data}
\label{sec:cmsod}

In this section, we describe the real and synthetic datasets used in our study, which derive from a public release by the CMS experiment~\cite{CMS:JetPrimary2011A}.
We list the observables used as inputs for topic modeling and demonstrate the challenge of sample dependence induced by the CMS detector.
Going further than previous studies with CMS Open Data, we perform full phase-space, central-value unfolding using the \OmniFold technique \cite{Andreassen:2019cjw,Andreassen:2021zzk}.
Our ability to validate the unfolding is limited due to our having only a single Monte Carlo (MC) dataset with CMS full detector simulation, but we check for detector-level closure in several distributions of interest and find that the unfolding works well at a qualitative level.

\subsection{Review of CMS 2011A Jet Primary Dataset}
\label{sec:jetprimary}

The CMS 2011 Open Data release corresponds to \SI{2.3}{fb^{-1}} of proton-proton collisions at center-of-mass energy $\sqrt{s}=\SI{7}{TeV}$~\cite{CMS2011Release}.
Related trigger streams~\cite{Khachatryan:2016bia} are grouped together into primary datasets, and our analysis is based on the \Jet Primary Dataset for CMS Run 2011A~\cite{CMS:JetPrimary2011A}, which includes single jet and dijet triggers.
Each event from CMS has full particle-flow reconstruction~\cite{CMS-PAS-PFT-09-001,CMS-PAS-PFT-10-001,Sirunyan:2017ulk}.
This dataset was extensively studied, processed, and simplified in \Ref{Komiske:2019jim}, which applied event geometry techniques based on the energy mover's distance \cite{Komiske:2019fks} to explore the space of jets in real collider data.
Our current study is based on the reprocessed ``MIT Open Data'' (MOD) formatted HDF5 files from \Ref{Komiske:2019jim}, which were publicly released on the \Zenodo platform~\cite{MOD:ZenodoCMS}.
See \Ref{Komiske:2022enw} for a recent analysis of multipoint correlators with this dataset.

The CMS 2011 release also has associated MC datasets for a variety of physics processes, simulated with a model of CMS based on \textsc{Geant} 4~\cite{Agostinelli:2002hh}.
Relevant for our analysis, there is a sample of hard QCD scattering events generated by \Pythia 6.4.25~\cite{Sjostrand:2006za} with tune Z2~\cite{Field:2011iq}.
These generated events are further processed using the CMS simulation and reconstruction software~\cite{CMS:QCDsim170-300,CMS:QCDsim300-470,CMS:QCDsim470-600,CMS:QCDsim600-800,CMS:QCDsim800-1000,CMS:QCDsim1000-1400,CMS:QCDsim1400-1800,CMS:QCDsim1800}.
Both generator-level truth particles and simulation-level reconstructed objects are available in the same MOD HDF5 format on \Zenodo~\cite{MOD:ZenodoMC170,MOD:ZenodoMC300,MOD:ZenodoMC470,MOD:ZenodoMC600,MOD:ZenodoMC800,MOD:ZenodoMC1000,MOD:ZenodoMC1400,MOD:ZenodoMC1800}.

For both the real and synthetic datasets, we keep up to \cor{the two jets in an event with the highest $p_T$}.
Jets are clustered by CMS using the anti-$k_t$ algorithm with $R = 0.5$~\cite{Cacciari:2008gp,Cacciari:2011ma}.
As a baseline, jets are selected if their transverse momentum and pseudorapidity satisfy:
\begin{equation}
p_T \in [475,525]~\GeV, \qquad |\eta| < 1.9.
\end{equation}
Jet substructure observables are computed using all truth particles (for generator-level data) or all particle flow candidates (PFCs, for simulation-level data).
For the measured dataset, we applied CMS-provided jet energy correction (JEC) factors~\cite{CMS:2016lmd} (which include area-median pileup subtraction~\cite{Cacciari:2008gn}) and imposed the ``medium'' jet quality criteria~\cite{CMS:2010xta,2011JInst...611002C} (quality $\geq 2$ in the MOD format).
For this simulated dataset, we only use events that correspond to generated parton-level $p_T$ greater than \SI{300}{GeV} and less than \SI{1800}{GeV}, divided into six independent event samples~\cite{MOD:ZenodoMC300,MOD:ZenodoMC470,MOD:ZenodoMC600,MOD:ZenodoMC800,MOD:ZenodoMC1000,MOD:ZenodoMC1400}.
Note that there is a very small tail of events from the $\hat{p}_T \in [170, 300]$ GeV and  $\hat{p}_T\ge1800$ GeV samples~\cite{MOD:ZenodoMC170,MOD:ZenodoMC1800} that satisfy our selection criteria but are ignored for this study.
Unlike the analysis in \Ref{Komiske:2019jim}, we place no further cuts on the $p_T$ values of the PFCs, nor do we separate out charged from neutral PFCs.

\subsection{Choice of Mixtures and Observables}
\label{subsec:observables}

To define the two mixtures for topic modeling, we split the $p_T \in [475,525]~\GeV$ jet sample into two mixtures of roughly equal sizes:
\begin{itemize}
\item Forward ($M_1$, quark-enriched):  $|\eta| \in [0.65,1.9]$,
\item Central ($M_2$, gluon-enriched):  $|\eta| < 0.65$.
\end{itemize}
A related strategy of considering the more forward/central jet was pursued by ATLAS in \Refs{ATLAS:2015rlw,ATLAS:2016vxz,ATLAS:2019rqw}.
As defined by \Pythia \cor{6.4.25} parton-level information, forward (central) jets have a quark fraction of:
\begin{equation}
	\label{eq:frac_exp_pythia}
	\text{\Pythia Parton:} \quad
	\begin{aligned}
		f_1 &\simeq 0.689, \\
		f_2 &\simeq 0.510.
	\end{aligned}
\end{equation}
This relatively modest difference in quark fraction is nevertheless sufficient for topic extraction.
For reference, we summarize the various quark fractions we encounter in this study in \Tab{tab:fractions}.

Following the analysis of \Ref{Komiske:2018vkc}, we consider a suite of six well-studied jet substructure observables as inputs to topic modeling:
\begin{itemize}
\item $N_{\rm const}$:  Number of jet constituents,
\item $N_{95}$:  Image activity~\cite{Pumplin:1991kc}; i.e.~number of pixels in a $33 \times 33$ jet image that contain 95\% of the jet $p_T$,
\item $p_T^D$:  Transverse momentum dispersion~\cite{CMS:2013kfa}; i.e.~$\sum_i p_{T,i}^2 / \left(\sum_i p_{T,i} \right)^2$,
\item $\tau_2$:  2-subjettiness  \cite{Thaler:2010tr, Thaler:2011gf} with $\beta = 1$ and $k_t$-axes,
\item $\tau_1$:  1-subjettiness with $\beta = 1$ and $k_t$-axes,
\item $m_{\rm jet}$:  Jet mass.
\end{itemize}
The first observable ($N_{\rm const}$) is known to be good quark/gluon discriminant.
The middle three observables ($N_{95}$, $p_T^D$, and $\tau_2$) yield moderate quark/gluon separation power.
The last two observables ($\tau_1$ and $m_{\rm jet}$) exhibit Casimir scaling in their quark/gluon separation~\cite{Larkoski:2013eya}, limiting their usefulness for jet topic modeling.

We also consider four machine-learned observables, as discussed in \Secs{subsection:classification}{subsection:regressions}:
\begin{itemize}
\item EFN$_{\rm cls}$:  classification of forward versus central jets with an EFN,  
\item PFN$_{\rm cls}$:  classification of forward versus central jets with a PFN,
\item EFN$_{\rm reg}$:  regression for jet pseudorapidity $|\eta|$ with an EFN,
\item PFN$_{\rm reg}$:  regression for jet pseudorapidity $|\eta|$ with a PFN.
\end{itemize}
For the PFNs, we input the full particle four-vector information, but exclude charge/flavor information since we do not wish to distinguish up-type from down-type quarks.
Apart from the input and output values, the network architectures are the same, with a latent space dimension of $\ell = 128$, 2 hidden layers of 100 nodes for the per-particle function $\Phi$, and 3 hidden layers of 100 nodes for the per-jet function $F$.
The activation functions are ReLU for classification and Leaky ReLU for regression.
Networks are built using the \textsc{EnergyFlow} package \cite{energyflow}  based on \textsc{Keras} \cite{chollet2015keras} and trained using the \textsc{Adam} optimizer~\cite{kingma2014adam}.

To avoid training complications associated with the real CMS datasets, we trained these four observables on the particle-level \Pythia 6.4.25 samples.
This is valid from the perspective of jet topics, since any observable (even a machine-learned observable trained on imperfect synthetic data) can be used as an input to the jet topics procedure.
\cor{Of course, such a suboptimal observable will not provide as good of an estimate of the true likelihood ratio, generically resulting in residual mixing between the categories.}
We checked that similar results could be obtained by training on the central-value-unfolded samples described below in \Sec{sec:cvu}, though the results were not as stable as we would have liked between different numbers of unfolding iterations.\cor{\footnote{\cor{Because topic modeling is sensitive to the very high and very low purity regions of phase space, instabilities can arise if the machine-learning algorithm picks up on small changes near the endpoints of the classifier.  These instabilities can be amplified in the presence of large weight variations, which we encounter in \Fig{fig:unfoldedweights} below.  It would be interesting to see if these instabilities could be reduced if the unfolding were explicitly regularized, as opposed to implicitly regularized by the number of unfolding iterations.}}}

\subsection{The Challenge of Sample Dependence}
\label{sec:sample_dependence}

\begin{figure*}[t]
\centering
\subfloat[]{\includegraphics[width=0.33\textwidth]{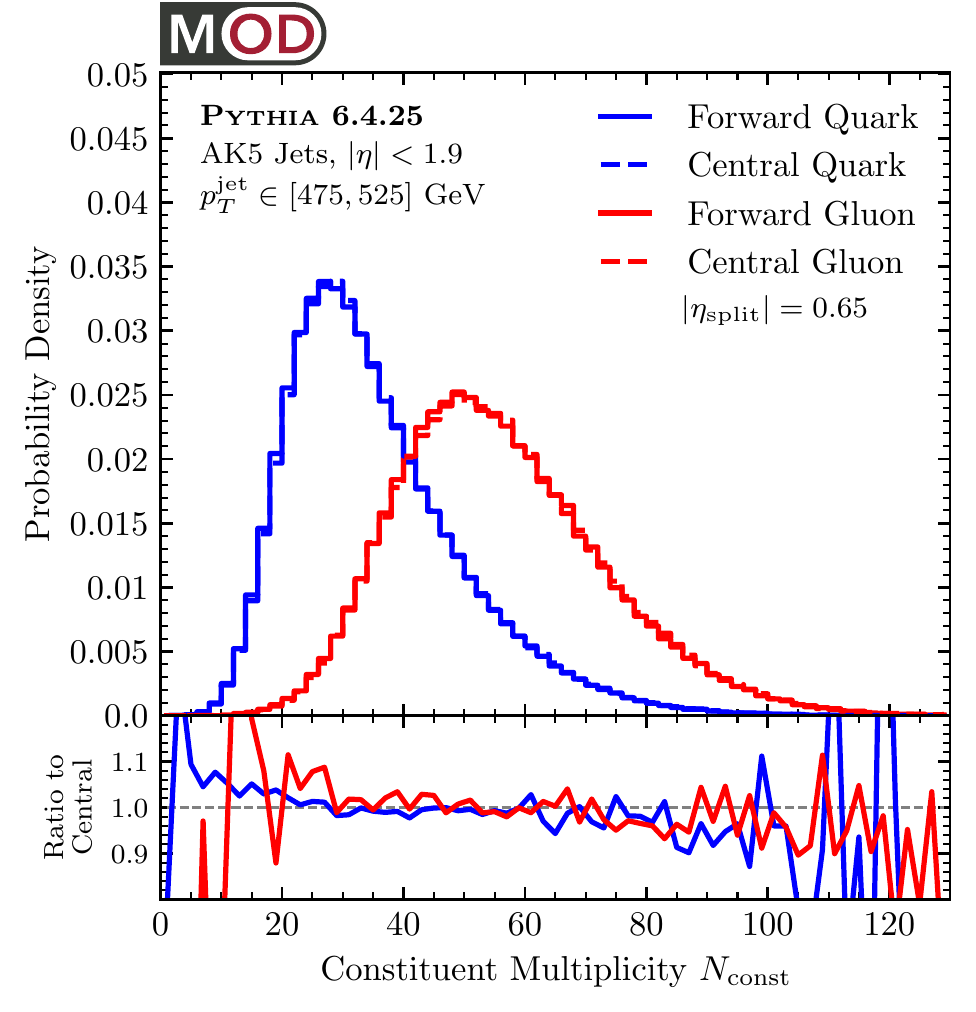}\label{fig:qq_and_gg_truth_mult}}
\subfloat[]{\includegraphics[width=0.33\textwidth]{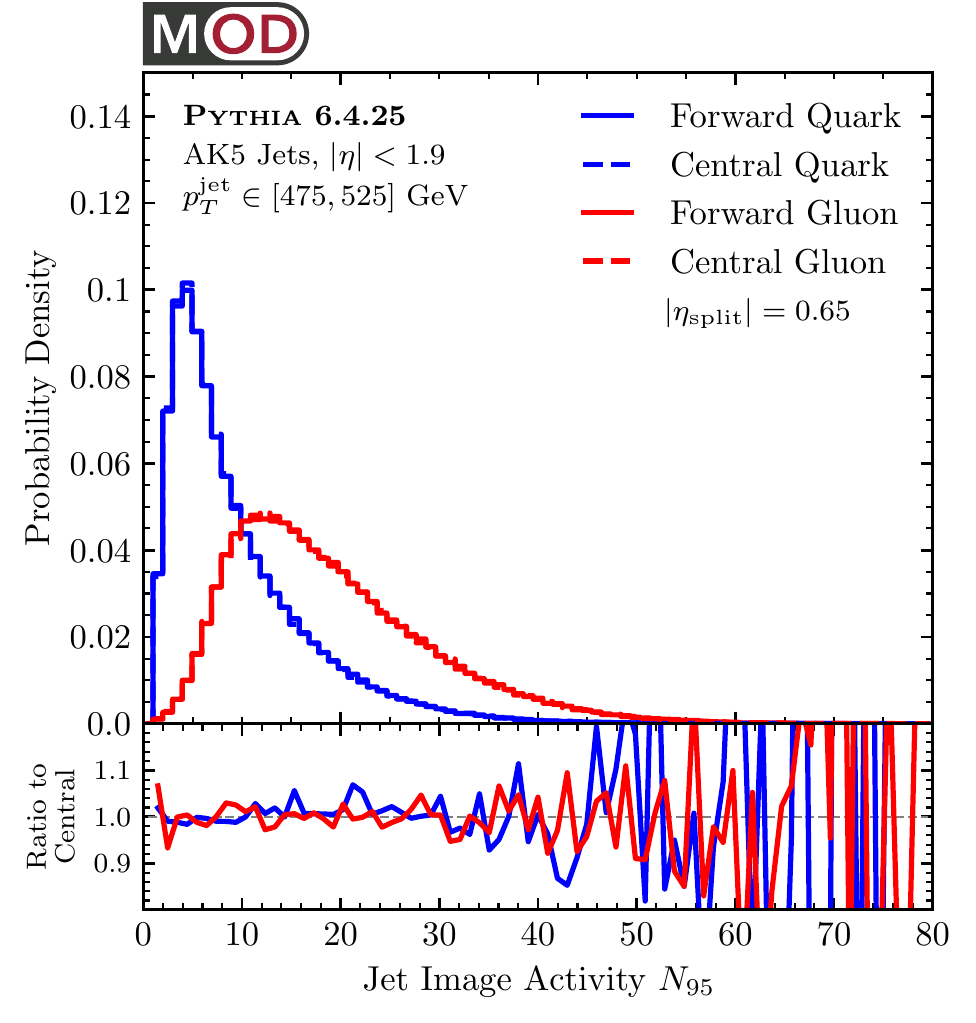}\label{fig:qq_and_gg_truth_n95}}
\subfloat[]{\includegraphics[width=0.33\textwidth]{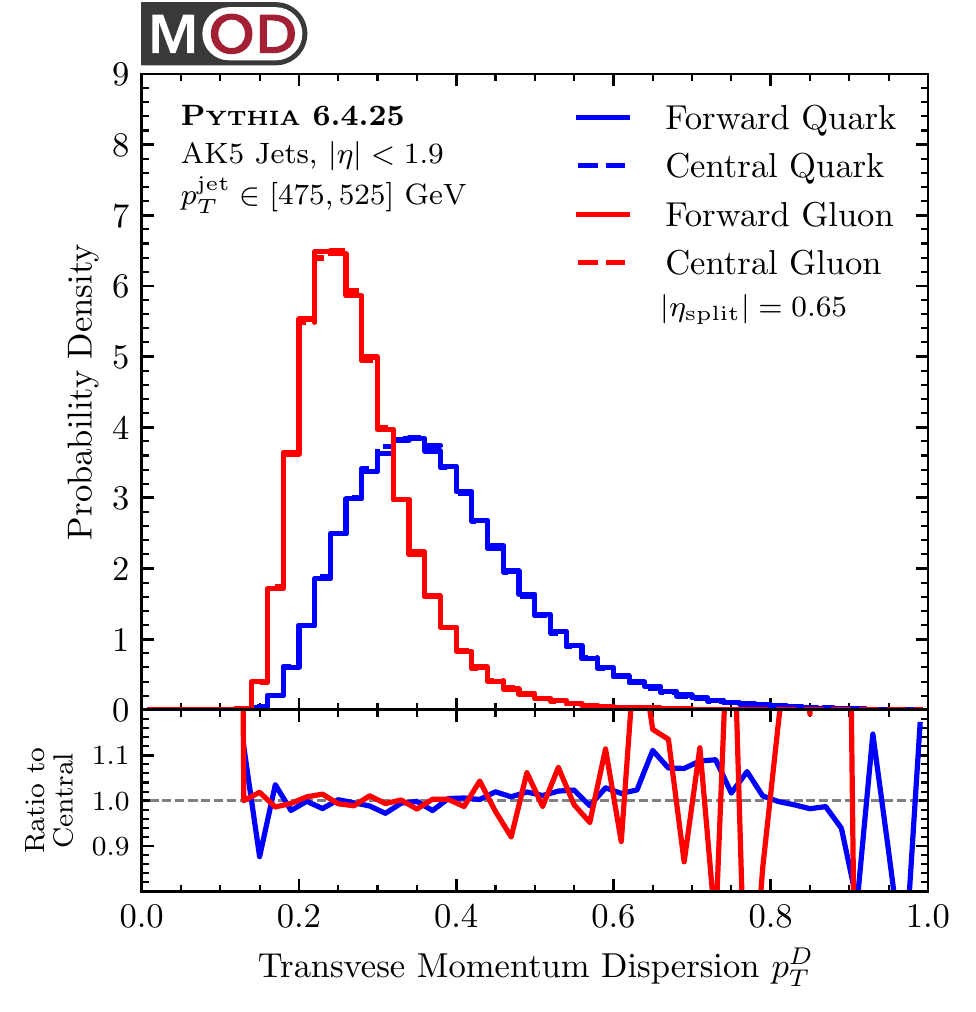}\label{fig:qq_and_gg_truth_ptd}}\\
\subfloat[]{\includegraphics[width=0.33\textwidth]{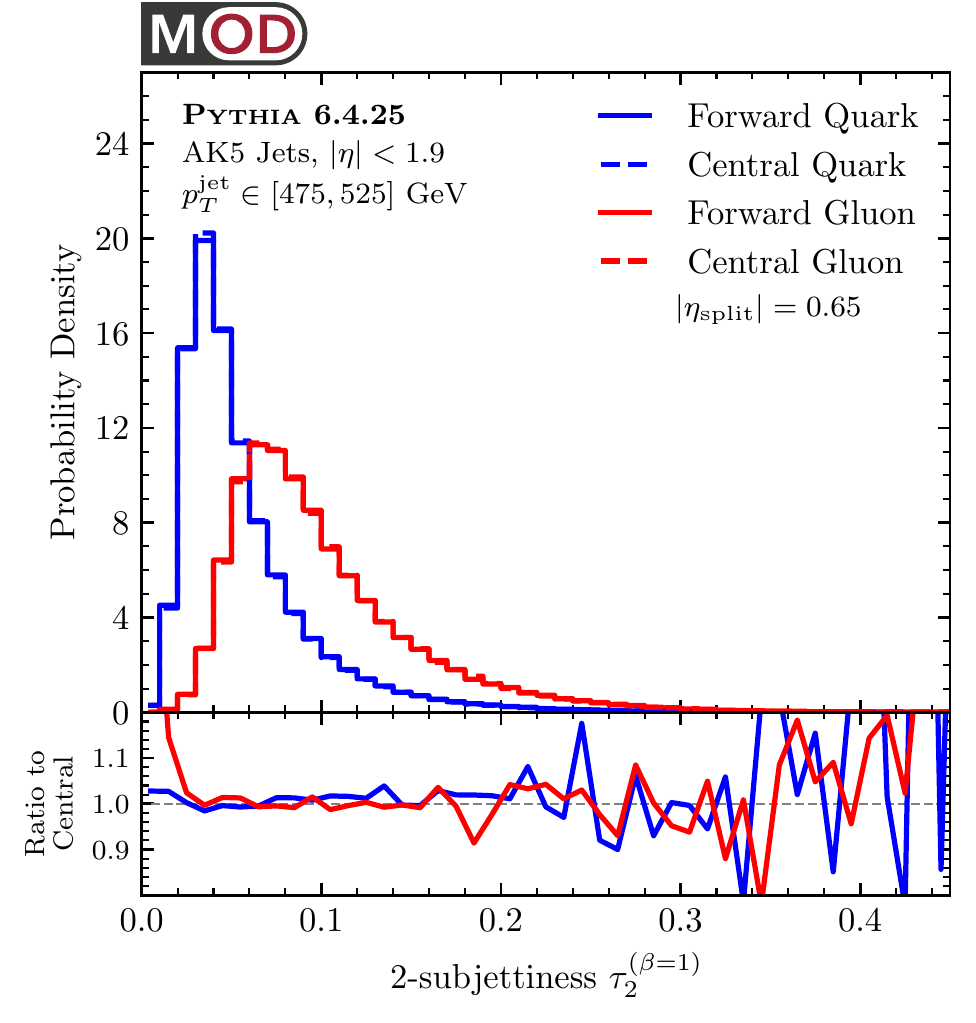}\label{fig:qq_and_gg_truth_nsub2}}
\subfloat[]{\includegraphics[width=0.33\textwidth]{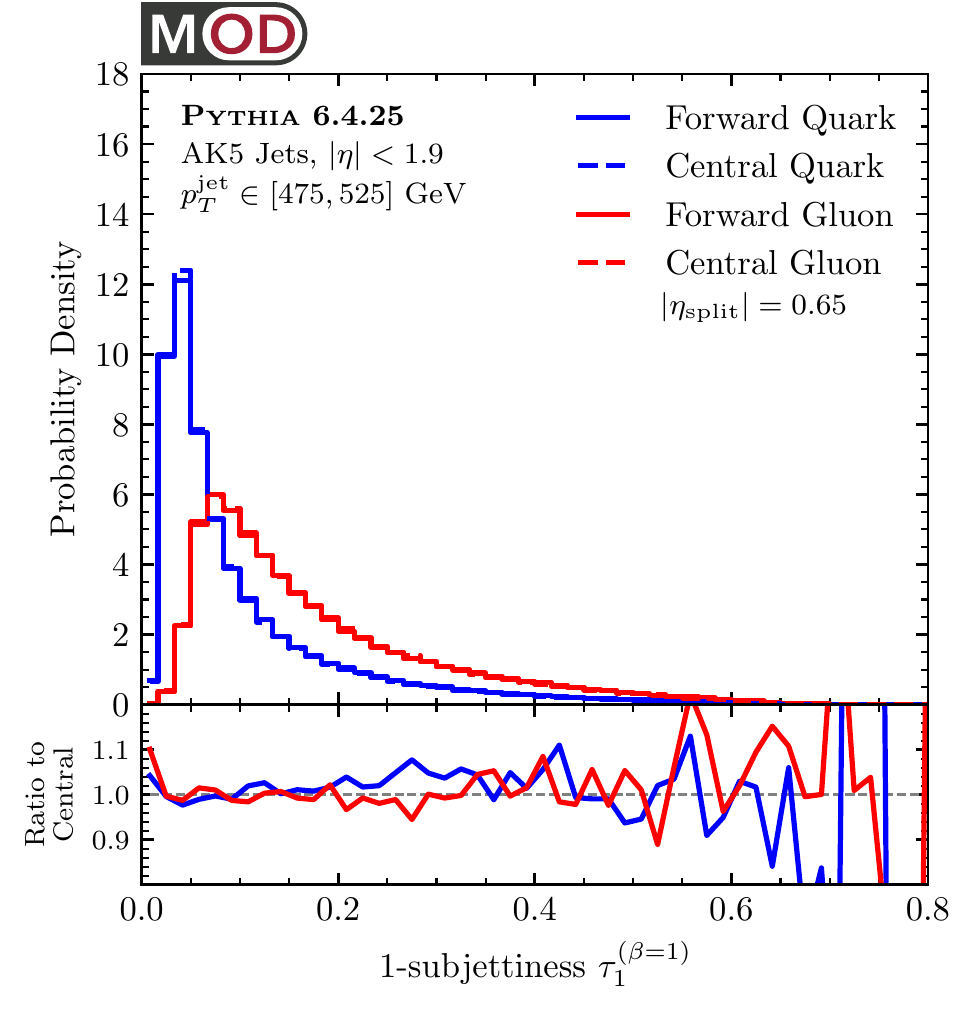}\label{fig:qq_and_gg_truth_nsub1}}
\subfloat[]{\includegraphics[width=0.33\textwidth]{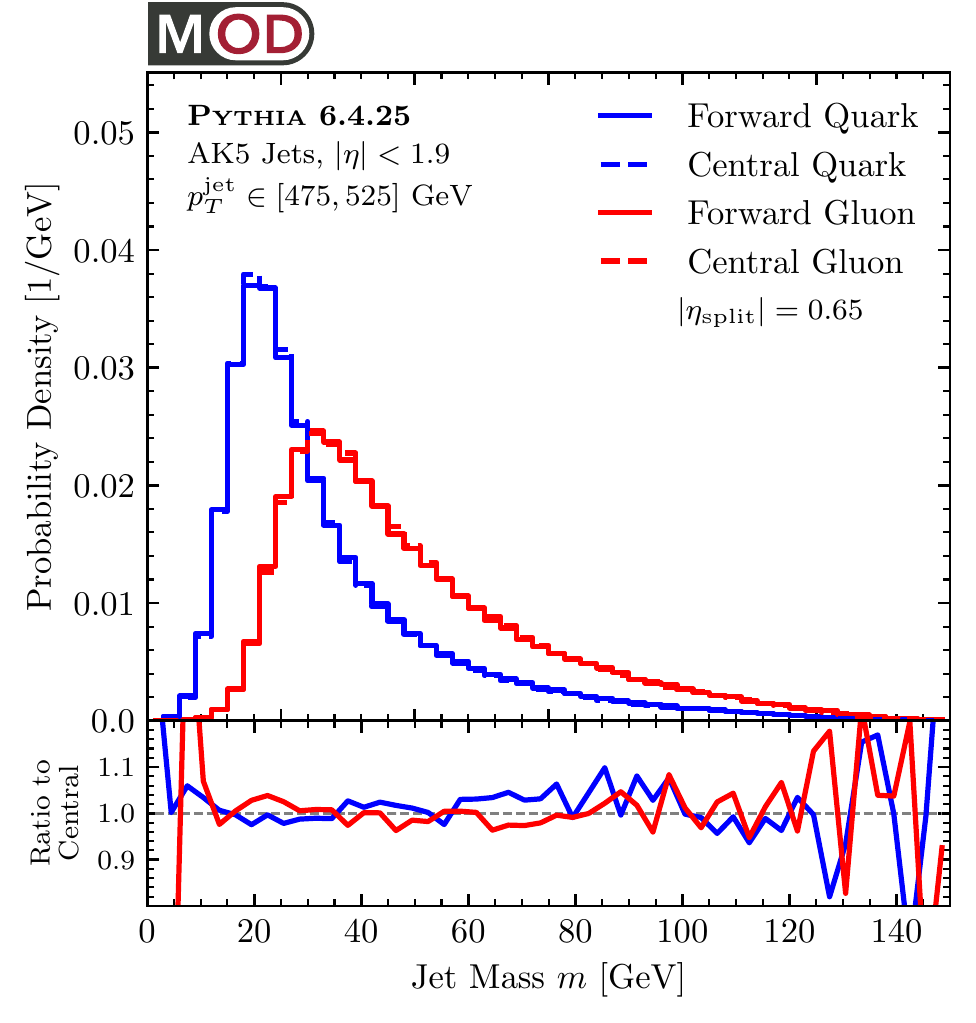}\label{fig:qq_and_gg_truth_mass}}
\caption{Distributions in \Pythia 6.4.25 for the six substructure observables from \Sec{subsec:observables}: (a) constituent multiplicity, (b) image activity, (c) momentum dispersion, (d) 2-subjettiness, (e) 1-subjettiness, and (f) jet mass.
Using parton truth labels from the \Pythia event record, the distributions for quark (blue) and gluon (red) are compared between forward jets (solid, $|\eta| \in [0.65,1.9]$) and central jets (dashed, $|\eta| < 0.65$).
As expected from leading-power factorization, the differences are modest such that sample independence holds to an excellent approximation.
}
\label{fig:qq_and_gg_truth_1}
\end{figure*}

\begin{figure*}[p]
\centering
\subfloat[]{\includegraphics[width=0.33\textwidth]{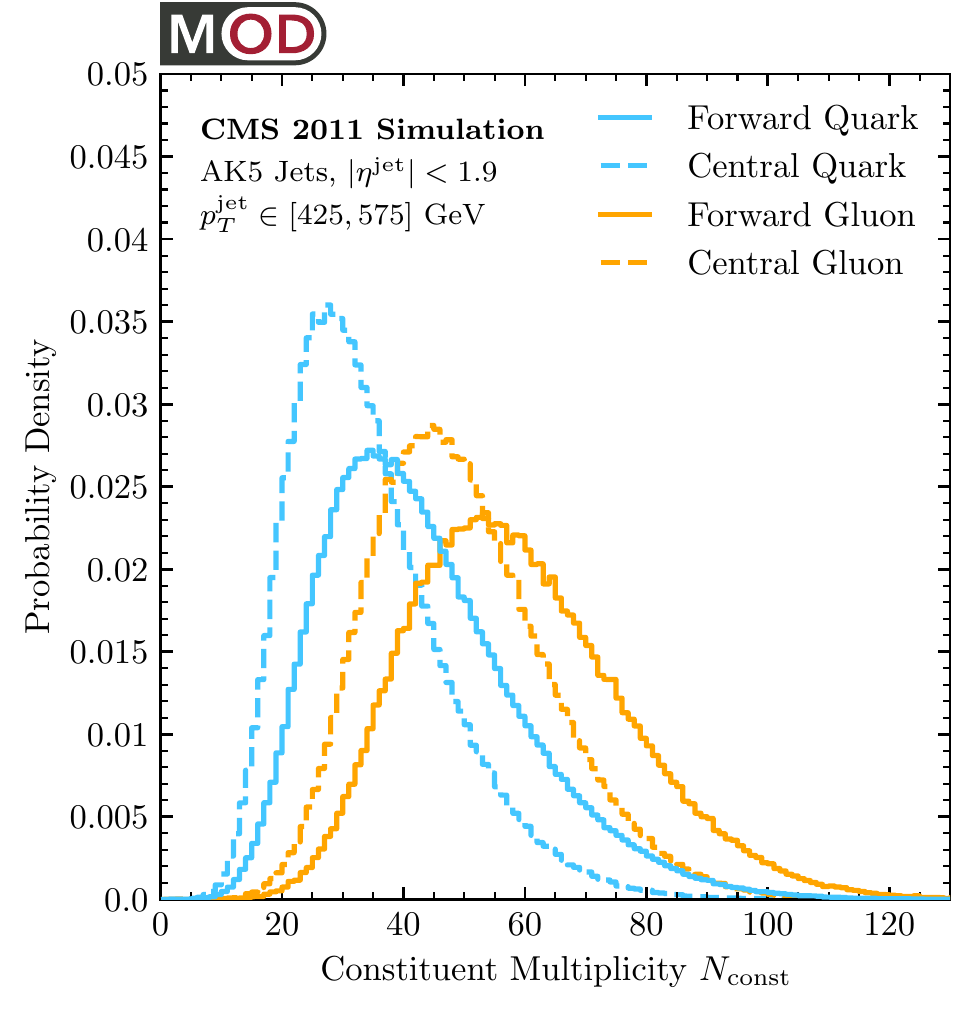}\label{fig:unfolding_mult_qq_gg}}
\subfloat[]{\includegraphics[width=0.33\textwidth]{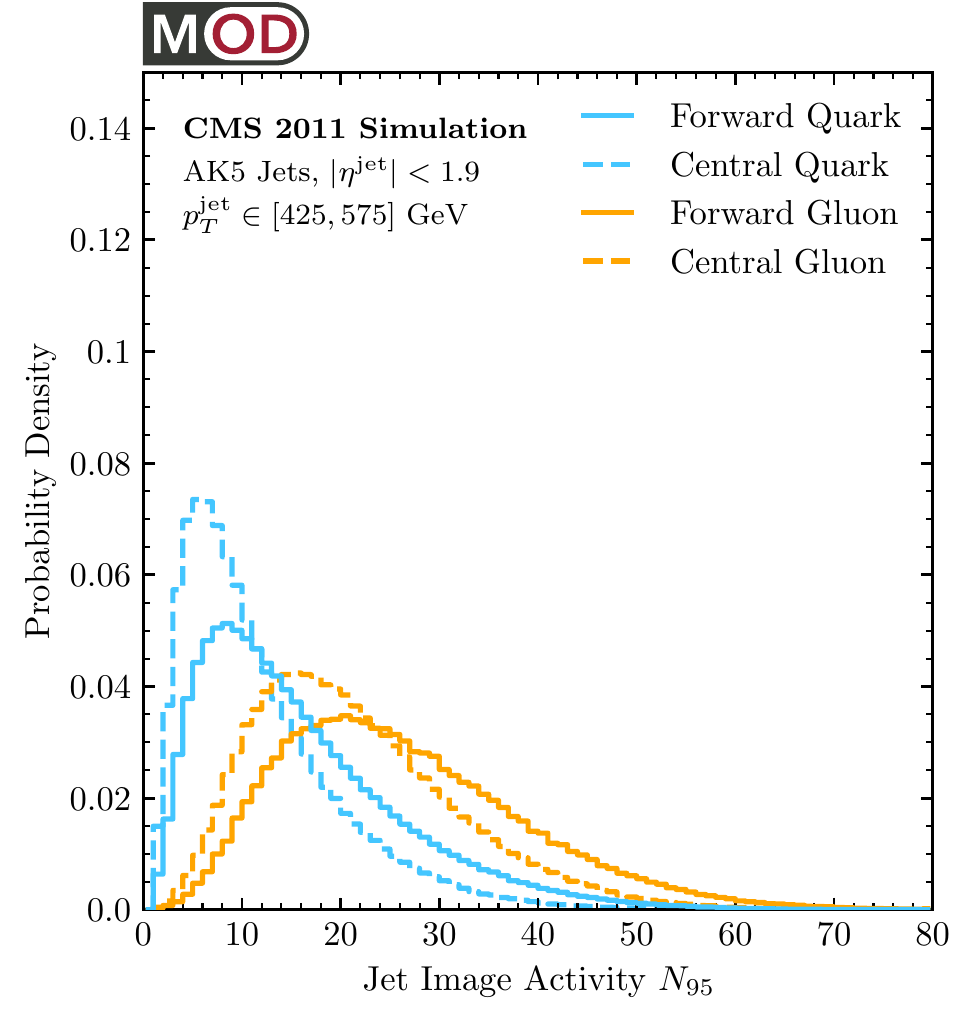}\label{fig:unfolding_n95_qq_and_gg}}
\subfloat[]{\includegraphics[width=0.33\textwidth]{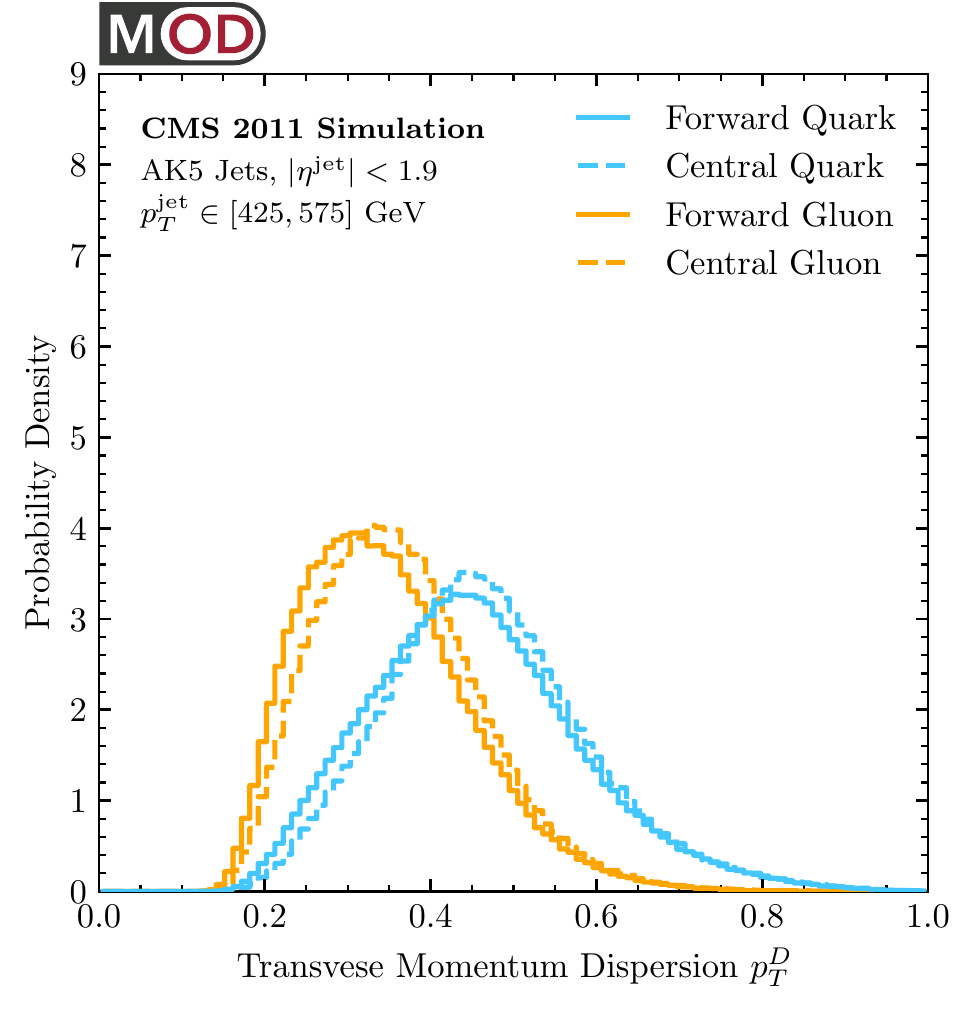}\label{fig:unfolding_ptd_qq_gg}}\\
\subfloat[]{\includegraphics[width=0.33\textwidth]{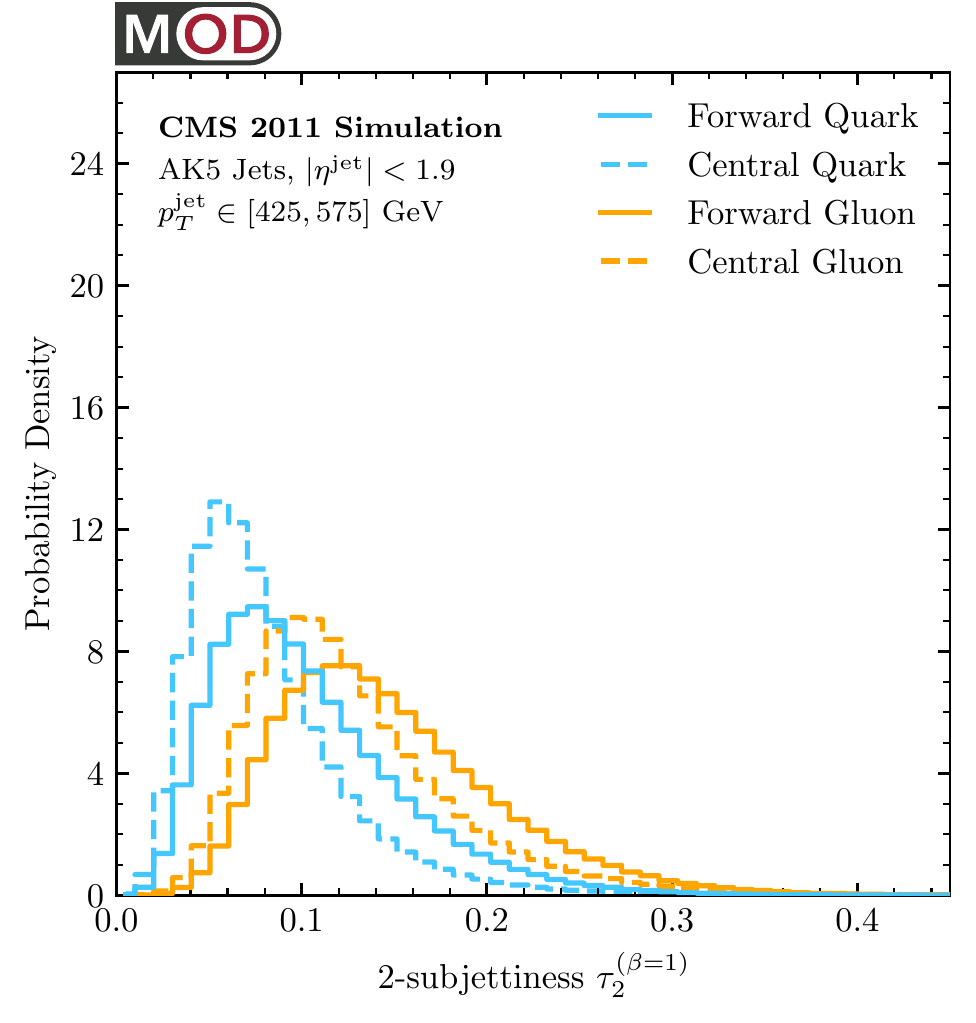}\label{fig:unfolding_nsub2_qq_and_gg}}
\subfloat[]{\includegraphics[width=0.33\textwidth]{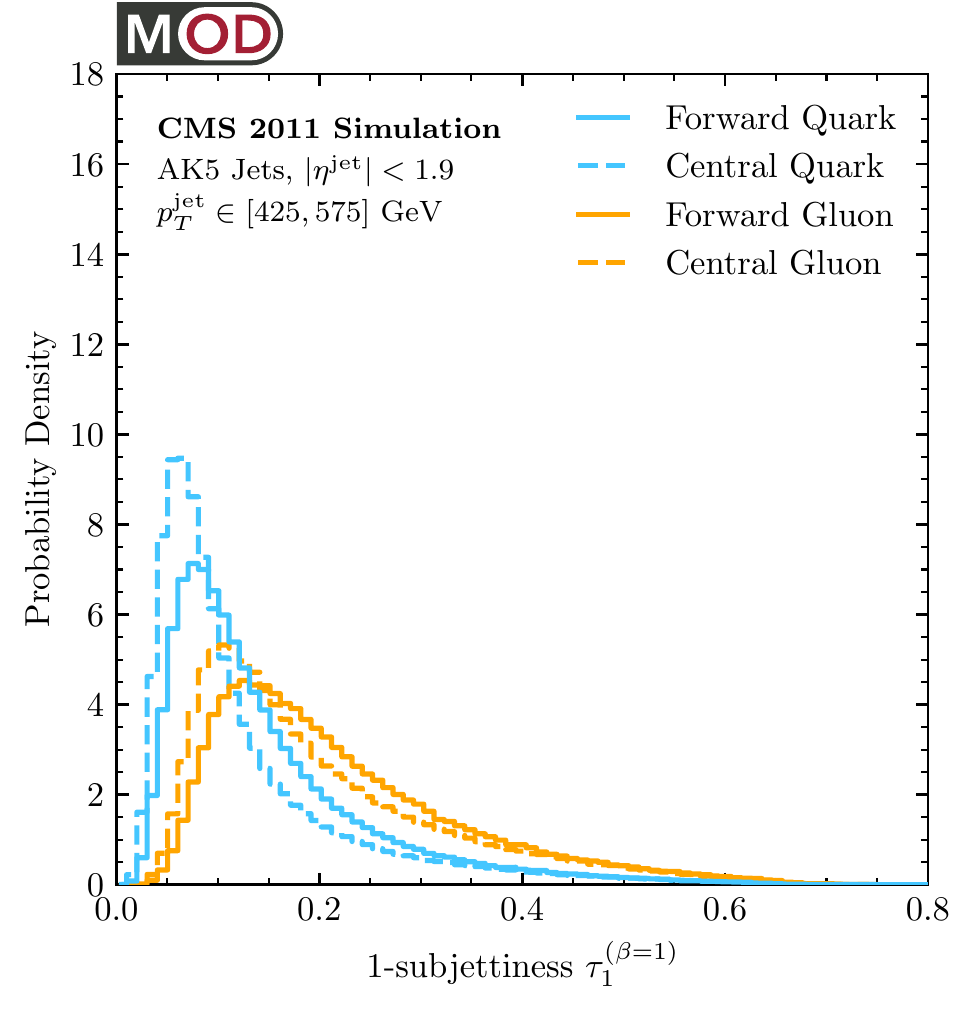}\label{fig:unfolding_nsub1_qq_and_gg}}
\subfloat[]{\includegraphics[width=0.33\textwidth]{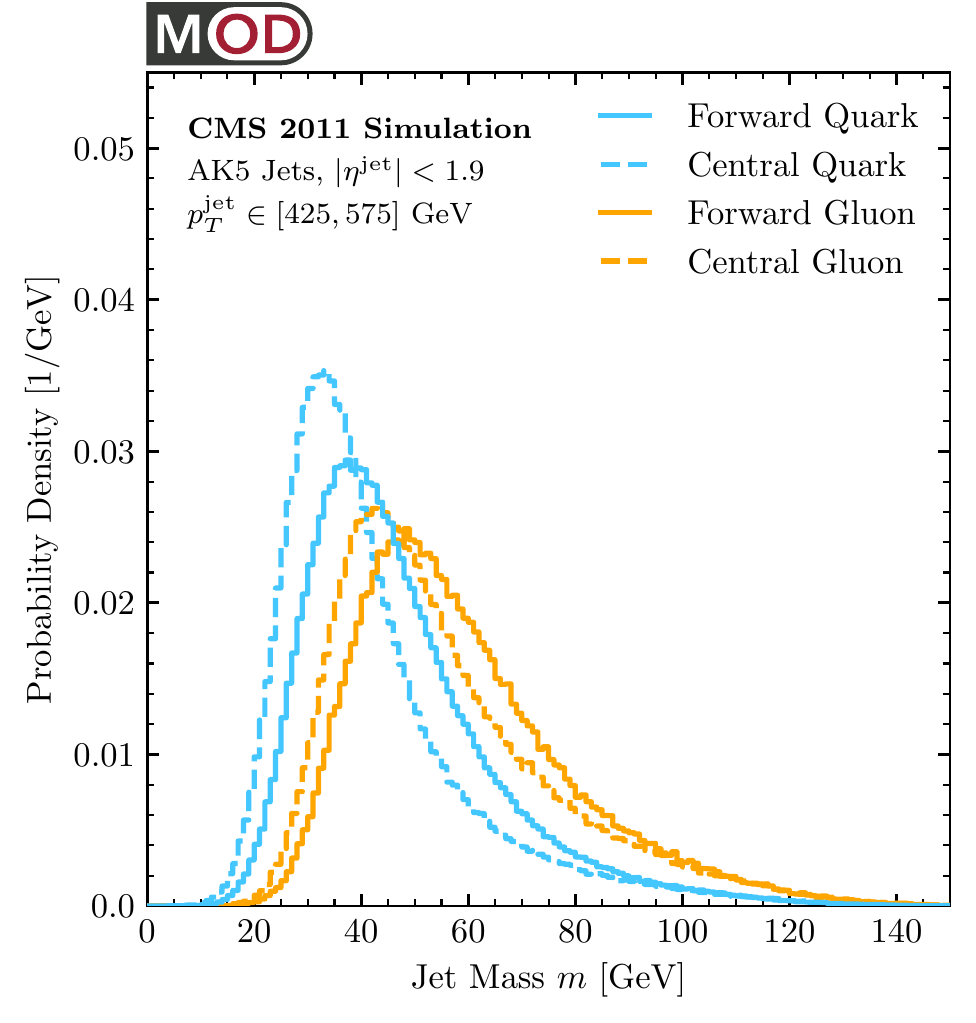}\label{fig:unfolding_mass_qq_and_gg}}
\\
\caption{Similar to \Fig{fig:qq_and_gg_truth_1} but now for detector-level distributions after \Pythia events are passed through the CMS detector simulation.
We see that quark jets (cyan) and gluon jets (orange) look different in the forward and central parts of the CMS detector, which violates the assumption of sample independence and necessitates the use of unfolding.
}
\label{fig:qq_and_gg_sim_1}
\end{figure*}

A key requirement for jet topic modeling is sample independence, which means that each mixture distribution should be a linear combination of the same underlying quark and gluon distributions.
To test this, we can use the MC samples, where we have quark/gluon labels based on \Pythia truth parton information.
As shown in \Fig{fig:qq_and_gg_truth_1}, sample independence holds to an excellent approximation for generator-level jets, with forward quarks and central quarks having nearly identical distributions, and similarly for gluons.

At detector-level, though, we see considerable sample dependence in \Fig{fig:qq_and_gg_sim_1}.
\cor{This sample dependence is due to a number of factors, including changes in the detector geometry and granularity, changes to object reconstruction efficiencies, and differing response to pileup and other sources of noise.}
In general, forward jets have more reconstructed constituents and therefore larger values of $N_{95}$, $\tau_2$, $\tau_1$, and $m_{\rm jet}$.
(More constituents typically implies smaller values of $p_T^D$.)
This dependence of the CMS detector response on the overall jet kinematics introduces undesirable sample dependence in our mixed samples, motivating the need for (central value) unfolding.

\subsection{Central Value Unfolding with \OmniFold}
\label{sec:cvu}

\begin{figure*}[t]
	\centering
	\subfloat[]{\includegraphics[width=0.45\textwidth]{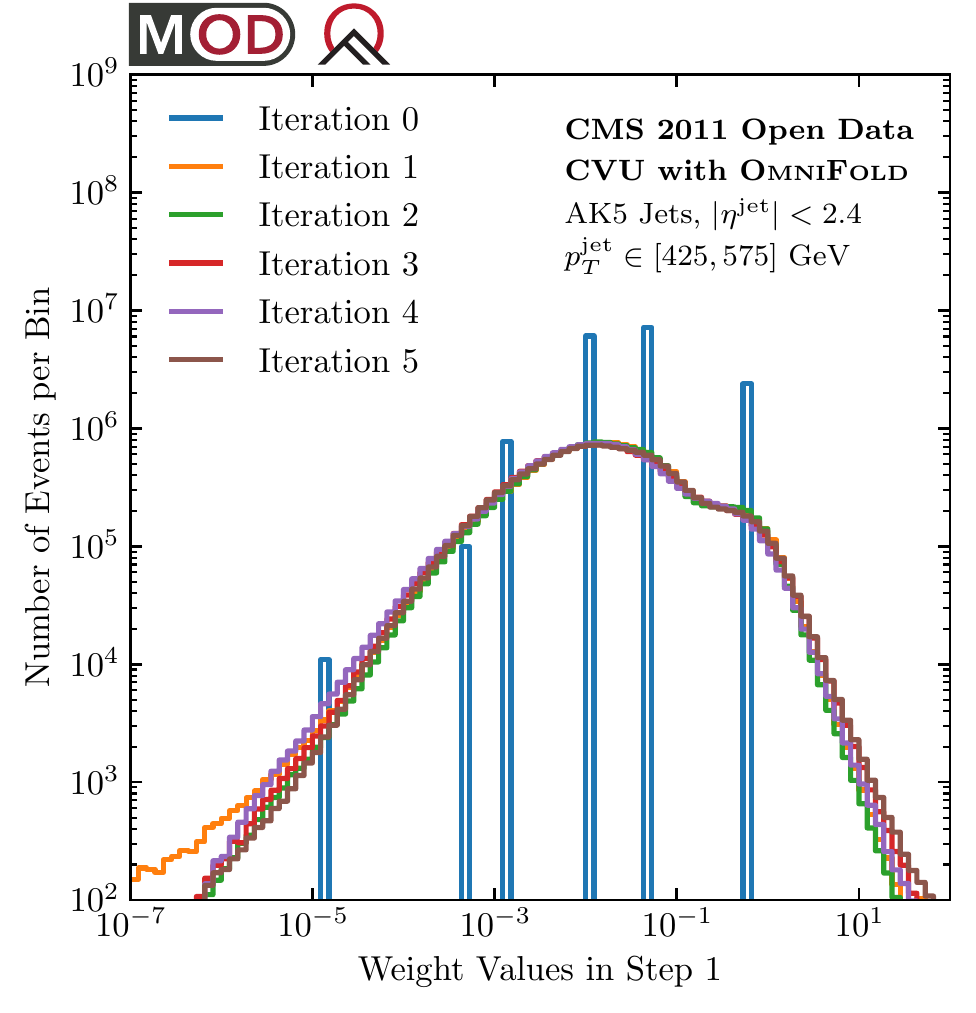}\label{fig:unfolding_step1weights}}
	$\quad$
	\subfloat[]{\includegraphics[width=0.45\textwidth]{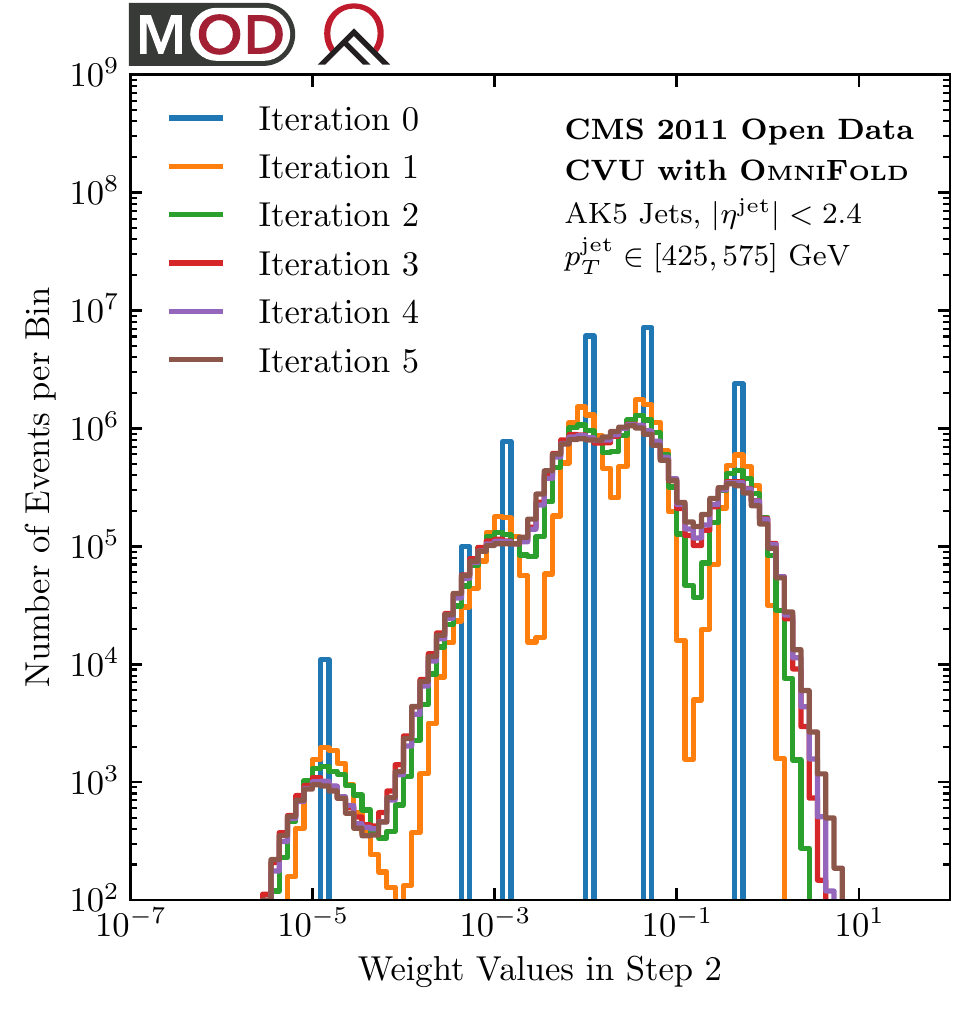}\label{fig:unfolding_step2weights}}
	\caption{
		The weight distribution produced by central value unfolding after (a) step 1 and (b) step 2 of the \OmniFold algorithm. 
		The initial weights from the CMS Open Data are shown in blue, where the spikes arise because there are six independent samples that have been individually unweighted.
		The results in this paper correspond to using weights from iteration 4, after step 2.
	}
	\label{fig:unfoldedweights}
\end{figure*}

In a full experimental analysis, one would perform detector unfolding to mitigate the impact of detector effects.
Such an unfolding would also account for various sources of systematic uncertainties, including the choice of generator used to extract the response matrix.
See \Ref{Kogler:2018hem} for a review of jet substructure measurements.
For this study, we perform \emph{central value unfolding} (CVU), where we correct detector effects based on the response matrix derived from a single MC sample without assigning systematic uncertainties.%
\footnote{We use the term ``central value'' to refer to the central value with respect to systematic variations.  This differs from the central value associated with statistical or initialization variations.}
This is sufficient to address the issues of sample dependence, though not sophisticated enough to derive distributions suitable for quantitative studies, e.g.~comparisons to theoretical calculations or tunings of event generators.
The statistical uncertainties on the unfolding are estimated in a naive way through the variance of the unfolded weights (i.e.~the usual way one would account for uncertainties in a weighted event sample).
We checked that these uncertainties are comparable in size to the statistical uncertainties associated with the prior distribution.

Because the absence of systematic uncertainties will be viewed as anathema to some readers (especially the authors of \Refs{ATLAS:2019rqw,CMS:2021iwu}), we want to explain more about our philosophy.
We will leverage a full phase space unfolding strategy called \OmniFold~\cite{Andreassen:2019cjw,Andreassen:2021zzk}, which assigns each MC event a weight at the truth particle level.%
\footnote{See \Ref{H1:2021wkz} for an application of \OmniFold by the H1 collaboration on an eight-dimensional phase space, with proper uncertainty quantification.  See \Ref{H1prelim-22-034} for a preliminary analysis by the H1 collaboration to unfold six jet substructure observables in step 2 of \OmniFold by using the full phase space at step 1.}
With these per-event weights, one can in principle compute the unfolded distribution for any observable, including ones not envisioned at the time of the unfolding.
Full phase space unfolding is a great opportunity for the field, and there are already discussions about how to release per-event unfolded data in a suitable format~\cite{Arratia:2021otl}.
It is also a great challenge, though, since it is not clear how to validate the unfolding for arbitrary down-stream analyses, nor is it clear what the best strategy would be to assign per-event systematic uncertainties.
Furthermore, \OmniFold is based on neural networks, which introduces additional sources of uncertainties from the initialization and training paradigms.
For these reasons, we feel it is prudent to perform a proof-of-concept study that shows how unfolding can indeed mitigate the issue of sample dependence, without attempting to make a quantitative claim about the degree of mitigation or the systematic uncertainties on the final distributions.

With that apologia, we apply the \OmniFold algorithm to all reconstructed jets in the range $p_T \in [375,700]~\GeV$ and $|\eta| < 2.4$.
\OmniFold can be viewed as an unbinned version of iterated Bayesian unfolding~\cite{DAGOSTINI1995487}.
In step 1 of the algorithm, event weights are computed such that the detector-level \cortwo{full phase-space density} of the simulated sample match that of the observed data.
In step 2 of the algorithm, the event weights are refined such that the weights can be expressed as a function of the particle-level inputs.
For this analysis, we do not account for migration of events out of the jet selection range~\cite{Andreassen:2021zzk}, though we later restrict our attention to a smaller $p_T$ range to avoid edge complications.
The weights distributions are shown in \Fig{fig:unfoldedweights}, where one can see that the starting weights for the six initial samples are smeared out in each \OmniFold iteration.
These weights are publicly available at \Ref{MOD:ZenodoCMSOmniFold}.

We use four iterations of \OmniFold as the default, and checked that the qualitative features were robust with one more or one fewer iteration.
\cor{More iterations lead to a better fit to the data, at the expense of larger fluctuations.}
The unfolding weights are computed using a PFN, as parametrized in \Eq{eq:PFN_def}.
The model takes as input a tuple of all particle 4-vectors, experimentally accessible particle-type information (``PFN-Ex", see \Ref{Komiske:2018cqr}), and four global features: jet $p_T$, jet $\eta$, jet $\phi$, and jet mass.
For step 1, we use a latent space with $\ell = 256$, where the per-particle $\Phi$ function has 2 hidden layers of 128 nodes each, and the per-jet $F$ function has 3 hidden layers of 128 nodes each.
For step 2, we use $\ell = 192$, where $\Phi$ has 2 hidden layers of 100 nodes, and $F$ has 3 hidden layers of 100 nodes.
The ReLU function is used as an activation function on each layer besides the output layer that uses softmax.

\begin{figure*}[t]
\centering
\subfloat[]{\includegraphics[width=0.33\textwidth]{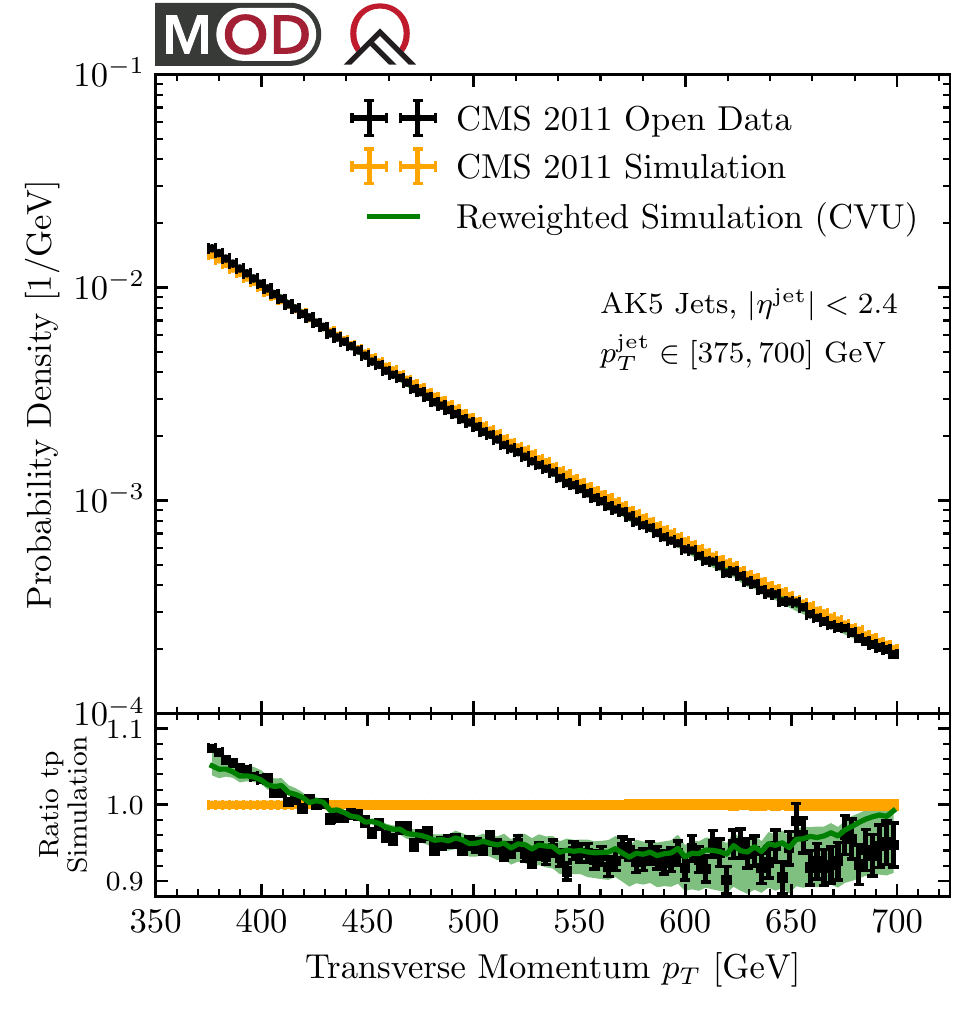} \label{fig:unfolding_pT}}
\subfloat[]{\includegraphics[width=0.33\textwidth]{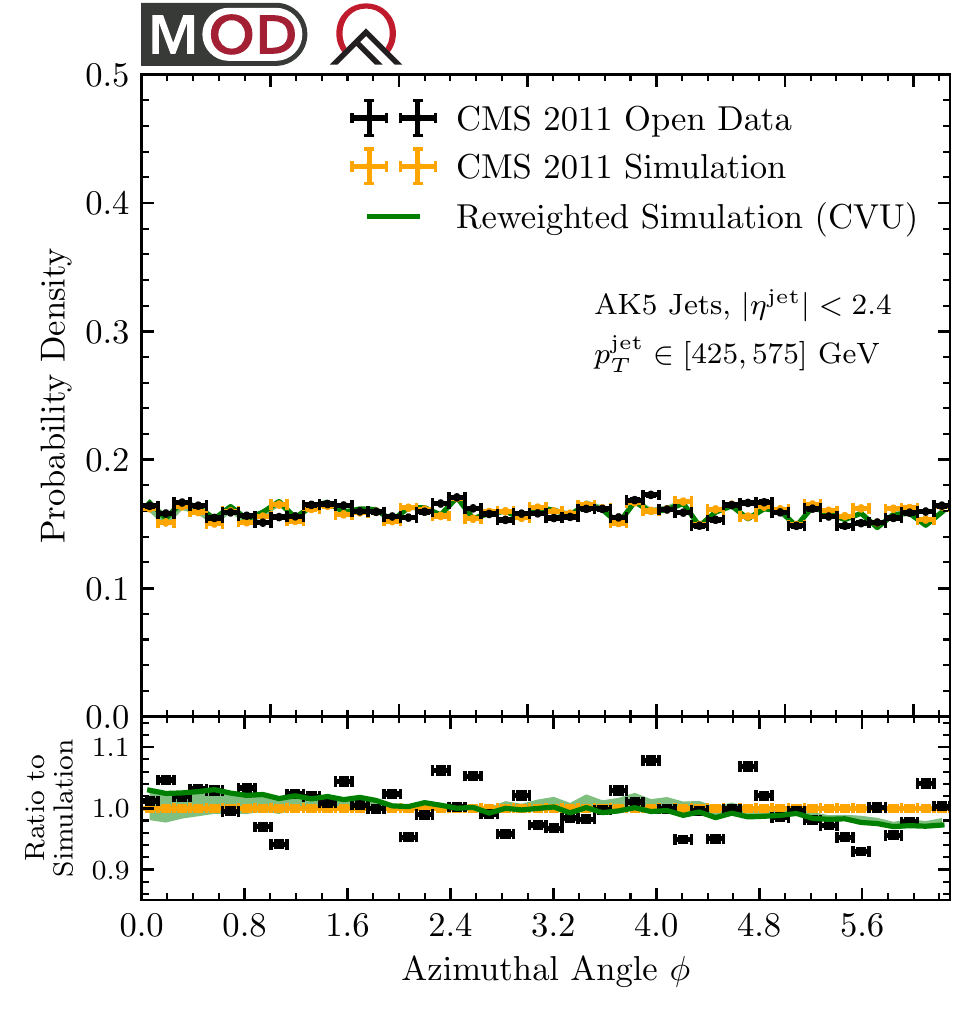} \label{fig:unfolding_phi}}
\subfloat[]{\includegraphics[width=0.33\textwidth]{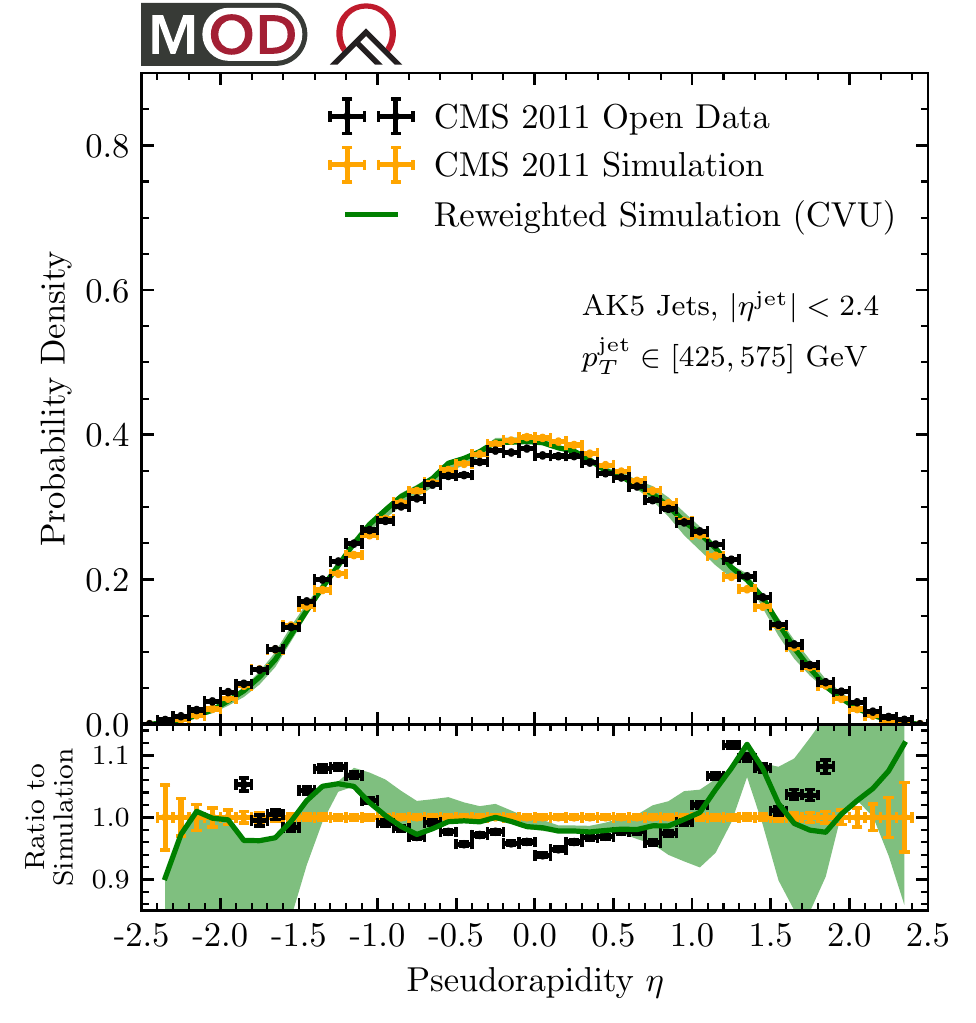} \label{fig:unfolding_eta}}
\caption{Validation of central value unfolding with \OmniFold applied to the CMS 2011 Open Data.
Shown are distributions for (a) jet $p_T$, (b) jet azimuth, and (c) jet pseudorapidity.
The detector-level distributions after 
\cor{using the unfolding weights} (green) match better to the CMS measured distributions (black) than to \Pythia run through the CMS detector simulation (orange).
Note the smaller jet $p_T$ range for the azimuth and pseudorapidity plots.
The error bars on the open data/simulation correspond to statistical uncertainties, while the green band corresponds to using 3 or 5 iterations of \OmniFold compared to the default of 4.
}
\label{fig:unfolding_jet}
\end{figure*}

We emphasize that, after step 2 of \OmniFold, the unfolded weights are assigned to the truth particle-level information.
Combining the weights with \Pythia parton-level information, we can derive the effective quark and gluon fraction of the central-value-unfolded sample:
\begin{equation}
	\label{eq:quark_fractions_CVU}
	\text{\Pythia Parton with CVU:} \quad
	\begin{aligned}
		f_1 \simeq 0.711,\\
		f_2 \simeq 0.543.
	\end{aligned}
\end{equation}
These fractions are unphysical for two reasons.
First, they correspond to parton-level information that only have meaning at leading logarithmic accuracy.
Second, the \OmniFold weights are derived from particle-level alone, without reference to parton-level information, so these fractions do not correspond to a true generative model in the sense of \Eqs{eq:mixture_M1}{eq:mixture_M2}.
Nevertheless, \Eq{eq:quark_fractions_CVU} is a useful benchmark to guide the eye in interpreting the jet topic modeling results.

\subsection{Unfolding Validation \cortwo{at Detector Level}}
\label{sec:unfolding_validation}

\begin{figure*}[p]
\centering
\subfloat[]{\includegraphics[width=0.33\textwidth]{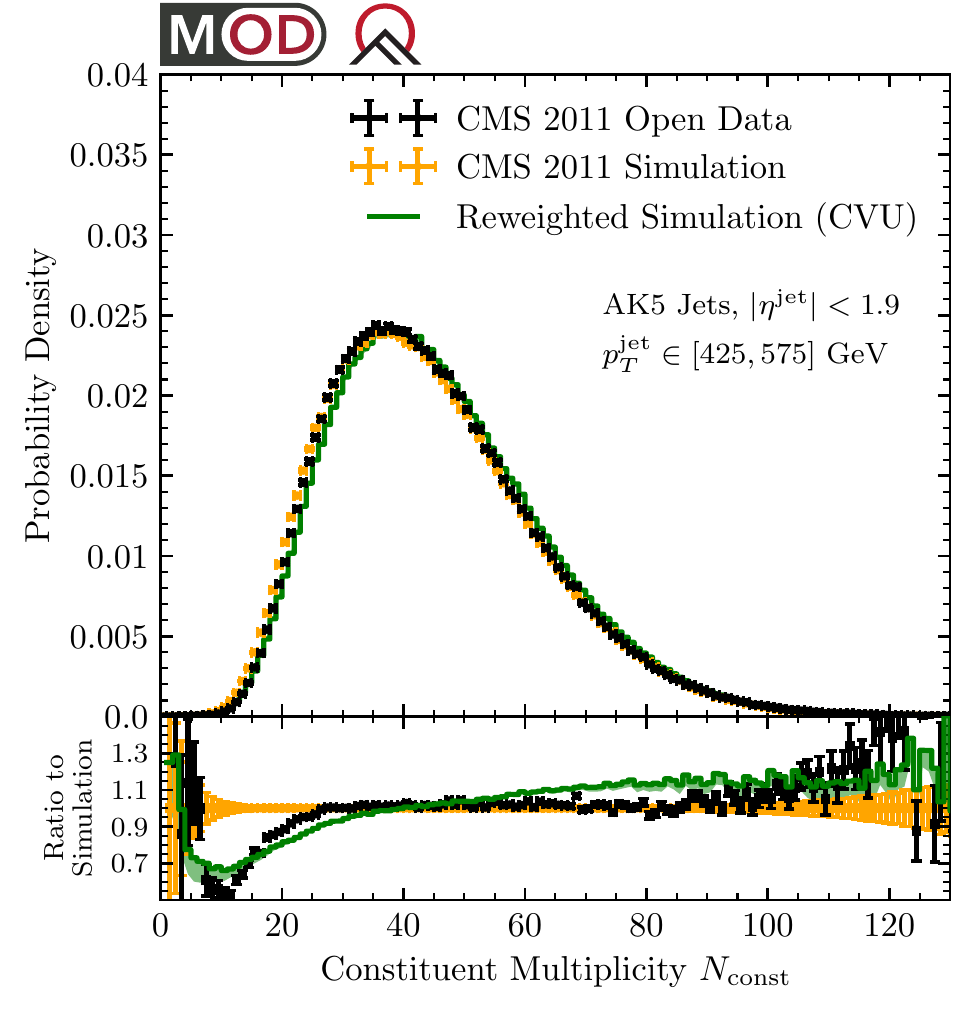}\label{fig:unfolding_mult}}
\subfloat[]{\includegraphics[width=0.33\textwidth]{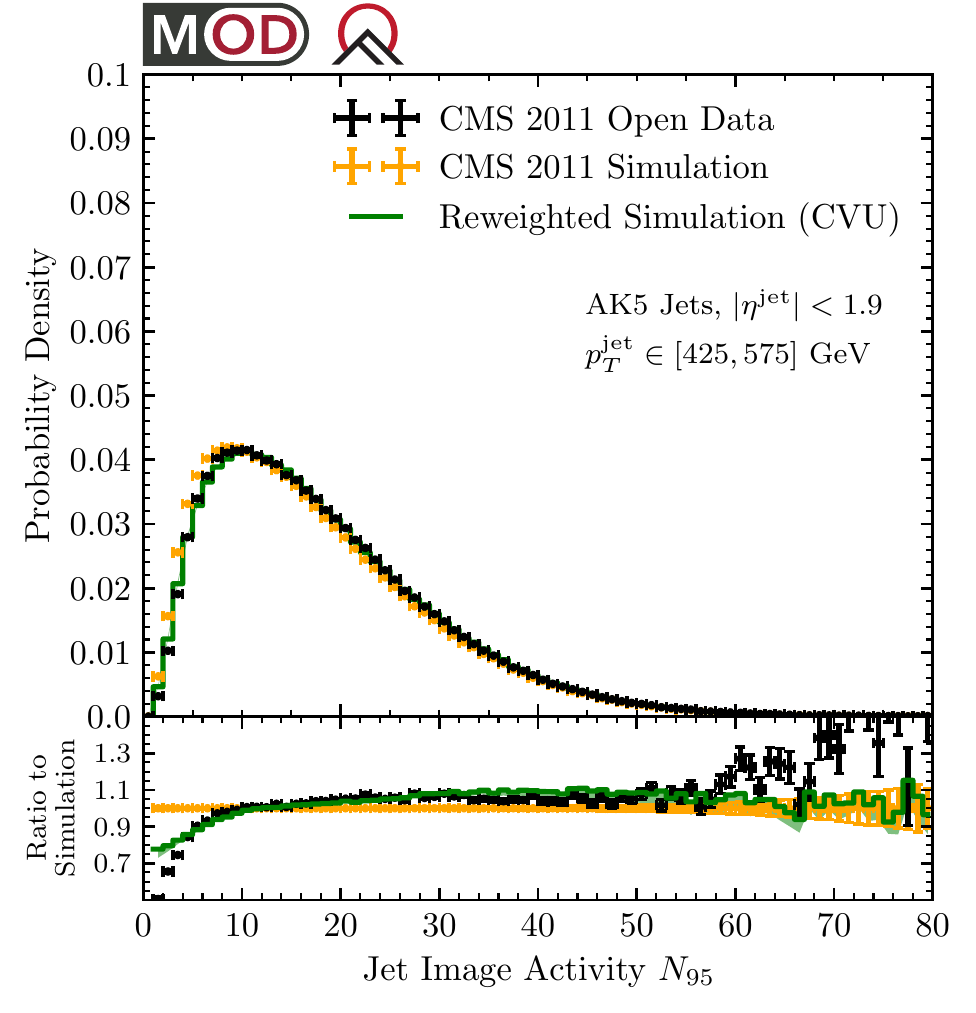}\label{fig:unfolding_n95}}
\subfloat[]{\includegraphics[width=0.33\textwidth]{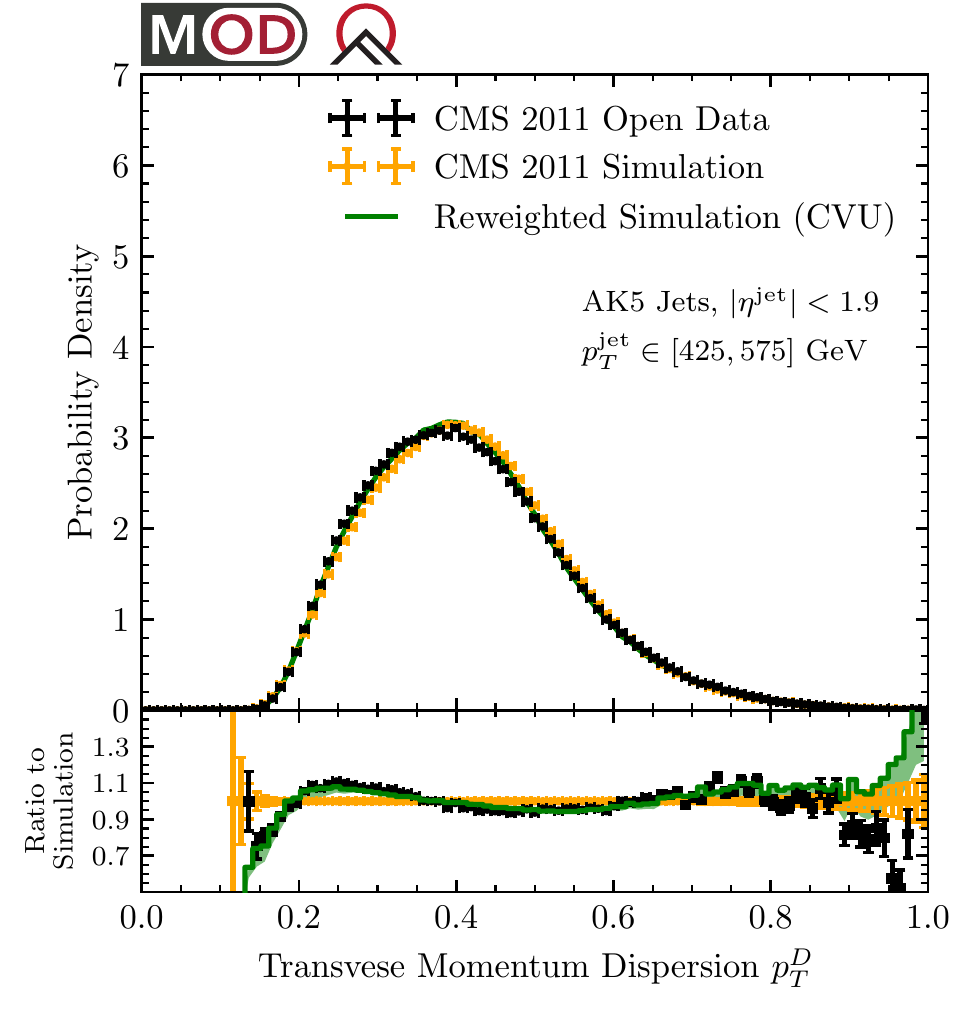}\label{fig:unfolding_pTd}}\\
\subfloat[]{\includegraphics[width=0.33\textwidth]{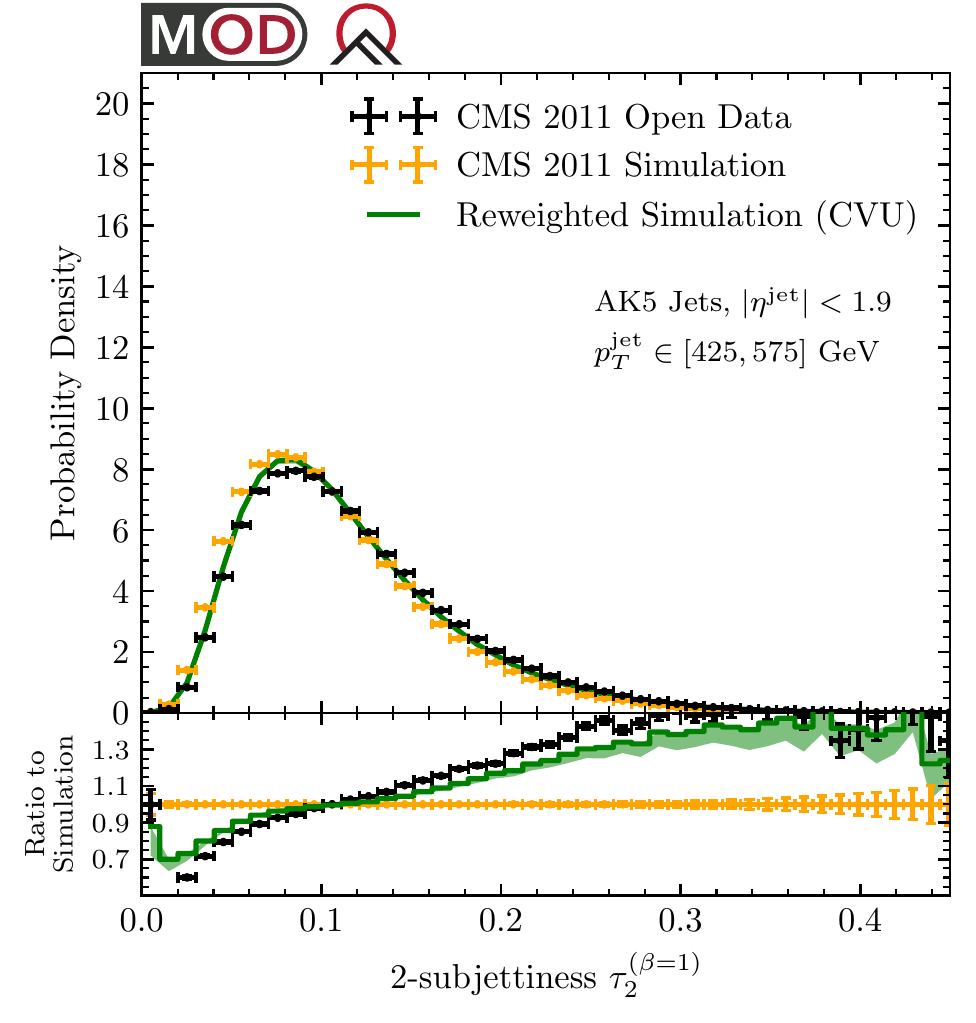}\label{fig:unfolding_nsub2}}
\subfloat[]{\includegraphics[width=0.33\textwidth]{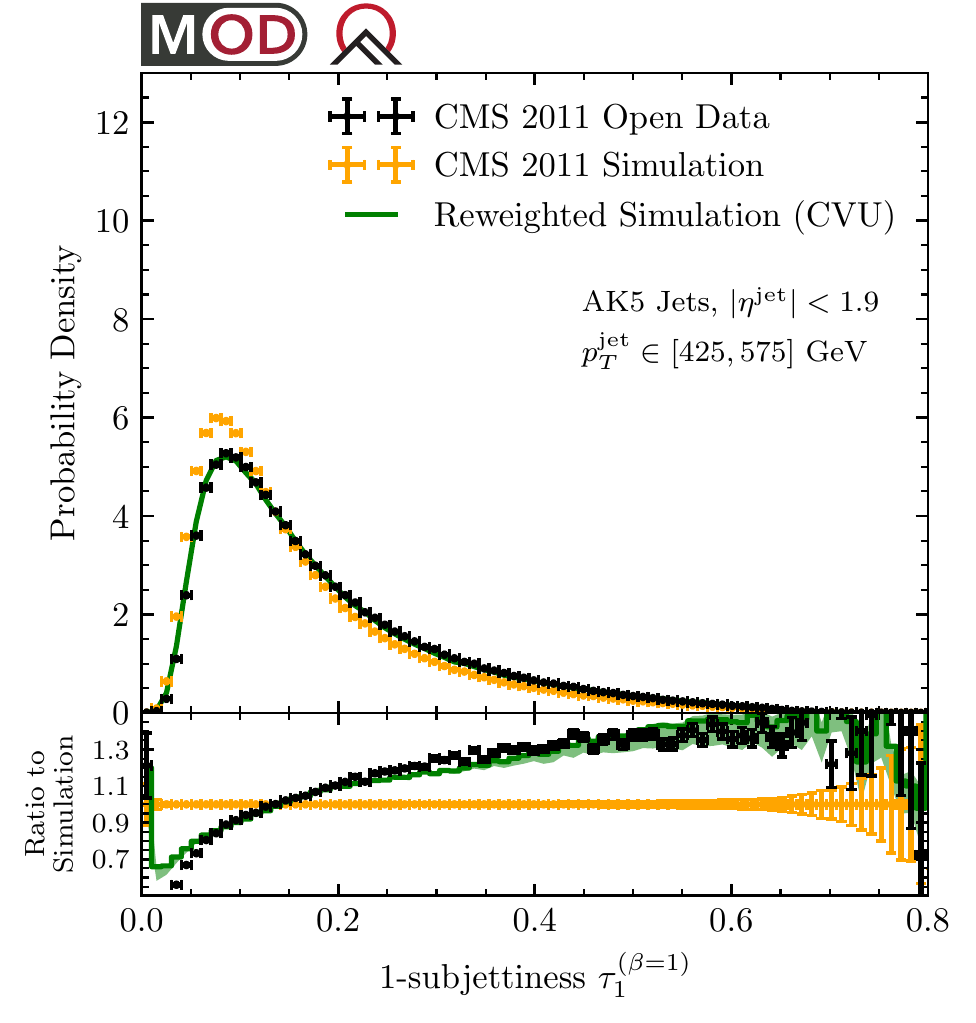}\label{fig:unfolding_nsub1}}
\subfloat[]{\includegraphics[width=0.33\textwidth]{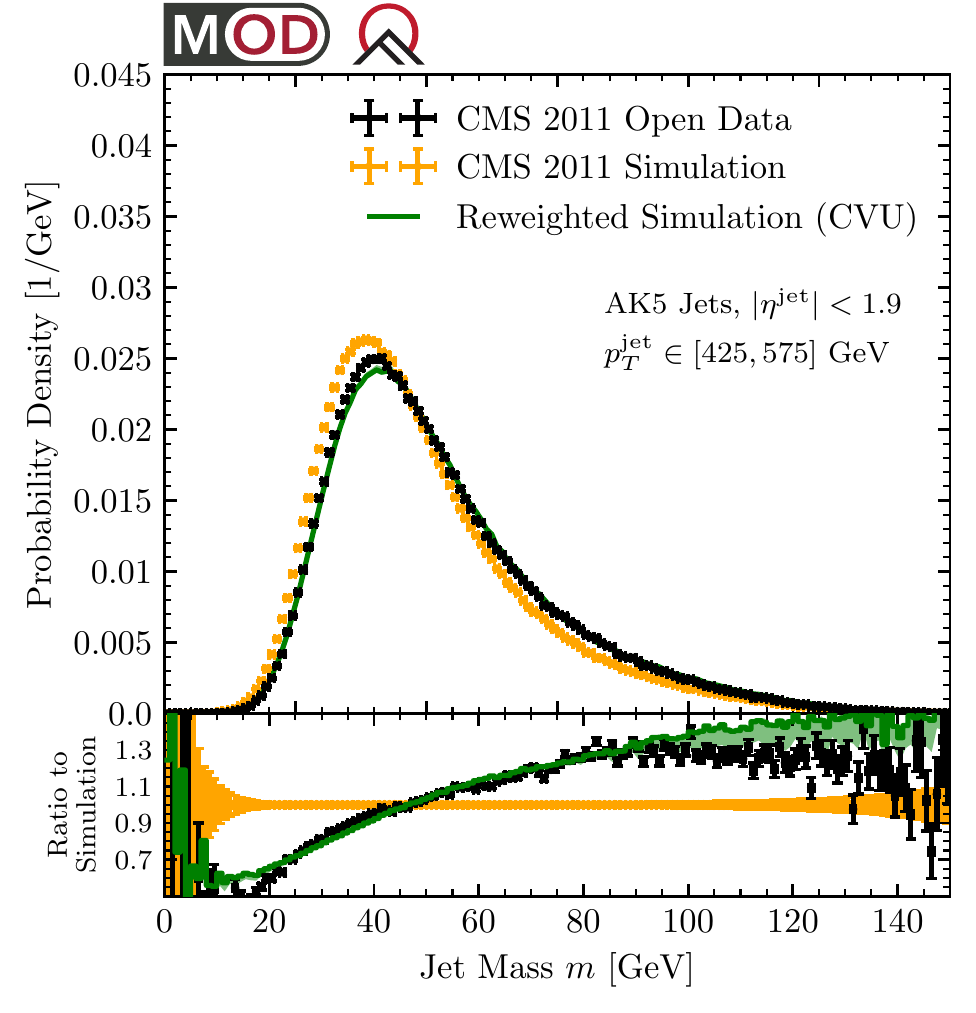}\label{fig:unfolding_mass}}
\caption{Similar to \Fig{fig:unfolding_jet}, but for the six substructure observables from \Sec{subsec:observables}.
Here, we are using a narrower jet $p_T$ and pseudorapidity range to mitigate edge effects in the unfolding.
}
\label{fig:unfoldedobs}
\end{figure*}

To validate the unfolding, we compare detector-level distributions before and after \OmniFold.
\cor{For all of the plots in this subsection, we emphasize that distributions are shown at detector level, and we only use unfolding to determine the event weights (particle-level distributions are shown later in \Secs{section:extracting_k}{sec:quark_and_gluon_dist}).}
In \Fig{fig:unfolding_pT}, we show the result of 
%the unfolding 
\cor{using the unfolding weights} on the overall jet transverse momentum, in the range of $p_T \in [375,700]~\GeV$ and $|\eta| < 2.4$.
The spectrum from the MC (orange dots) differs from that of the CMS data (black dots) by upwards of 10\%, but the %unfolded spectrum 
\cortwo{simulation} \cor{spectrum with unfolding weights} (green line) successfully matches the spectrum \cortwo{seen in the CMS data}.
The shaded green region is the envelope from considering three, four, or five \OmniFold iterations, with four being the default.
\cortwo{The weights are obtained after step 2 of \OmniFold.}

To reduce $p_T$ edge effects, we further restrict to $p_T \in [425,575]~\GeV$ for the remaining distributions.
%
%Unfolding 
\cor{Using the unfolding weights} has a negligible effect on the azimuthal spectrum in \Fig{fig:unfolding_phi}, which suggests that the CMS simulation does not match the detailed azimuthal structure of the CMS detector.

For the pseudorapidity spectrum in \Fig{fig:unfolding_eta}, there is a slight cliff-like feature at around $|\eta| \approx 1.4$, which is roughly where the detector barrel meets the detector endcap.
%
%The unfolding 
\cor{Using the unfolding weights} adjusts the truth-level pseudorapidity spectrum in the vicinity of this cliff such that the detector-level pseudorapidity structure better matches the data.
This is an example of a potential systematic issue that requires a detailed understanding of jet reconstruction at CMS.
If the barrel/endcap interface is incorrectly modeled in the CMS simulation \cor{or if there are sources of contamination not captured in simulation and not mitigated by the jet quality criterion,} then unfolding could erroneously modify the truth-level pseudorapidity spectrum.
This is particularly important for our analysis, because quark/gluon composition of the samples are affected by the pseudorapidity spectrum.
Note that this issue is potentially present with any unfolding algorithm, not just \OmniFold, and requires expert experimental knowledge to assess.

\begin{figure*}[t]
\centering
\subfloat[]{\includegraphics[width=0.33\textwidth]{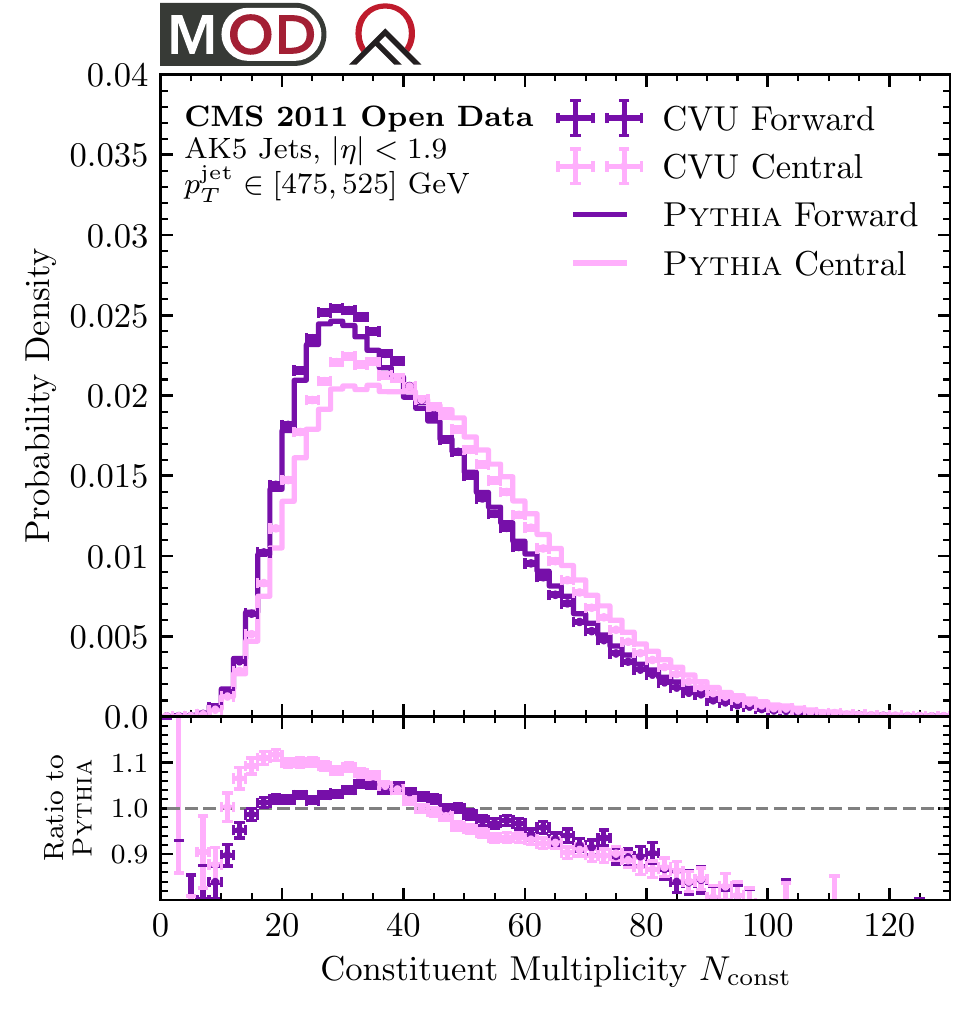}\label{fig:q_and_g_truth_mult}}
\subfloat[]{\includegraphics[width=0.33\textwidth]{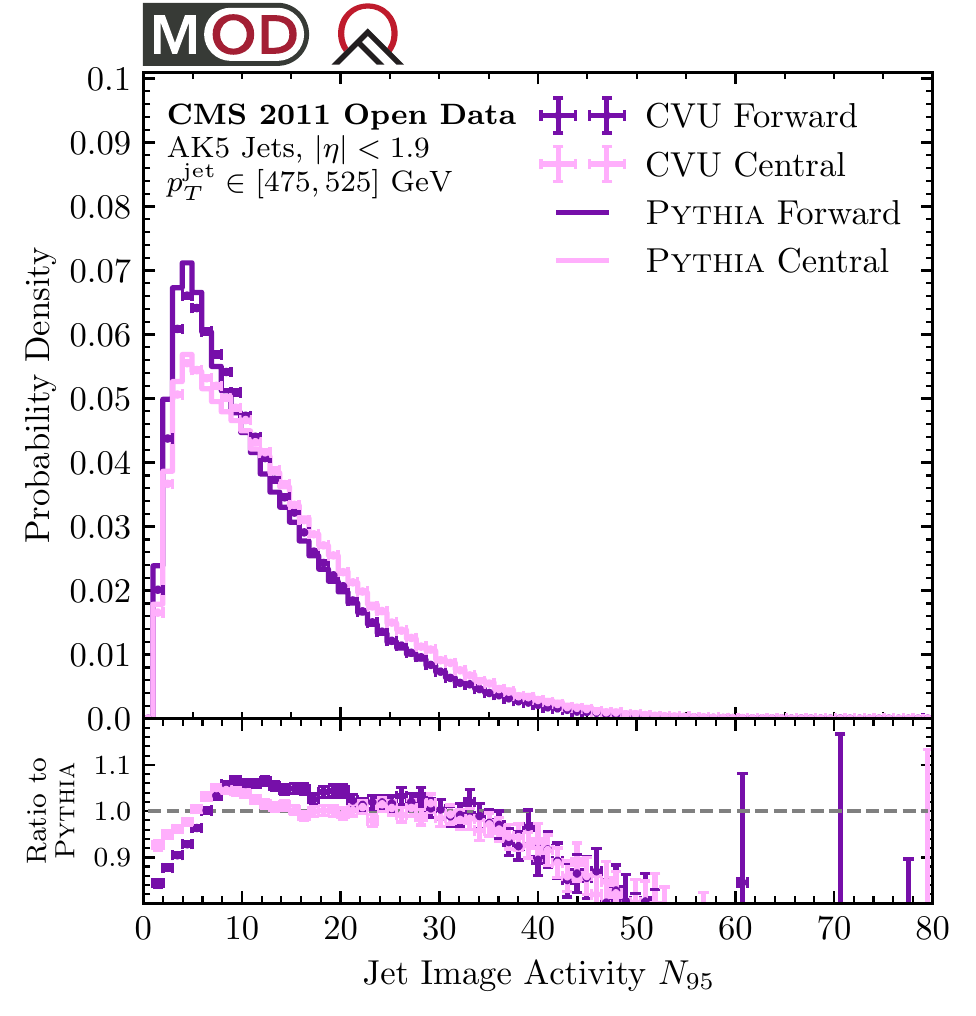}\label{fig:q_and_g_truth_n95}}
\subfloat[]{\includegraphics[width=0.33\textwidth]{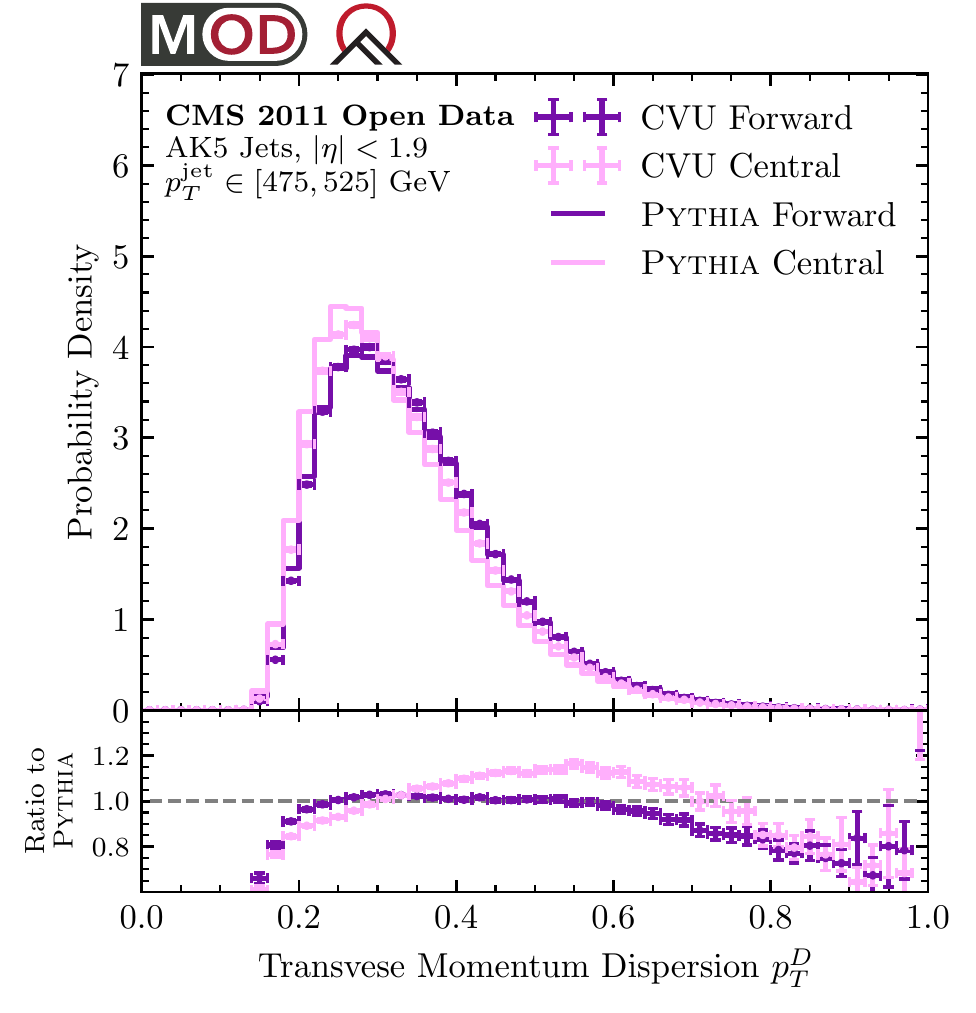}\label{fig:q_and_g_truth_pTd}}
\\
\subfloat[]{\includegraphics[width=0.33\textwidth]{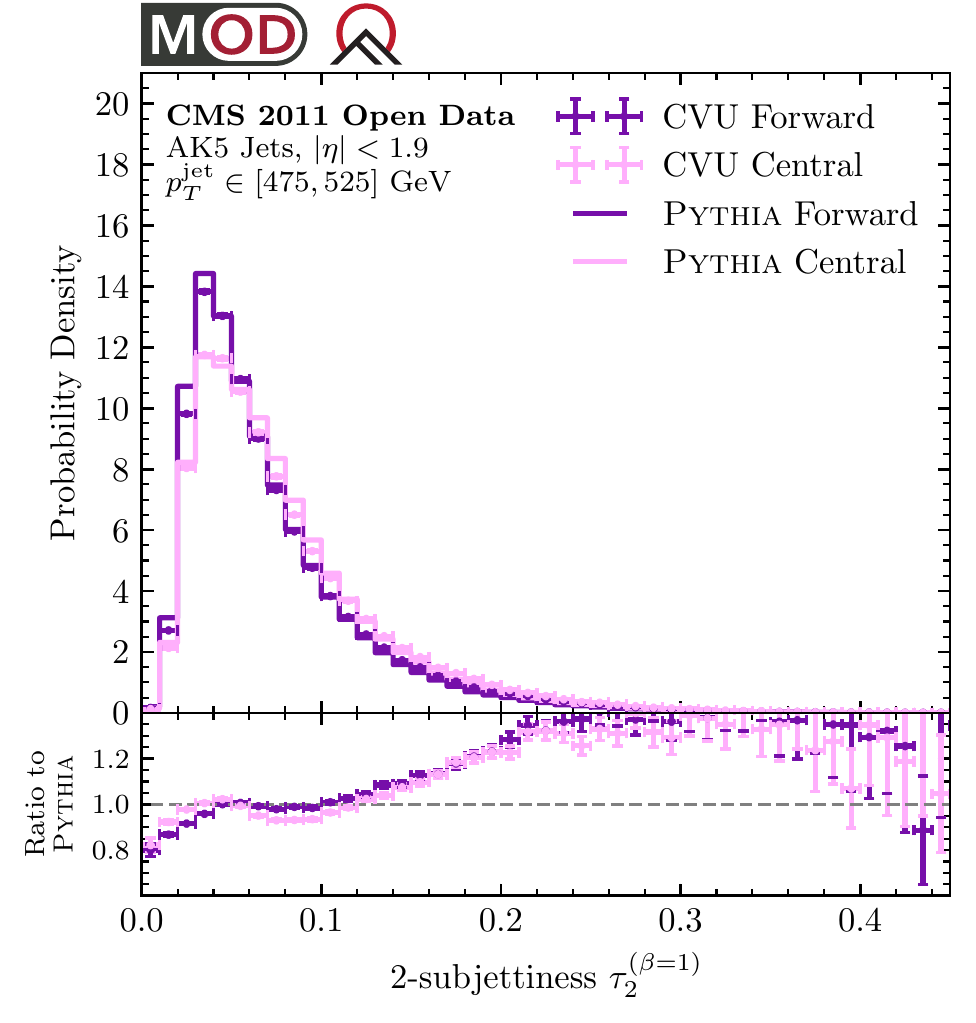}\label{fig:q_and_g_truth_nsub2}}
\subfloat[]{\includegraphics[width=0.33\textwidth]{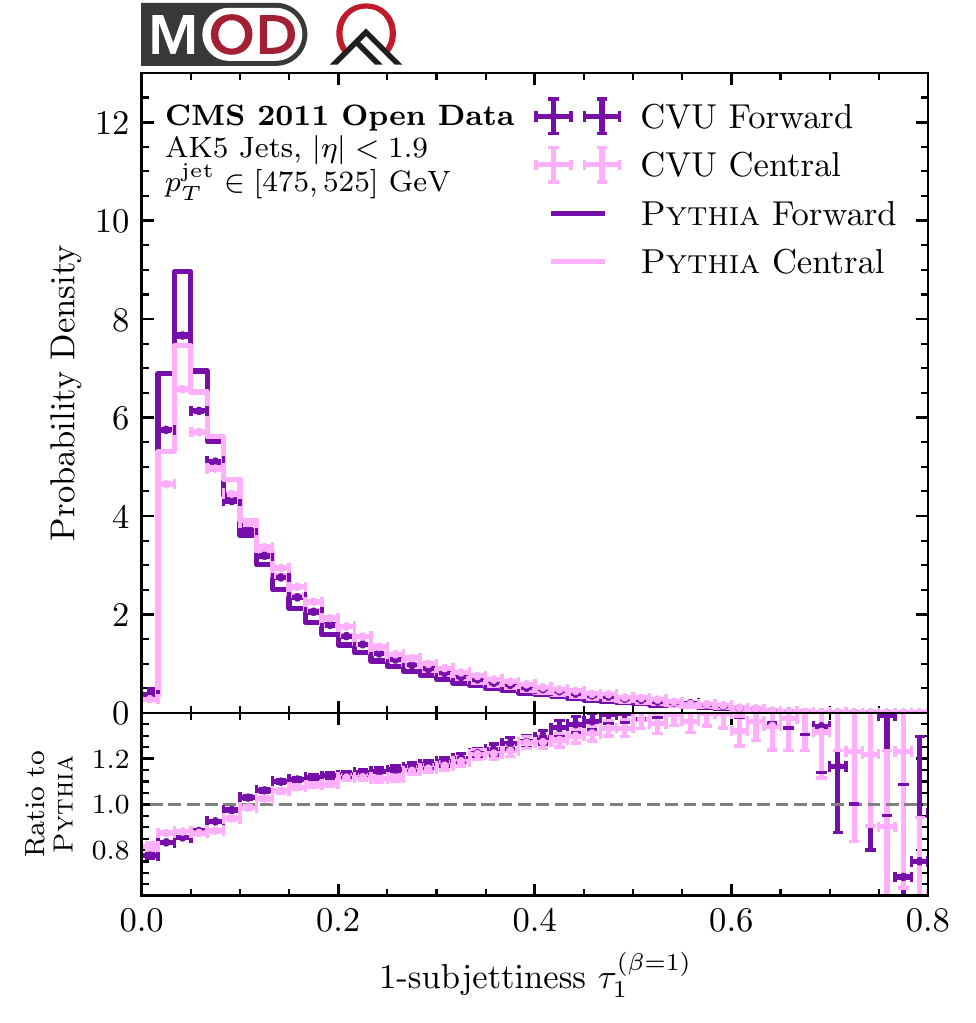}\label{fig:q_and_g_truth_nsub1}}
\subfloat[]{\includegraphics[width=0.33\textwidth]{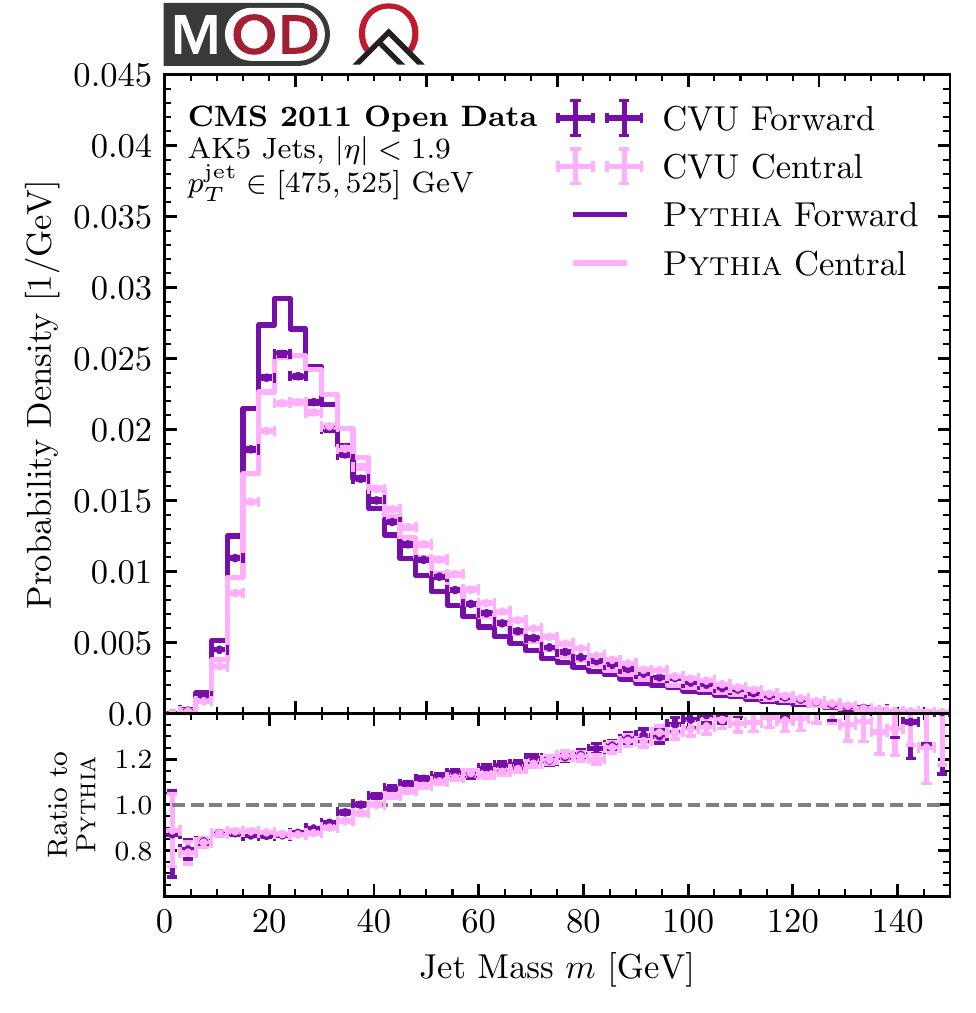}\label{fig:q_and_g_truth_mass}}
\caption{Particle-level distributions for forward (purple) and central (pink) jets for the six substructure observables from \Sec{subsec:observables}.
We compare central-value-unfolded results from the CMS 2011 Open Data (data points) to \Pythia 6.4.25 (solid curve), where the differences are subtle but noticeable.
The CVU distributions are inputs for the subsequent jet topics analyses.
See \App{sec:pythia_only_analysis} for results obtained from the \Pythia distributions.
The error bars correspond to statistical uncertainties only.
}
\label{fig:forw_vs_cent_gen}
\end{figure*}

To study the substructure of jets, we now restrict our attention to the detector-level jet kinematic range of
\begin{equation}
	p_T \in [475,525]~\GeV, \qquad |\eta| < 1.9.
\end{equation}
The tightening of the pseudorapidity range is because the tracker extends only out to $|\eta| < 2.4$, so we want to make sure that all charged particles within the jet radius of $R = 0.5$ have a chance to be reconstructed.
In \Fig{fig:unfoldedobs}, we show detector-level distributions for the six observables from \Sec{subsec:observables}.
Overall, unfolding improves the agreement between the detector-level distributions in regions of phase space with sufficient numbers of events, particularly for $N_{95}$, $p_T^D$, $\tau_1$, and $m_{\rm jet}$.
%
%The unfolding 
\cor{Using the unfolding weights} improves the agreement for $\tau_2$, though there are still noticeable discrepancies.
%
%The unfolding 
\cor{Using the unfolding weights} makes comparatively little impact on $N_{\rm const}$, %suggesting that this observable may not be perfectly modeled by the CMS simulation.
\cor{suggesting that the response of this observable may not be perfectly modeled by the CMS simulation or that there are correlations in the Pythia sample that cannot be corrected via unfolding.}

At a qualitative level, we conclude that %unfolding
\cor{using the unfolding weights} satisfies this basic detector-level closure test.
In the context of a complete measurement, deviations from closure would be assessed as a systematic uncertainty.
Though not shown, we checked that the particle-level distributions exhibit the same degree of sample independence as seen in \Fig{fig:qq_and_gg_truth_1}.

%%%%%%%%%%%%%%%%%%%%%%%%%%%%%%%%%%%%%%%%%%%%%%%%%%%

\begin{figure*}[p]
	\centering
	\subfloat[]{\includegraphics[width=0.4\textwidth]{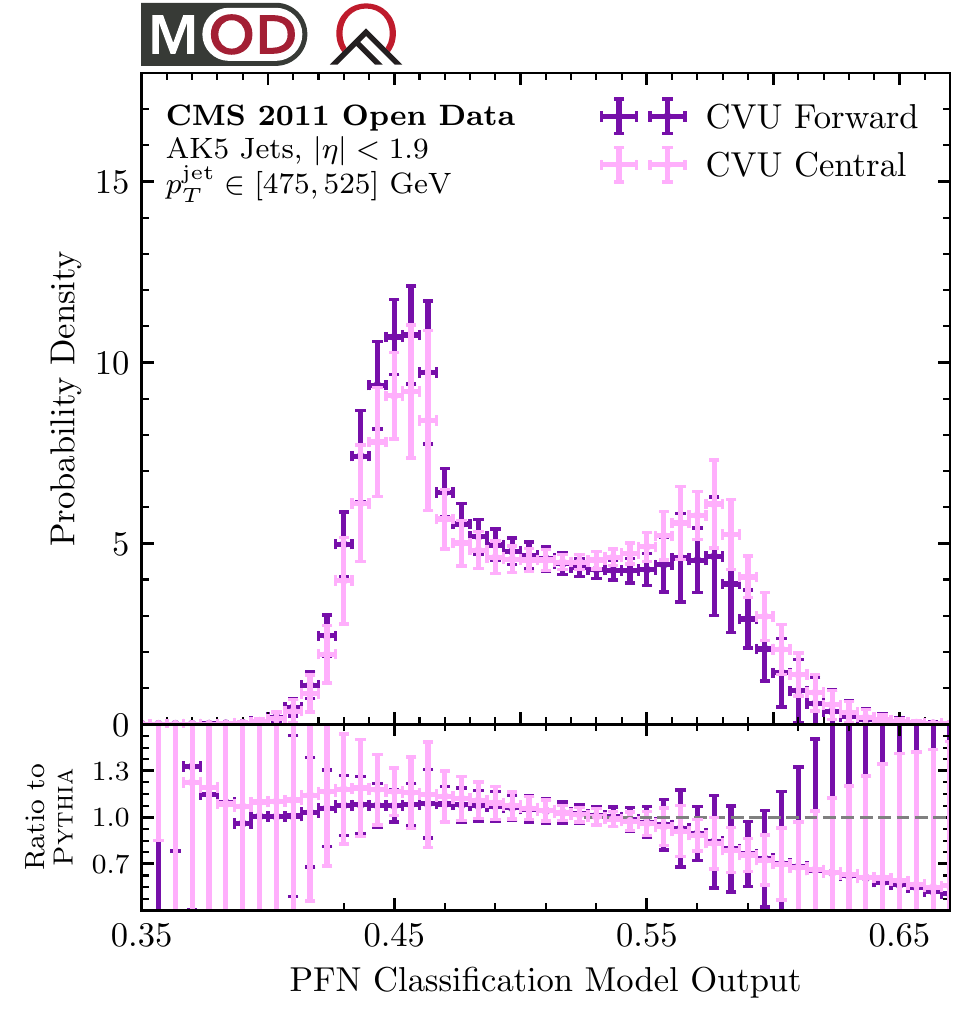}\label{fig:f_vs_c_PFN}}
	$\quad$
	\subfloat[]{\includegraphics[width=0.4\textwidth]{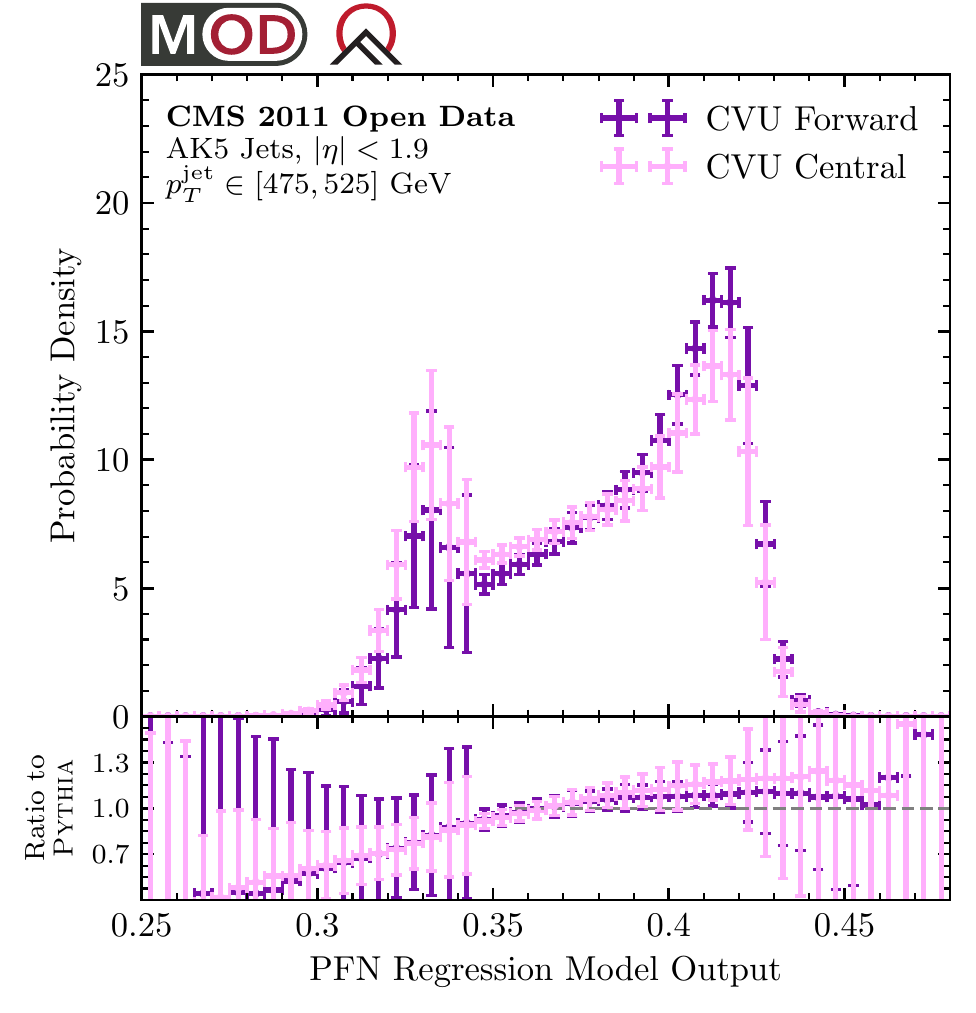}\label{fig:f_vs_c_PRNreg}}
	\\
	\subfloat[]{\includegraphics[width=0.4\textwidth]{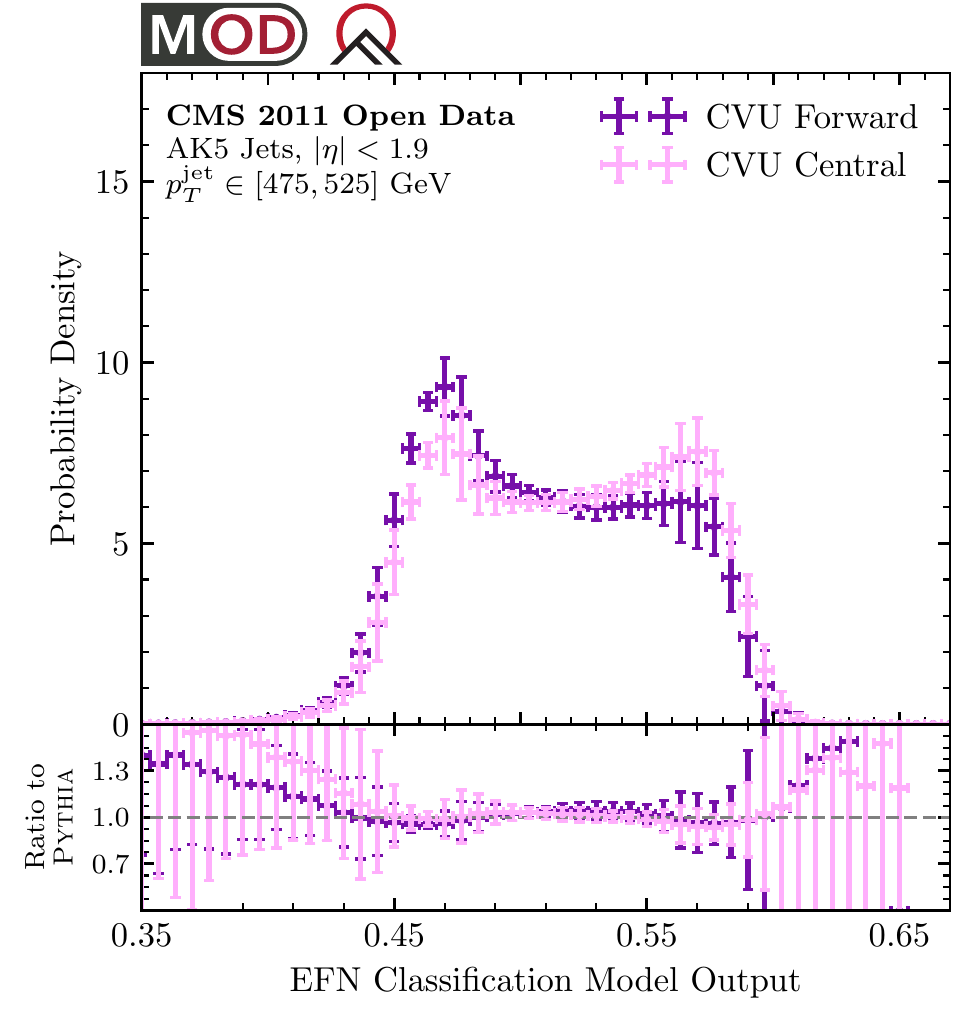}\label{fig:f_vs_c_EFN}}
	$\quad$
	\subfloat[]{\includegraphics[width=0.4\textwidth]{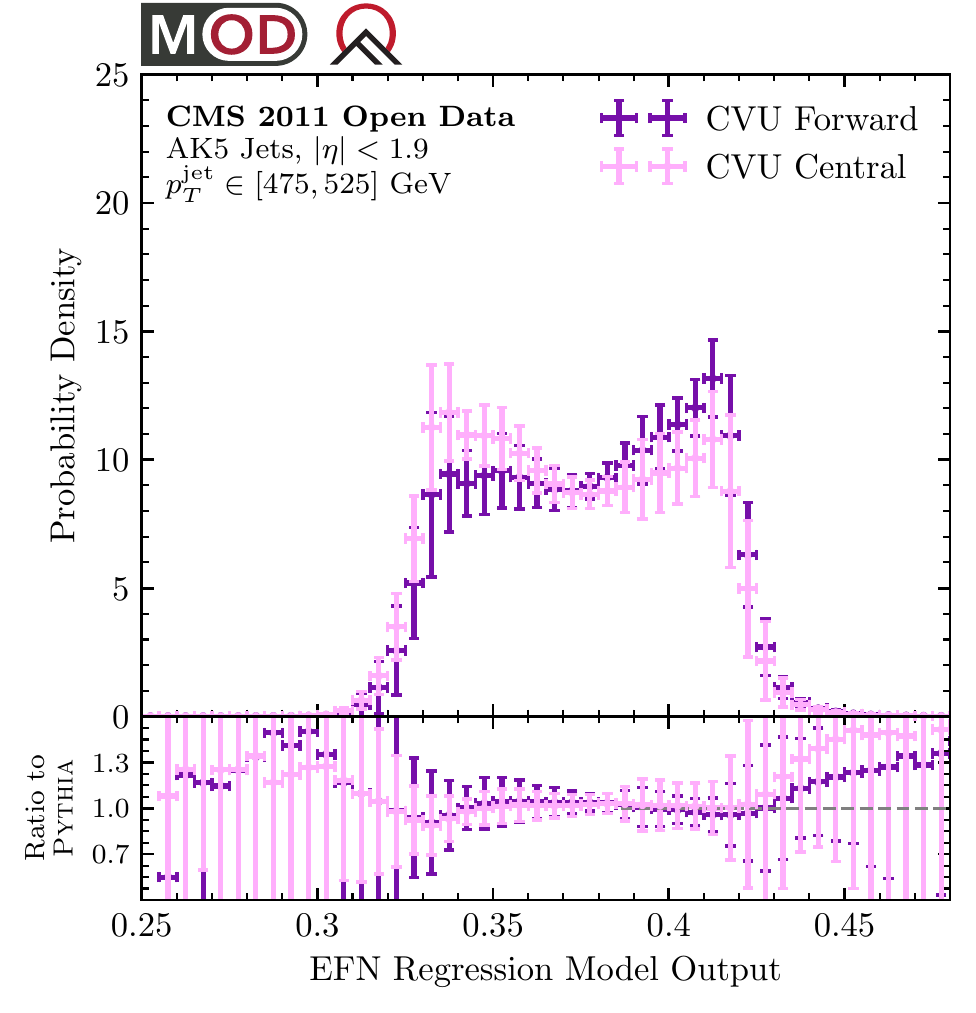}\label{fig:f_vs_c_EFNreg}}
	\caption{Similar to \Fig{fig:forw_vs_cent_gen}, but for the four machine-learned observables from \Sec{subsec:observables}:  (a) PFN classification, (b) PFN regression, (c) EFN classification, and (d) EFN regression.
The uncertainties include fully correlated uncertainties from the training.
For an individual training, there is better separation power than implied here.
For visual clarity, we have omitted the \Pythia distributions from the top panels.
	}
	\label{fig:forw_vs_cent_ML_obses}
\end{figure*}

\section{Extracting Reducibility Factors}
\label{section:extracting_k}

In this section, we apply the methods described in \Sec{sec:theory} to extract reducibility factors and determine the quark fractions of the forward/central CMS Open Data jet samples.
All studies are based on the CVU dataset obtained in \Sec{sec:cvu}.
\cor{All plots in the remainder of this paper are based on particle-level information.}

\subsection{Unfolded Forward/Central Distributions}

The starting point for the jet topics analysis is particle-level observable distributions.
After central value unfolding, we have a sample of forward jets ($|\eta| \in [0.65,1.9]$) and central jets ($|\eta| < 0.65$) with full particle-level kinematic information.
We restrict our attention to the particle-level jet kinematic range of:
\begin{equation}
	p_T \in [475,525]~\GeV, \qquad |\eta| < 1.9.
\end{equation}
This yields a sample of 3.25 millions jets for analysis.

In \Fig{fig:forw_vs_cent_gen}, we plot the six observables from \Sec{subsec:observables} for these two samples.
While the forward and central distributions are rather similar, the differences are enough to yield a non-trivial result from jet topic modeling.
As expected, the forward jets are more quark-like than the central jets, with fewer jet constituents and smaller invariant masses.

The unfolding induces 10\%-20\% differences from the baseline \Pythia distributions, which will have a noticeable impact on the results below.
The unfolding primarily affects the constituent multiplicity and jet mass distributions, with more modest differences in the other observables.
In \App{sec:pythia_only_analysis}, we repeat the jet topics analysis on the \Pythia samples for comparison.

In \Fig{fig:forw_vs_cent_ML_obses}, we show distributions for the four machine-learned observables from \Sec{subsec:observables}.
We emphasize that these observables were trained on the \Pythia samples but the distributions in the top panel are derived from the CVU samples.
The uncertainties in these distributions include both statistical uncertainties and fully-correlated variance from doing 10 different training runs.
While the training uncertainties are large in the bulk of the distribution, they are small (and correlated) near the endpoints, which is why we get relatively small uncertainties on the extracted reducibility factors.
We again see 10\%-20\% differences from the \Pythia baseline in the core of the distribution, which will show up in the jet topics results.

%%%%%%%%%%%%%%%%%%%%%%%%%%%%%%%%%%%%%%%%%%%%%%%%%%%%%%%%%%%%%%%%%%%%%%%%%%%

\subsection{Anchor Bin Method}
\label{sec:anchor}

As our first method to extract reducibility factors, we apply the anchor bin method from \Sec{subsection:bin_methods}.
This method requires choosing a histogram binning for the observable distributions $p_{M_a}(\figO)$. 
We use a quantile-based binning to address the statistical noise of the samples. 
Specifically, we chose the bins to minimize the number of events per bin, while simultaneously ensuring that no fewer than 1.0\% of central jets and 1.0\% of forward jets are contained in each bin. 
For continuous observables, this variable binning yields approximately the same number of events per bin.
For discrete observables like constituent multiplicity, this forces larger bin widths in the endpoint regions where statistical noise would otherwise dominate.

With around 15k events from each sample per bin, the Poisson uncertainty in the bin occupancy is around $1\%$.
To compute the statistical uncertainties, we use the standard formulae for weighted events.
Letting $B_i$ be the set of all events assigned to the $i$-th histogram bin and $p_i$ be the probability in that bin, we have:
\begin{align}
	\label{eq:mean_loglike}
	{\overline{p}}_i &= \sum_{x\in B_i}\omega(x),\\
	\label{eq:err_loglike}
	\text{Var}[p_i] &= \sum_{x\in B_i}{\omega(x)^2},
\end{align}
where $x$ represents the full kinematics, and $\omega(x)$ is the event weight.
These uncertainties are computed for the forward and central mixtures separately.
For the log-likelihood ratio $f(\mathcal{O}) = \ln{L_{M_2/M_1}(\mathcal{O})}$, we use standard error propagation in the Gaussian limit for each bin:
\begin{equation}
	\Delta f(\figO) = \sqrt{\left(\frac{\Delta p_{M_1}(\figO)}{p_{M_1}(\figO)}\right)^2+\left(\frac{\Delta p_{M_2}(\figO)}{p_{M_2}(\figO)}\right)^2}.
	\label{eq:delta_f}
\end{equation}

\begin{figure}[t]
	\includegraphics[width=0.45\textwidth]{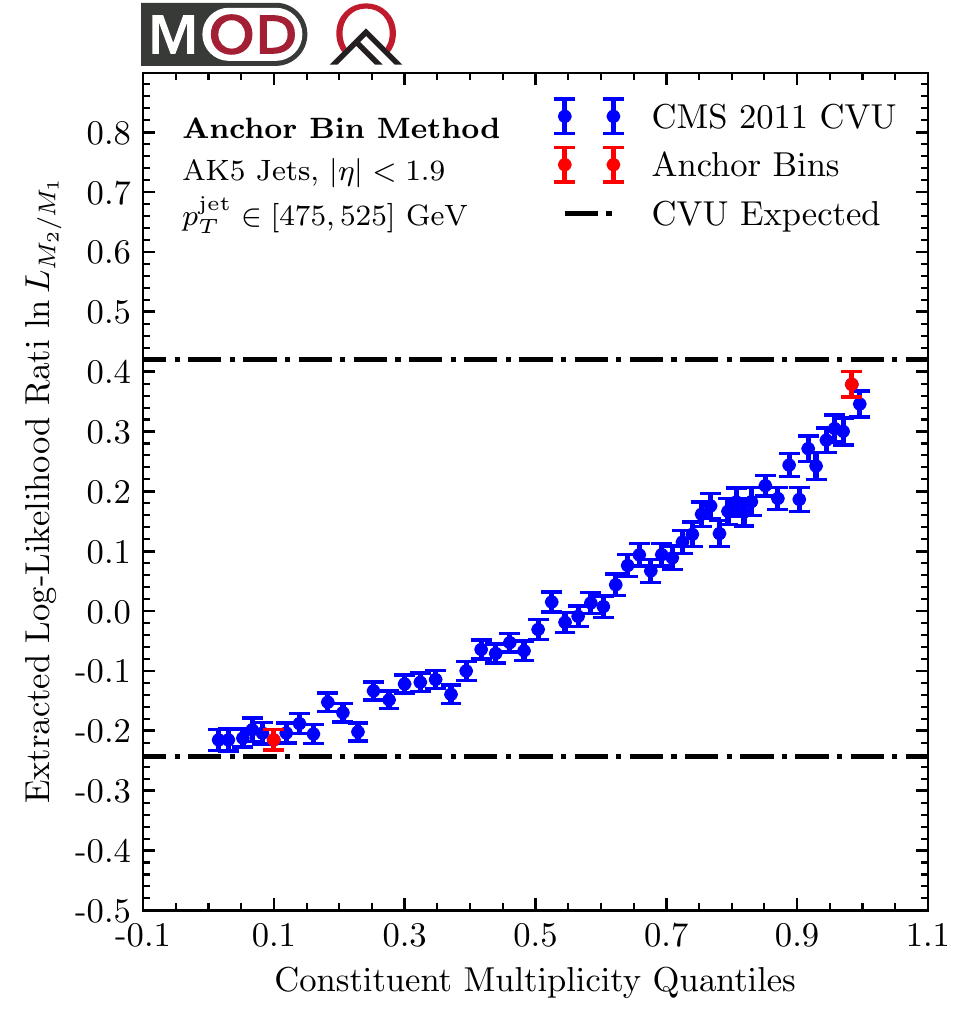}
	\caption{
	Demonstration of the anchor bin method on the CMS 2011 Open Data with central value unfolding.
	After binning the data into quantiles, we plot the log-likelihood ratio between the central and forward samples.
	The anchor bins, shown in red, are at the extrema of this ratio after accounting for uncertainties.
	The horizontal dashed lines correspond to the \Pythia parton-level expectation in \Eq{eq:quark_fractions_CVU}.
	The error bars correspond to statistical uncertainties only.
	}
	\label{fig:loglikelihood_example}
\end{figure}

In \Fig{fig:loglikelihood_example}, we show a typical log-likelihood ratio, in this case for constituent multiplicity.
To account for statistical noise, we use a slightly different procedure to determine the anchor bins from the one described in \Sec{subsection:bin_methods}.
Following \Refs{Metodiev:2018ftz,Komiske:2018vkc}, instead of directly searching for minimum and maximum values of $f(\figO)$, we choose the anchor bins based on values adjusted for uncertainty:
\begin{align}
	\kappa_{21} = \min_\figO \left(f(\figO)+\Delta f(\figO)\right), \\
	\kappa_{12} = \max_\figO \left(f(\figO)-\Delta f(\figO)\right).
\end{align}
The red dots in \Fig{fig:loglikelihood_example} correspond to the anchor bins.
Their values match reasonably well to the horizontal dashed lines, which are the expected reducibility factors given the quark fractions estimated in \Eq{eq:quark_fractions_CVU}.
Full results from the anchor bin method are summarized in \Fig{fig:all_results} below.

%%%%%%%%%%%%%%%%%%%%%%%%%%%%%%%%%%%%%%%%%%%%%%%%%%%%%%%%%%%%%%%%%%%%%%%%%%%

\begin{figure*}[t]
	\subfloat[]{
	\includegraphics[width=0.45\textwidth]{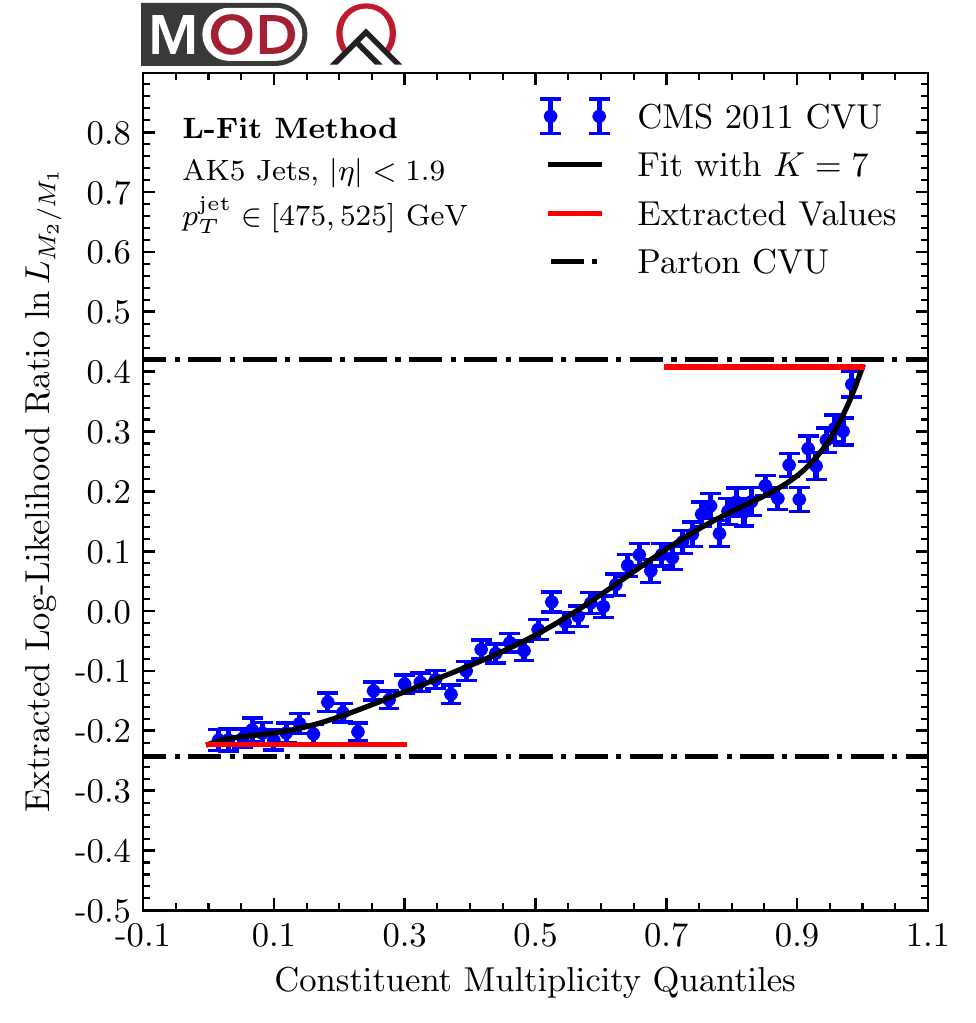}
	\label{fig:pdf_fit}
	}
	$\quad$
	\subfloat[]{
		\includegraphics[width=0.45\textwidth]{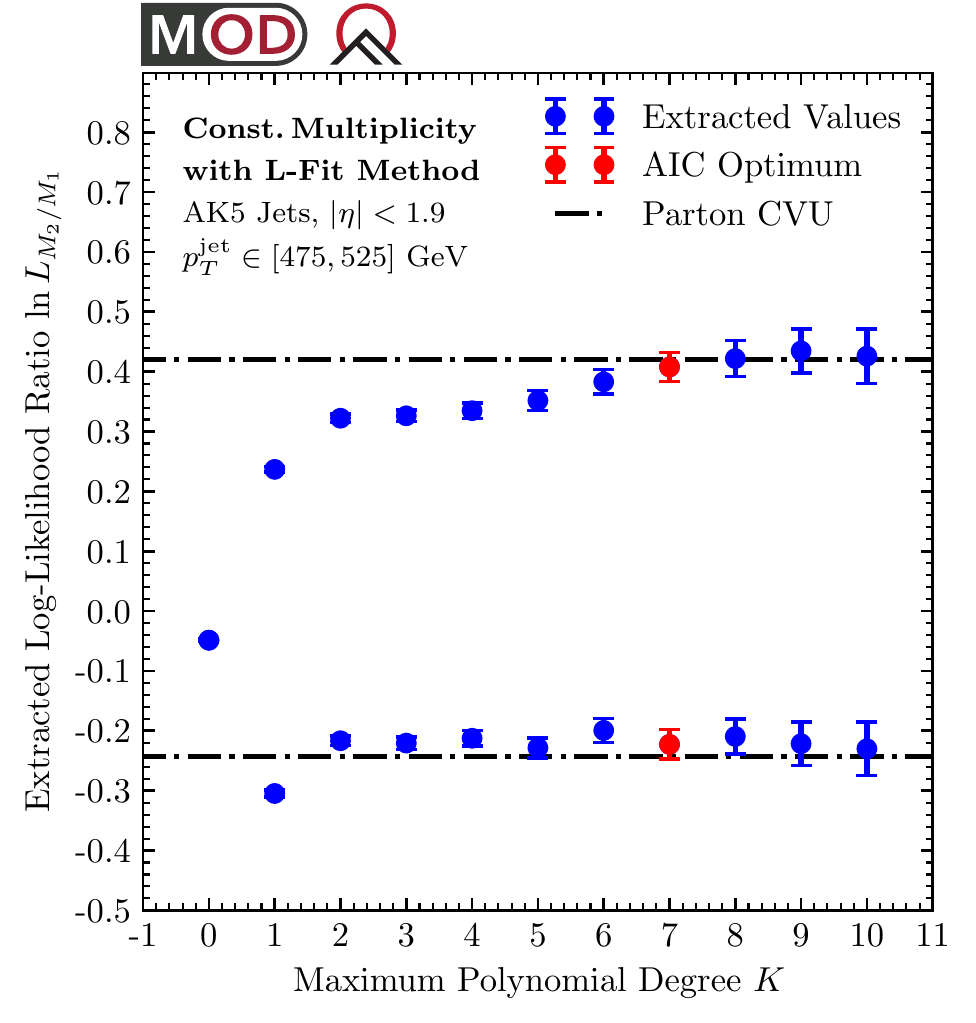}
		\label{fig:pdf_anchor_mult}
	}
	
	\caption{Demonstration of the L-fit method on the CMS 2011 Open Data with central value unfolding.
	(a) The same likelihood ratio as \Fig{fig:loglikelihood_example} but now fit to a $K = 7$ degree polynomial.  The anchor values (red) are at the extrema of the fitted curve.
	(b) Extracted anchor values as a function of the degree of the fit polynomial.  The red point indicates the optimum determined from the AIC in \Eq{eq:chi2_lfit}.
	The error bars correspond to statistical uncertainties only.
	}
	\label{fig:Lfit_example}
\end{figure*}

\subsection{Log-Likelihood Ratio Fit Method}
\label{sec:lfit}

We now turn to the log-likelihood ratio fit (L-fit) method from \Sec{subsection:l-fit}.
Like the anchor bin method, it is based on the log-likelihood ratio of the two mixtures, but it uses curve fitting to a functional form.

This method still has some dependence on the choice of binning, though the binning impact is much smaller than for the anchor bin method.
Because of the reduced binning sensitivity, we use finer bins than in \Sec{sec:anchor}, such that each bin contains at least 0.1\% of jets from the central and forward mixtures.
We use the same central value and statistical uncertainty estimates from \Eqs{eq:mean_loglike}{eq:err_loglike}.

When expressed in quantiles, the log-likelihood ratio $f(\mathcal{O}^\mathrm{quant}) = \ln{L_{M_2/M_1}(\mathcal{O}^\mathrm{quant})}$ is fit to a polynomial of the form:
\begin{equation}
f(\mathcal{O}^\mathrm{quant}; b) = \sum_{k = 0}^K b_k \, \figL_k(\mathcal{O}^\mathrm{quant}),
\end{equation} 
where $K$ is the maximum degree used in the fit and $\figL_k(\mathcal{O}^\mathrm{quant})$ is a Legendre polynomial of degree $k$ as a function of $(2 \, \mathcal{O}^\mathrm{quant}-1)$, such that the Legendre polynomial domain $[-1, 1]$ is mapped to the quantile domain $[0, 1]$.
The fit coefficients are determined by minimizing the $\chi^2$ statistic:
\begin{equation}
	\label{eq:chi2_l-fit}
	\chi^2 = \sum_{i, j = 1}^{|B|} \big(\bar{f}_i - f(\figO_i; b)\big) \, W_{ij} \, \big(\bar{f}_j - f(\figO_j; b)\big),
\end{equation}
where $\bar{f}_i$ is the bin value, $W_{ij}$ is the inverse covariance matrix, and $|B|$ is the total number of bins.
Because the bins are uncorrelated (up to normalization effects), the covariance matrix is diagonal, with diagonal elements corresponding to the bin variances:
\begin{equation}
	W_{ij} = \frac{1}{\Delta f_i^2} \delta_{ij}.
\end{equation}
To implement $\chi^2$ minimization, we use the \texttt{curve\_fit()} algorithm from the \textsc{SciPy} \textsc{Python} library \cite{virtanen_scipy_2020} to perform gradient descent on the $b_k$ coefficients.

\begin{figure*}[t]
	\subfloat[]{
	\includegraphics[width=0.45\textwidth]{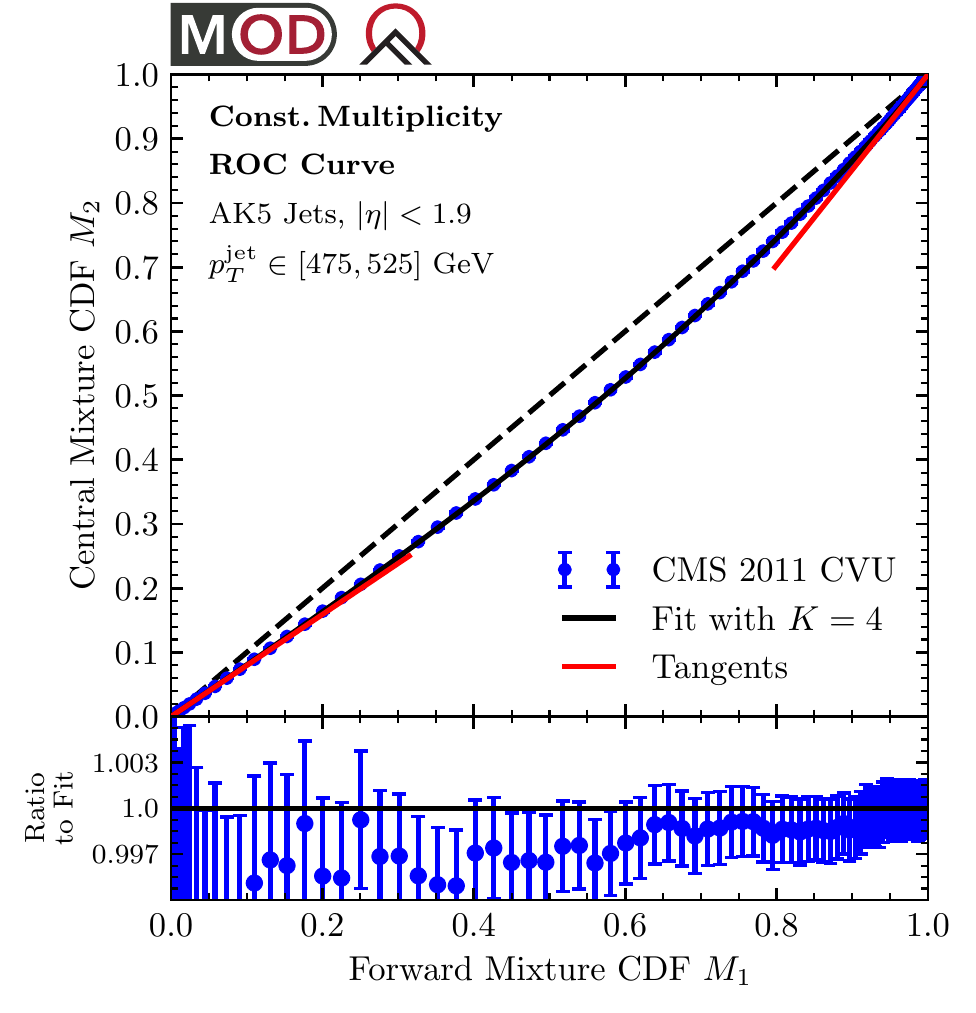}
	\label{fig:roc_curve}
	}
	$\quad$
	\subfloat[]{
		\includegraphics[width=0.45\textwidth]{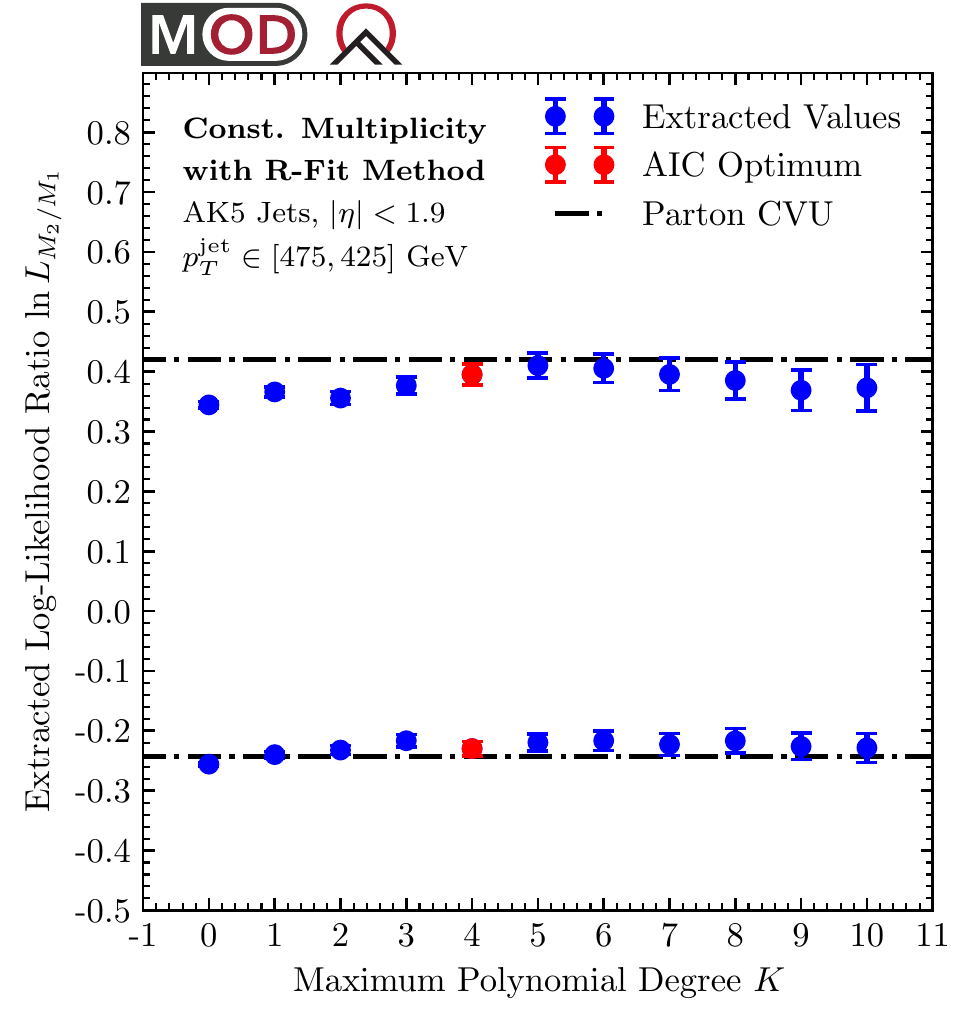}
	\label{fig:log_likelihood_roc}
	}
	\caption{Demonstration of the R-fit method on the CMS 2011 Open Data with central value unfolding.
	(a) ROC curve for distinguishing the central from forward samples, along with a polynomial fit of modified degree $K$.  The tangent lines (red) determine the anchor values.
	(b) Extracted anchor values as a function of the degree of the fit polynomial.  The red point indicates the optimum determined from the AIC in \Eq{eq:chi2_lfit}.
	The error bars correspond to statistical uncertainties only.
}
	\label{fig:r_fit_method_all}
\end{figure*}

The L-fit method depends on the maximum polynomial degree $K$ of the fit.
Following \Ref{Williams:2017gwf}, we use the Akaike Information Criterion (AIC) \cite{AIC_original} to choose the optimal model that minimizes:
\begin{equation}
	\chi^2 + 2 \, (K + 1),
	\label{eq:chi2_lfit}
\end{equation}
where $K + 1$ is the number of fit parameters.
To compute the reducibility factors, we evaluate the optimal model at the endpoints.
Because the rescaled Legendre polynomials are either $1$ or $-1$ at $\figO^{\rm quant}_{\min} = 0$ and $\figO^{\rm quant}_{\max} = 1$, this yields relatively simple expressions for the anchor values: 
\begin{align}
	\label{eq:kappa21_final_lfit}
	a_{21} &=\sum_{k=0}^{K}{b^{\text{opt}}_k (-1)^k},\\
	\label{eq:kappa12_final_lfit}
	a_{12} &= \sum_{k=0}^{K}b^{\text{opt}}_k,
\end{align}
which can be converted into reducibility factors using \Eq{eq:kappasanchor}.

The $\chi^2$ minimization method not only allows us to extract optimal coefficients $b^{\text{opt}}$, but also their covariance matrix near the optimum point $\text{Cov}[b^{\text{opt}}_i, b^{\text{opt}}_j]$.
This matrix contains information about the statistical uncertainties associated with the method and allows us to compute the uncertainties of the anchor values via:
\begin{align}
		\label{eq:kappa21_final_lfit_err}
	%\Delta a_{21}^2 &= \sum_{i,j = 0}^K b^{\text{opt}}_i b^{\text{opt}}_j (-1)^{i+j} \text{Cov}[b^{\text{opt}}_i, b^{\text{opt}}_j],
	\Delta a_{21}^2 &= \sum_{i,j = 0}^K  (-1)^{i+j} \text{Cov}[b^{\text{opt}}_i, b^{\text{opt}}_j],
	\\
	\label{eq:kappa12_final_lfit_err}
	%\Delta a_{12}^2 &= \sum_{i,j = 0}^K b^{\text{opt}}_i b^{\text{opt}}_j \text{Cov}[b^{\text{opt}}_i, b^{\text{opt}}_j].
	\Delta a_{12}^2 &= \sum_{i,j = 0}^K  \text{Cov}[b^{\text{opt}}_i, b^{\text{opt}}_j].
\end{align}

An example of the L-fit method is shown in \Fig{fig:Lfit_example}.
In \Fig{fig:pdf_fit}, we show the optimal polynomial fit for the case of multiplicity.
The anchor values, which are indicated by the red horizontal line, are comparable to the estimated values from \Eq{eq:quark_fractions_CVU}
In \Fig{fig:pdf_anchor_mult}, we show the extracted anchor values as a function of the maximum degree $K$, where the red dots indicate the choice that minimizes the AIC expression in \Eq{eq:chi2_lfit}.

In addition to the statistical uncertainty, we assess a systematic uncertainty from model selection.
Specifically, we minimize two alternative version of \Eq{eq:chi2_lfit}, where the proportionality coefficient of 2 in front of $K+1$ is replaced by either 1 or 4.
The half-difference between the minimum and maximum of the anchor value produced by these three models is used to estimate the systematic uncertainty, which is added to the statistical uncertainty in quadrature.
Typically the anchor values are quite robust to this coefficient change.
Full results from the L-fit method are summarized in \Fig{fig:all_results} below.

%%%%%%%%%%%%%%%%%%%%%%%%%%%%%%%%%%%%%%%%%%%%%%%%%%%%%%%%%%%%%%%%%%%%%%%%%%%

\subsection{ROC Curve Fit Method}
\label{sec:roc}

Our last method for extracting reducibility factors is the ROC curve fit (R-fit) method from \Sec{subsection:roc}.
In this approach, we build a ROC curve using observable cumulative probability distributions and then fit this curve to a functional form to determine its slope at the endpoints.
While the R-fit method does not require binning the data, for reasons of computational efficiency we use the same binning procedure as in \Sec{sec:lfit} for continuous observables.
For discrete observables, we use the full set of bins.

Unlike for the probability distribution function (PDF), mean values for the cumulative distribution function (CDF) are correlated. 
Therefore, to assess statistical uncertainties, we need to compute a full covariance matrix for the CDF.
For cumulative bin $C_i$, the cumulative probability $\overline{P}_i$ is estimated in the standard way: 
\begin{equation}
	\label{eq:mean_roc}
	\overline{P}_i = \sum_{x\in C_i}\omega(x).
\end{equation}
The covariance between cumulative bins $C_i$ and $C_j$ with $i \le j$ (i.e.~bin $i$ is fully contained by bin $j$) is:
\begin{equation}
	\label{eq:cov_roc}
	\text{Cov}[P_i,P_j] = \sum_{x \in C_i} \omega^2(x) - \frac{1}{N} \overline{P}_i \overline{P}_j,
\end{equation}
where $N$ is the total number of events.
Note that the symmetry between $i$ and $j$ is broken because the sum in the first term only goes over entries in bin $i$.
Unlike for \Eq{eq:err_loglike}, here we are keeping track of the $1/N$ correction to the covariance.

\begin{figure*}[t]
	\subfloat[]{
	\includegraphics[width=0.45\textwidth]{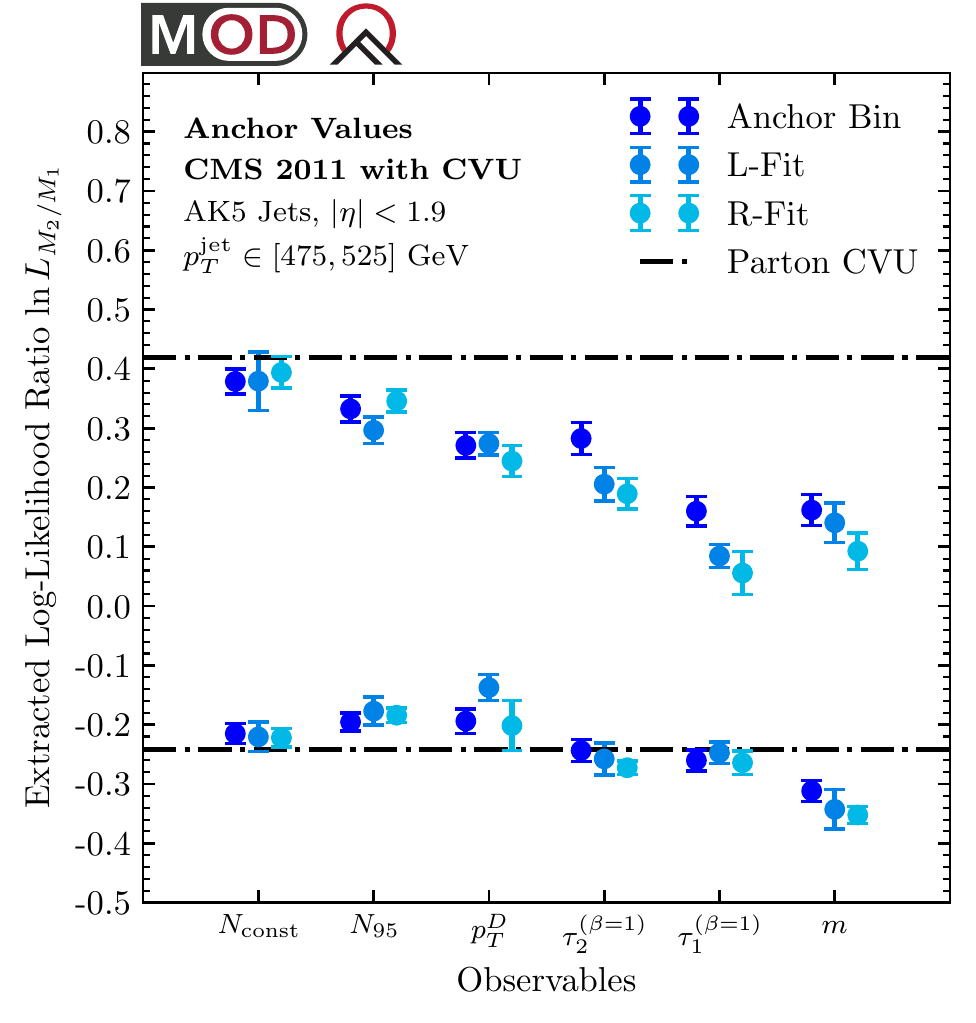}
	\label{fig:obs_results}
	}
	$\quad$
	\subfloat[]{
	\includegraphics[width=0.45\textwidth]{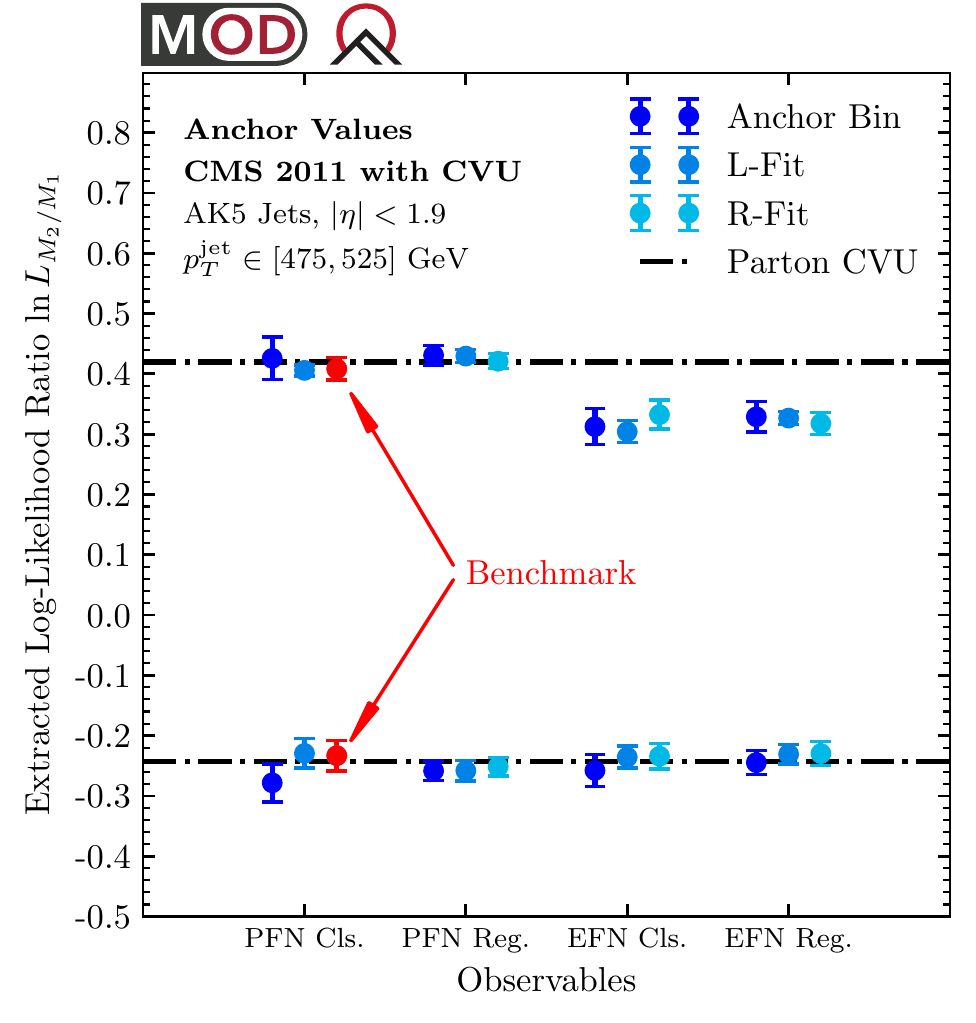}
	\label{fig:ML_results}
	}
	\caption{Comparing the extracted anchor values between the anchor bin method (blue), L-fit method (azure), and R-fit method (cyan).
	Shown are results for the (a) six substructure and (b) four machine-learned observables from \Sec{subsec:observables}, applied to the CMS 2011 Open Data with central value unfolding.
	Our benchmark method using the R-fit method with PFN classification is indicated in red.
	The error bars correspond to statistical uncertainties combined with the AIC uncertainty in quadrature.
	}
	 \label{fig:all_results}
\end{figure*}

Evaluating the statistical uncertainty for a ROC curve is more complex than for the methods in \Secs{sec:anchor}{sec:lfit}.
The ROC curve expresses the CDF for $M_2$ as a function of the CDF for $M_1$.
Thus, there are statistical uncertainties in the $x$-coordinate of the ROC curve that need to be converted to an additional effective noise in the $y$-coordinate in order to do error propagation on the fit form.
Let $r_i$ be the $y$ value of the ROC curve associated with the $i$-th cumulative bin, and let superscripts ${}^{(a)}$ refer to the mixture $M_{a}$.
The covariance between $r_i$ and $r_j$ is:
\begin{equation}
	\label{eq:COV_ROC_full}
	\text{Cov}[r_i, r_j] = \text{Cov}[P^{(2)}_{i}, P^{(2)}_{j}] +  \frac{\overline{p}^{(2)}_{i}\overline{p}^{(2)}_{j}}
	{\overline{p}^{(1)}_{i}\overline{p}^{(1)}_{j}}
	\text{Cov}[P^{(1)}_{i}, P^{(1)}_{j}],
\end{equation} 
where $p^{(a)}$ are the PDFs of the mixture distributions (i.e.~the slopes of the CDFs).
With this derivation, we are assuming that there are no correlations between the two mixtures.

To extract the anchor values, we fit the ROC curve to 
\begin{equation}
	f(r) = r + r (1-r)\left(\sum_{k=0}^{K}{b_k\, \mathcal{L}_k(r)}\right),
	\label{eq:fit_form}
\end{equation}
where $\mathcal{L}_k\left(x\right)$ is again a rescaled Legendre polynomial of degree $k$ defined on the interval $[0, 1]$.
Like in \Sec{sec:lfit}, we perform $\chi^2$ minimization with covariances given by \Eq{eq:COV_ROC_full}, with the optimal maximum degree $K$ and associated fit coefficients $b^{\text{opt}}_k$ determined by the AIC in \Eq{eq:chi2_lfit}.

As shown in \Eqs{eq:Rfit_kappa21}{eq:Rfit_kappa12}, the reducibility factors are related to the slopes at the endpoint of the ROC curve:
\begin{align}
	\label{eq:kappa12_final_roc}
	\kappa_{21} &= \left. \frac{df}{dr}\right|_{r = 0}=1+\sum_{k=0}^{K}{b^{\text{opt}}_k\, (-1)^k},\\
	\label{eq:kappa21_final_roc}
	\frac{1}{\kappa_{12}} &= \left. \frac{df}{dr}\right|_{r = 1}=1+\sum_{k=0}^{K}b^{\text{opt}}_k.
\end{align}
The associated uncertainties are:
\begin{align}
	\label{eq:kappa12_final_roc_err}
	\Delta \kappa_{12} &= \sum_{i,j = 0}^K (-1)^{i+j} \text{Cov}[b^{\text{opt}}_i, b^{\text{opt}}_j]
	\\
	\label{eq:kappa21_final_roc_err}
	\Delta \left( \frac{1}{\kappa_{21}} \right) &= \sum_{i,j = 0}^{K} \text{Cov}[b^{\text{opt}}_i, b^{\text{opt}}_j],
\end{align}
where $\text{Cov}[b^{\text{opt}}_i, b^{\text{opt}}_j]$ is determined from the $\chi^2$ fit.

An example of the R-fit methods is shown in \Fig{fig:r_fit_method_all}.
In \Fig{fig:roc_curve} we show an example ROC curve, with the fitted form and the extracted slope values.
The change in reducibility factors as a function of $K$ is shown in \Fig{fig:log_likelihood_roc}, where the red dot is the one corresponding to the AIC optimum.
Full results from the R-fit method are summarized in \Fig{fig:all_results} below.

\subsection{Summary of Results}
\label{sec:summary}

We now summarize and compare results obtained from the anchor bin, L-fit, and R-fit methods.
In \Fig{fig:all_results}, we show the extracted reducibility factors and their statistical uncertainties for the six fixed substructure observables and the four machine-learned observables.
We also checked that the same statistical uncertainties could be obtained through the bootstrap procedure \cite{EfroTibs93} applied to 10k pseudo-experiments.

For a fixed observable, we see that the three methods yield similar results with comparable uncertainties on the reducibility factors.
That said, this conclusion is sensitive to our choice of binning, and the agreement can be significantly worse with different binning schemes.
The L-fit method is robust to making the bins finer, but the results degrade substantially with coarser bins.
For the anchor bin method, the results degrade if the bins are either too fine or too coarse.
So while one can adjust the binning in the L-fit and anchor bin methods to get sensible results, the R-fit method works well out of the box, as long as the bins are fine enough to yield a reasonably smooth ROC curve.
Thus, our recommended method is the R-fit method, since it yields results that almost entirely independent of the binning choice.

As expected, the reducibility factors depend sensitively on the choice of observable used for jet topic modeling.
%
%In \Fig{fig:obs_results}, we see that the better the observable is for quark/gluon discrimination, the larger the difference in reducibility factors.
\cor{In \Fig{fig:obs_results}, we see that the better the observable is for quark/gluon discrimination, the larger the difference is between the two extracted reducibility factors.}
Among the fixed observables, the best quark/gluon discriminant is constituent multiplicity, so it is encouraging that the reducibility factors match the expectations from \Eq{eq:quark_fractions_CVU}.
Mass is a poor quark/gluon discriminant, so not surprisingly, it yields relatively poor results for jet topics.

In \Fig{fig:ML_results}, we show the results from the machine-learned observables.
Here, the uncertainties include both the statistical variance within each training run and the variance from 10 different training runs, which are added in quadrature.
The PFNs, which can in principle exploit all of the information available in the jet, yield excellent separation power, for both the classification and regression strategies.
The EFNs, which restrict their attention to IRC-safe information, yield worse separation power, in agreement with the expectations from \Ref{Komiske:2018vkc}.
Note that the PFNs and EFNs agree on the value of $\kappa_{21}$, which is a general expectation from the fact that quarks are irreducible with respect to gluons (but not vice versa), even when considering just IRC-safe information~\cite{Metodiev:2018ftz,Larkoski:2019nwj}.

For the results in \Sec{sec:quark_and_gluon_dist}, we focus on PFN classification with the R-fit method as our benchmark approach for extracting reducibility factors.
See \Fig{fig:pythia_red_fac_results} in \App{sec:pythia_only_analysis} for reducibility factors extracted from the \Pythia samples.

%%%%%%%%%%%%%%%%%%%%%%%%%%%%%%%%%%%%%%%%%%%%%%%%%%%%%%%%%%%%%%%%%%%%%%%%%%%

\section{Quark and Gluon Fractions and Distributions}
\label{sec:quark_and_gluon_dist}

\begin{figure*}[p]
	\includegraphics[width=0.45\textwidth]{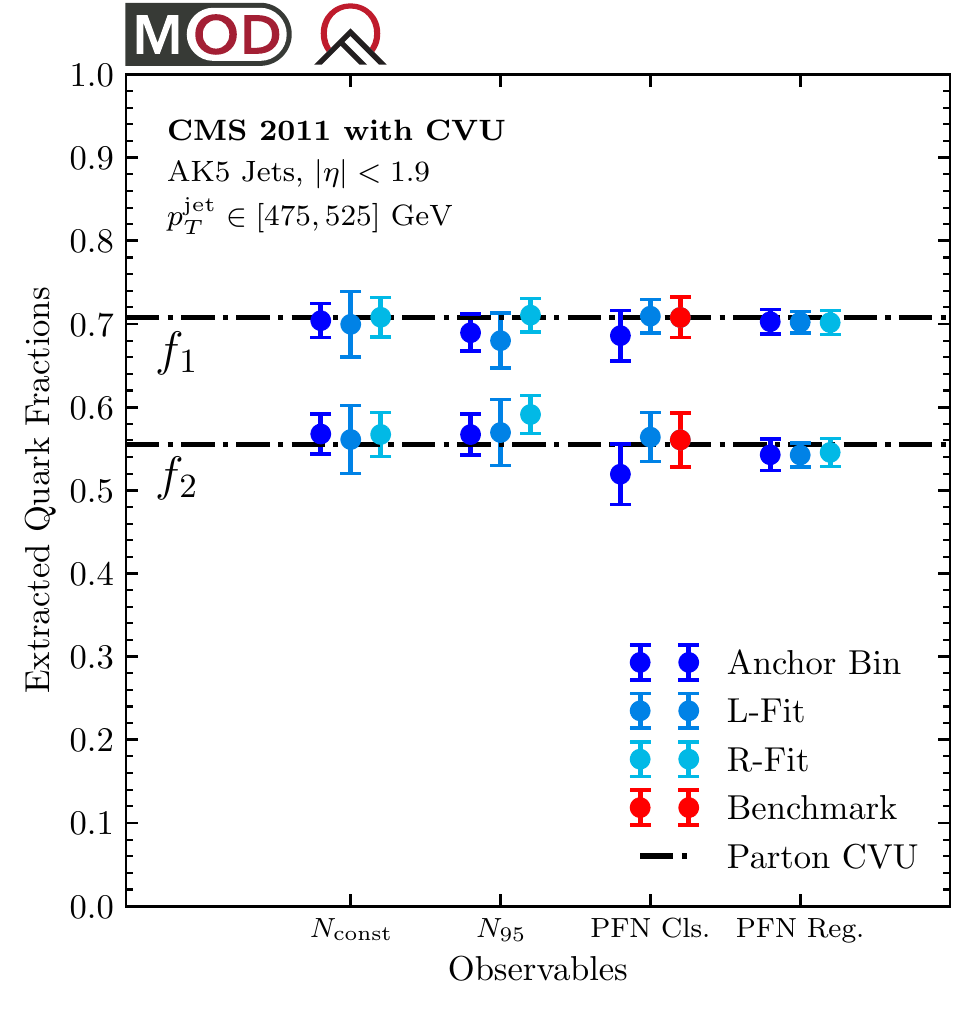}
	\caption{
	Extracted quark fractions from the CMS 2011 Open Data with central value unfolding in the forward ($M_1$) and central ($M_2$) samples.
	We focus on the four observables with the best quark/gluon classification performance and compare the anchor bin (blue), L-fit (azure), and R-fit (cyan) methods.
	The benchmark method is indicated in red.
	}
	\label{fig:all_fractions}
\end{figure*}

\begin{figure*}[p]
	\subfloat[]{
	\includegraphics[width=0.45\textwidth]{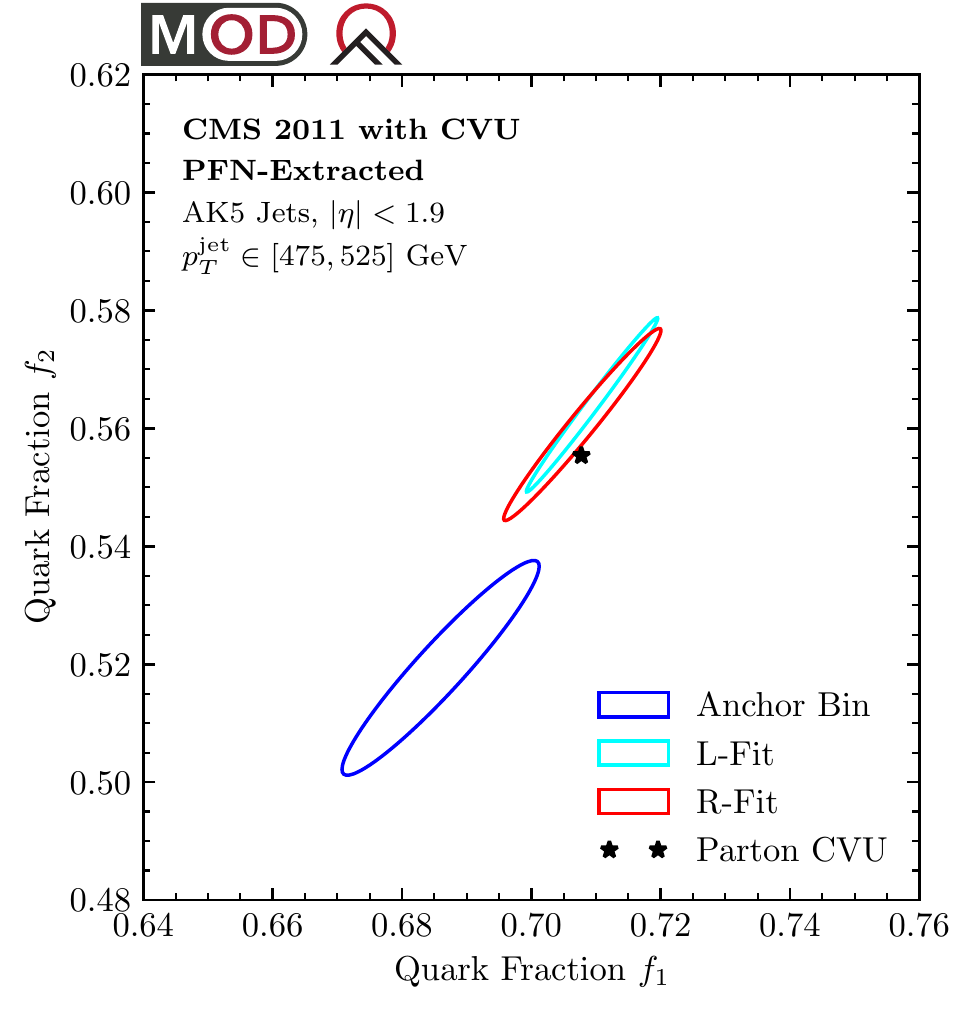}
		\label{fig:PFN_fractions}
	}
	$\quad$
	\subfloat[]{
	\includegraphics[width=0.45\textwidth]{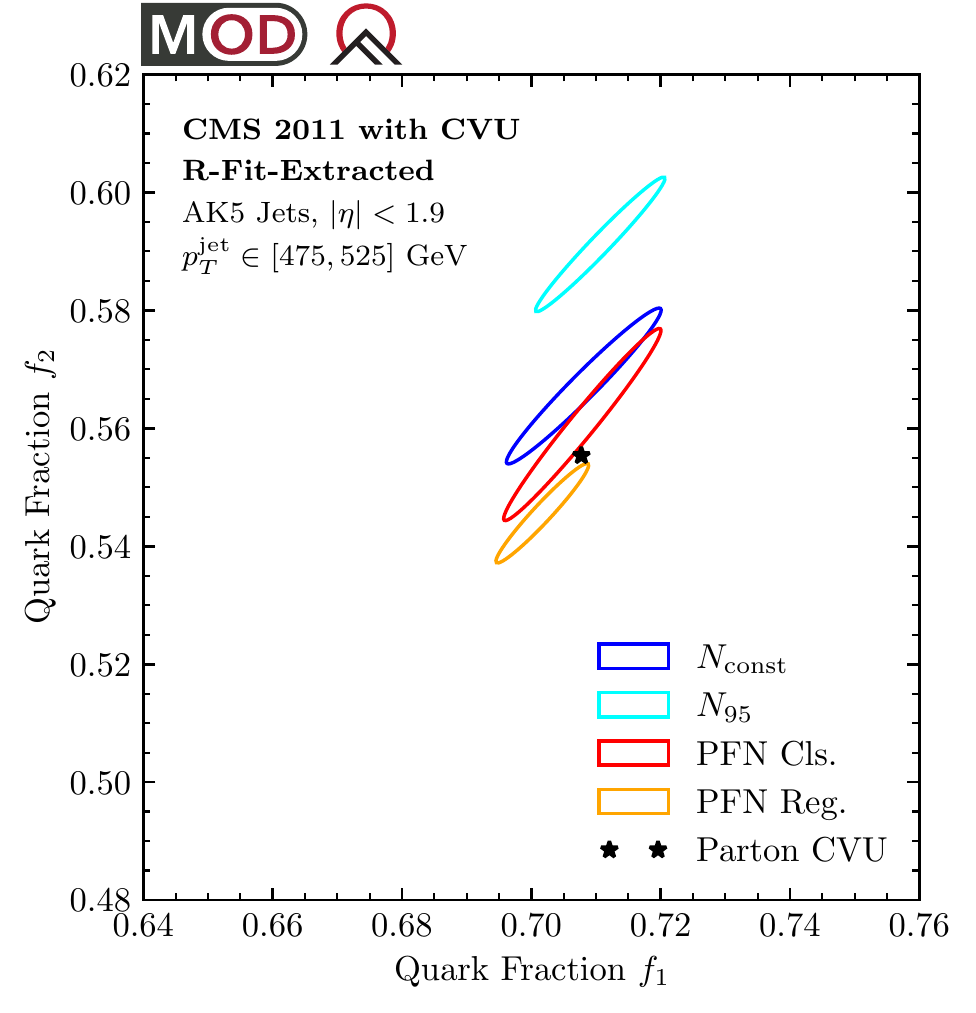}
	\label{fig:roc_fractions}
	}
	\caption{Statistical correlations for the extracted quark fractions by (a) fixing the PFN observable and sweeping over the reducibility factor method and (b) fixing the R-fit method and sweeping over the four observables in \Fig{fig:all_fractions}.  The diagonal alignment of the ellipses is primarily due to the correlations induced from the $\{\kappa_{12}, \kappa_{21}\} \to \{f_1, f_2\}$ mapping in \Eq{eq:fracs}.
	For the benchmark method in red, the extracted fractions match the CVU expectation in \Eq{eq:quark_fractions_CVU} within the statistical uncertainties (combined with AIC). 
	}
	\label{fig:ellipse_cvu}
\end{figure*}

With reducibility factors in hand, we can now isolate separate ``quark'' and ``gluon'' properties from the 2011 CMS Open Data.
The reason for the quotes is that our definition of jet flavor is based on the operational procedure described in \Sec{section:extracting_k}, which may differ from other definitions of jet flavor that have been proposed (see \Ref{Gras:2017jty} for a review).

\subsection{Quark and Gluon Fractions}

We first need to convert reducibility factors into quark fractions of the forward and central mixtures.
This conversion is straightforward using \Eq{eq:fracs}, and we use standard Gaussian error propagation to translate the uncertainties on the reducibility factors into uncertainties on the quark fractions.
The results are shown in \Fig{fig:all_fractions} for the four best discriminants:  constituent multiplicity, image activity ($N_{95}$), PFN classification, and PFN regression.
Interestingly, even though the reducibility factors for $N_{95}$ in \Fig{fig:obs_results} do not match the CVU expectations from \Eq{eq:quark_fractions_CVU}, the quark fractions approximately do, within uncertainties.

In \Fig{fig:ellipse_cvu}, we show the statistical correlations between extracted $f_1$ and $f_2$ values for selected methods.
By construction, the anchor bin method assumes no correlation between the anchor bin values, so the tilt of the ellipses is driven entirely by the correlations in \Eq{eq:fracs}.
The L-Fit and R-fit methods have a non-trivial correlation in the anchor bin values, but this correlation is relatively small compared to those induced by the $\{\kappa_{12}, \kappa_{21}\} \to \{f_1, f_2\}$ mapping.
See \Figs{fig:pythia_quark_frac_results}{fig:ellipses_pythia} in \App{sec:pythia_only_analysis} for the quark factors extracted from the \Pythia samples.

\begin{figure*}[p]
\centering
\subfloat[]{\includegraphics[width=0.33\textwidth]{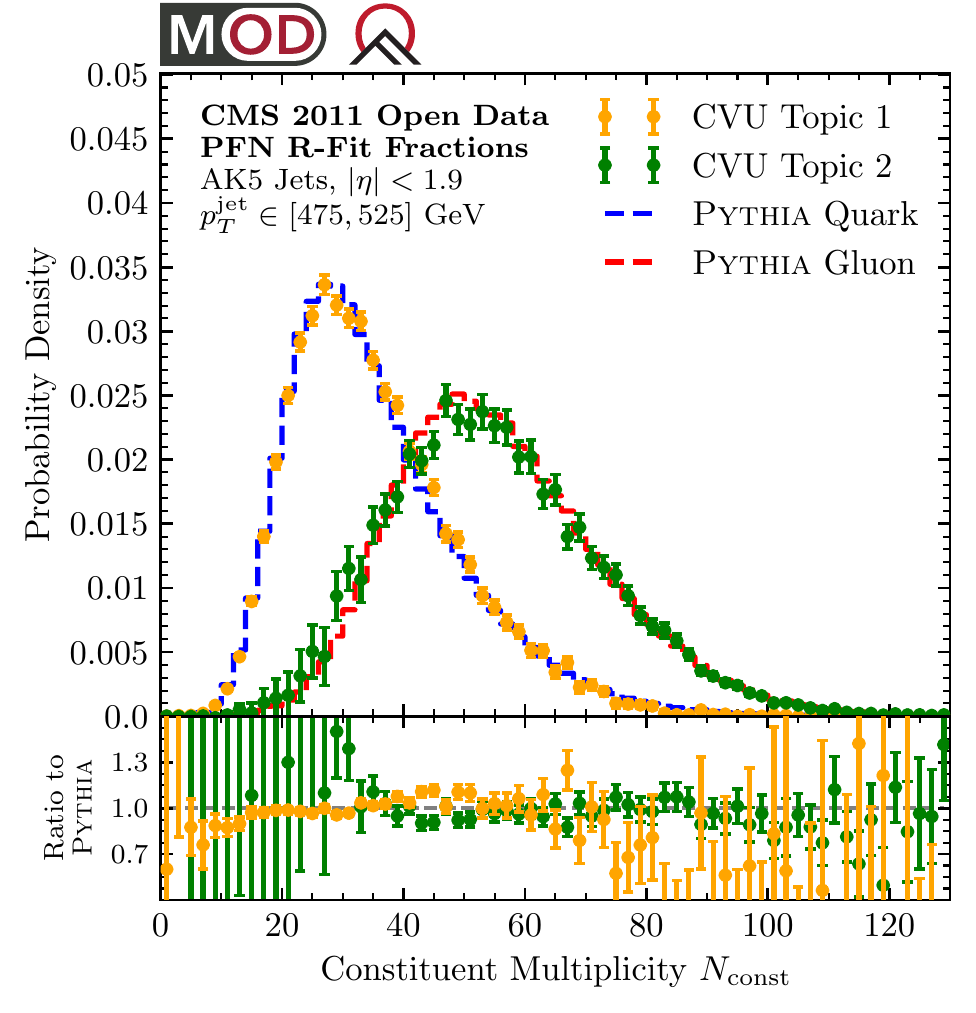}\label{fig:qg_mult}}
\subfloat[]{\includegraphics[width=0.33\textwidth]{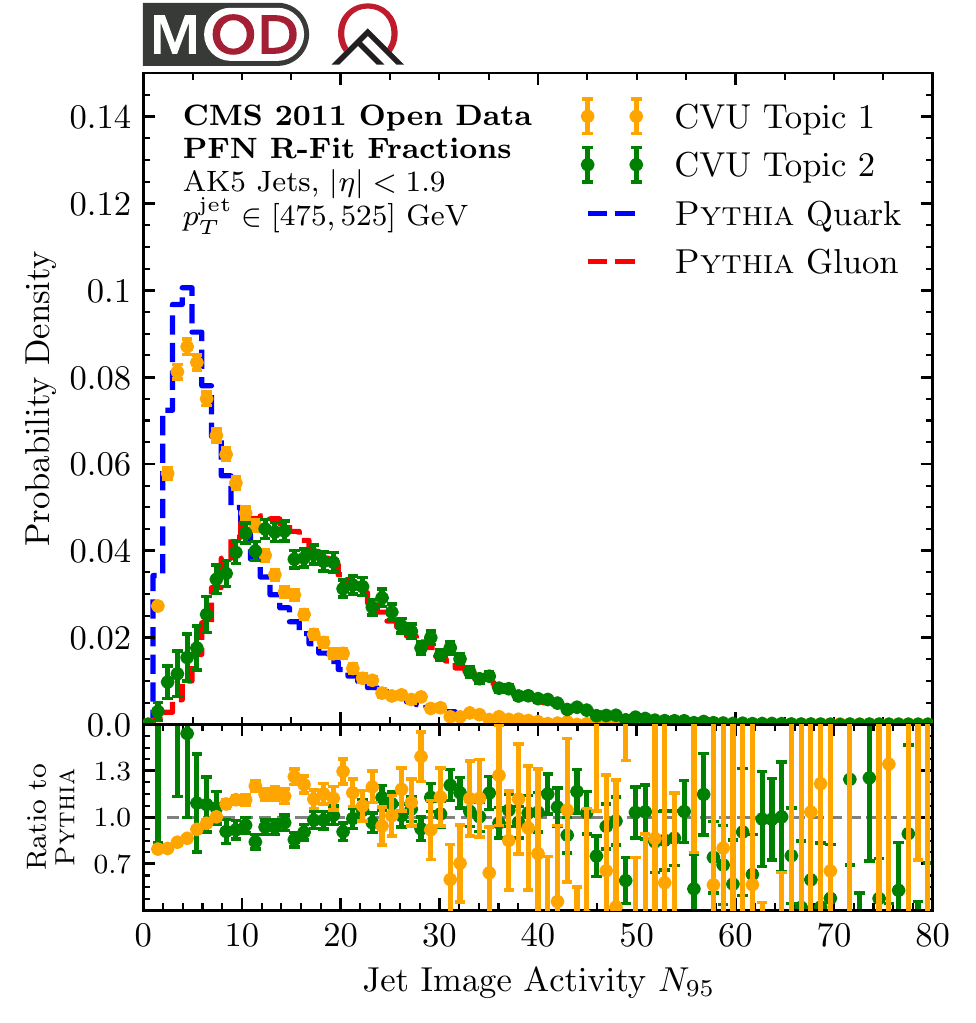}\label{fig:qg_n95}}
\subfloat[]{\includegraphics[width=0.33\textwidth]{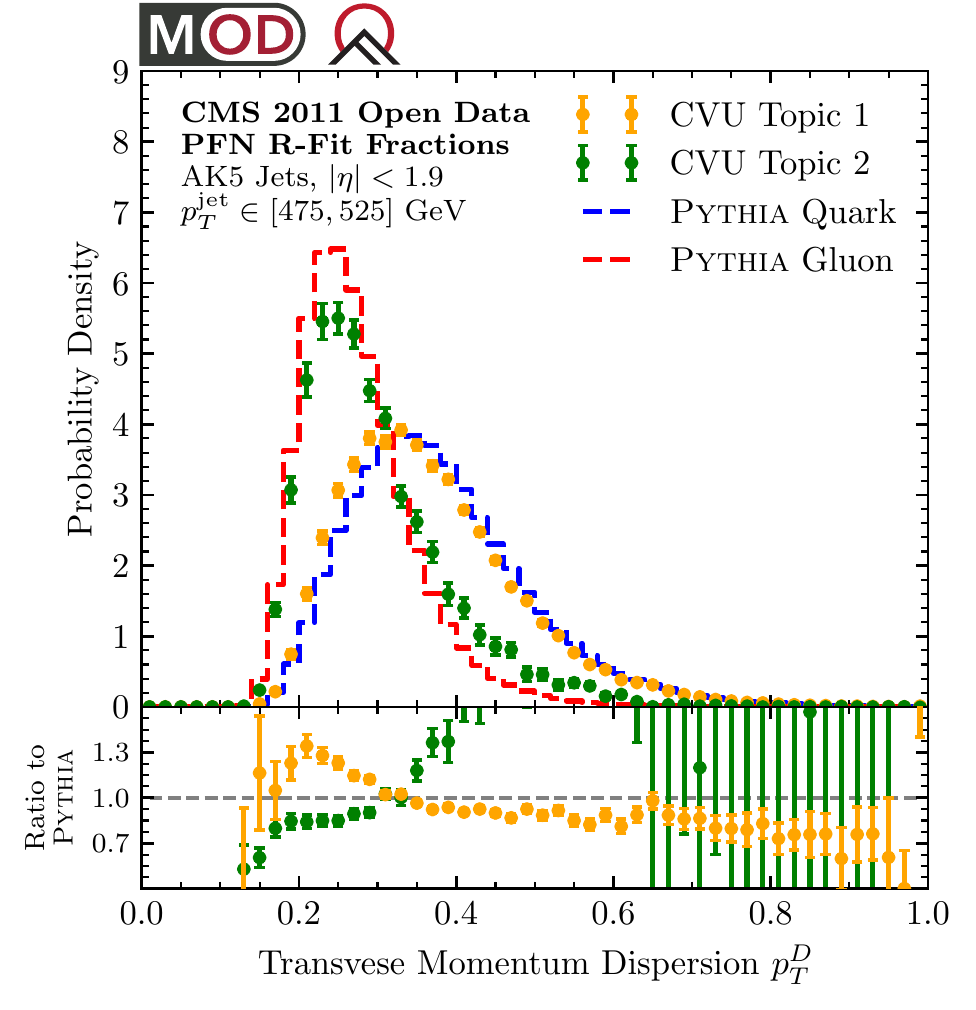}\label{fig:qg_ptd}}
\\
\subfloat[]{\includegraphics[width=0.33\textwidth]{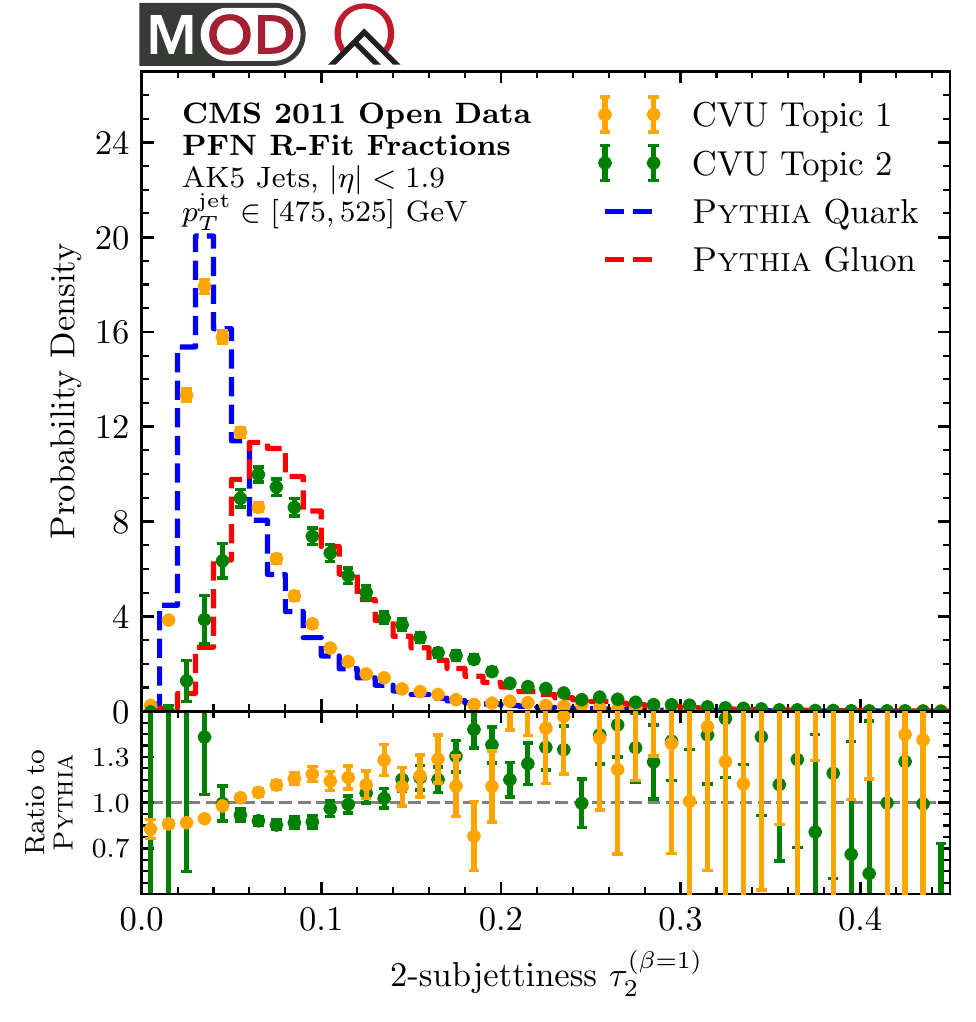}\label{fig:qg_nsub2}}
\subfloat[]{\includegraphics[width=0.33\textwidth]{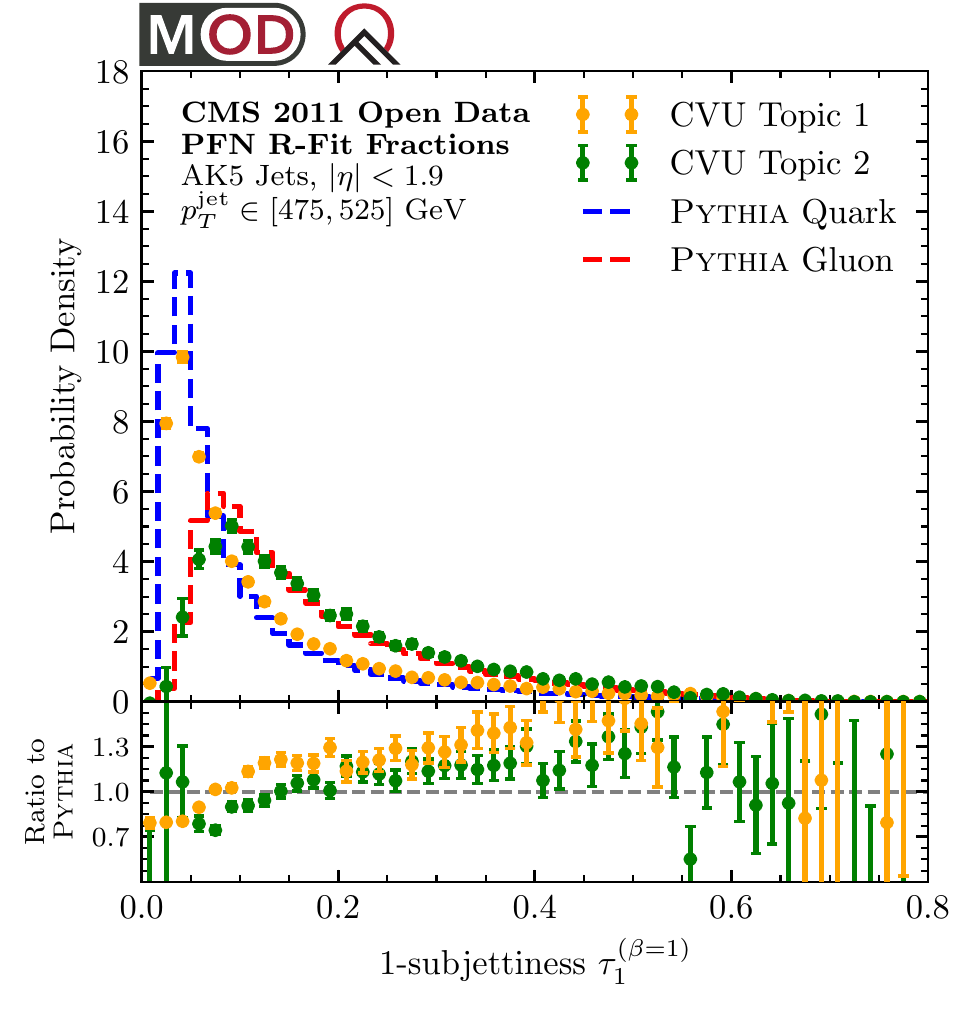}\label{fig:qg_nsub1}}
\subfloat[]{\includegraphics[width=0.33\textwidth]{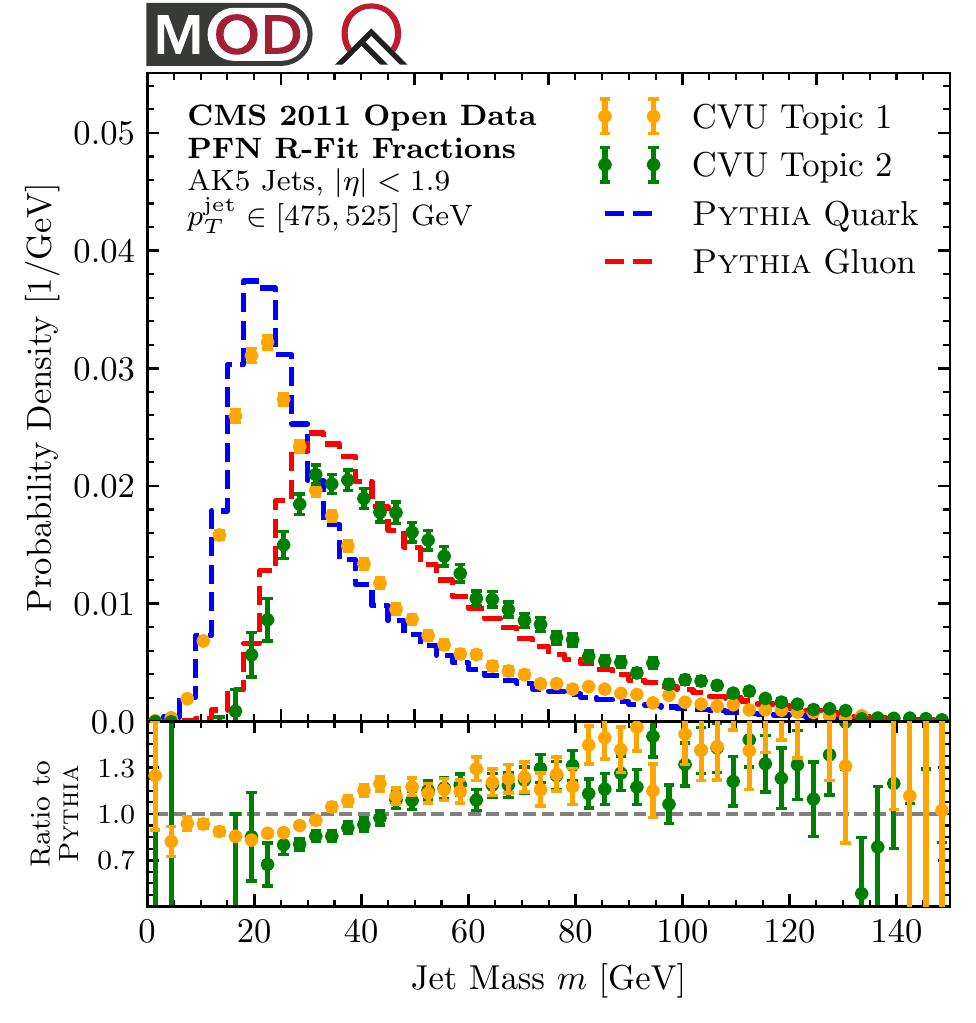}\label{fig:qg_mass}}
\caption{
Extracted ``quark'' (yellow, topic 1) and ``gluon" (green, topic 2) distributions for the six substructure observables from \Sec{subsec:observables}.
These are obtained by mixing the CMS 2011 Open Data CVU distributions from \Fig{fig:forw_vs_cent_gen} according the quark fractions from \Eq{eq:quark_fractions_PFN}.
For comparison, we show the same \Pythia truth-parton-labeled quark (blue dashed) and gluon (red dashed) distributions from \Fig{fig:forw_vs_cent_gen}.
The error bars correspond to statistical uncertainties on the distributions combined with the reducibility factor uncertainties in quadrature.
}
\label{fig:simqg}
\end{figure*}

With all of the methods in \Fig{fig:all_fractions} showing similar behavior, we choose the R-fit method with the PFN classifier as our benchmark.
As already discussed, the R-fit method is the most statistically robust, and the PFN method has somewhat more conservative uncertainties, due to fluctuations from the different machine learning training runs.
The extracted fractions are:
\begin{equation}
	\label{eq:quark_fractions_PFN}
	\text{CMS 2011 CVU Jet Topics:} \quad
	\begin{aligned}
		f_1 &\simeq 0.708 \pm 0.025, \\
		f_2 &\simeq 0.561 \pm 0.033.
	\end{aligned}
\end{equation}
We emphasize that these uncertainties do not include any experimental systematics \cor{nor} uncertainties from the unfolding.

%%%%%%%%%%%%%%%%%%%%%%%%%%%%%%%%%%%%%%%%%%

\subsection{Individual Substructure Distributions}

Given the quark fractions, we can use \Eqs{eq:p_q}{eq:p_g} to extract the ``quark''
and ``gluon'' distributions for any observable.
We emphasize that we do not need to use the same observable for extracting fractions and for deriving distributions.
We use standard error propagation to compute the uncertainties on individual flavor distributions, which depends on the uncertainties on the quark fractions from \Eq{eq:quark_fractions_PFN} and the statistical uncertainties on the binned mixture distributions.

We show jet topics results on the CMS 2011 CVU sample for six substructure distributions in \Fig{fig:simqg}.
These are compared against distributions from parton truth labels in \Pythia.
The constituent multiplicity distributions are remarkably similar between the CMS 2011 CVU jet topics analysis and \Pythia. %, suggesting that \Pythia may have been tuned to match this observable.
For image activity, the gluon jets look similar, but our extracted quark jets exhibit more radiation.
For momentum dispersion, the quark and gluon jets change in opposite directions.
For 2-subjettiness, both types of jets show a suppression of the Sudakov peak.
For 1-subjettiness and jet mass, both types of jets exhibit more radiation.
We conclude that a full analysis, with unfolding uncertainties, has the potential to tell us detailed information about modeling quark and gluon jets in parton showers.

In \App{sec:pythia_only_analysis}, we repeat the jet topics procedure on \Pythia directly.
Because jet topics is an operational procedure, there is no guarantee that the resulting distributions will match those from parton truth labels. 
Indeed, we find that the extracted quark fraction from the \Pythia samples is systematically lower than the estimate from parton truth labels.

\subsection{Tagging Performance}
\label{sec:ROC_extraction}

\begin{figure}[t]
\centering
\includegraphics[width=0.45\textwidth]{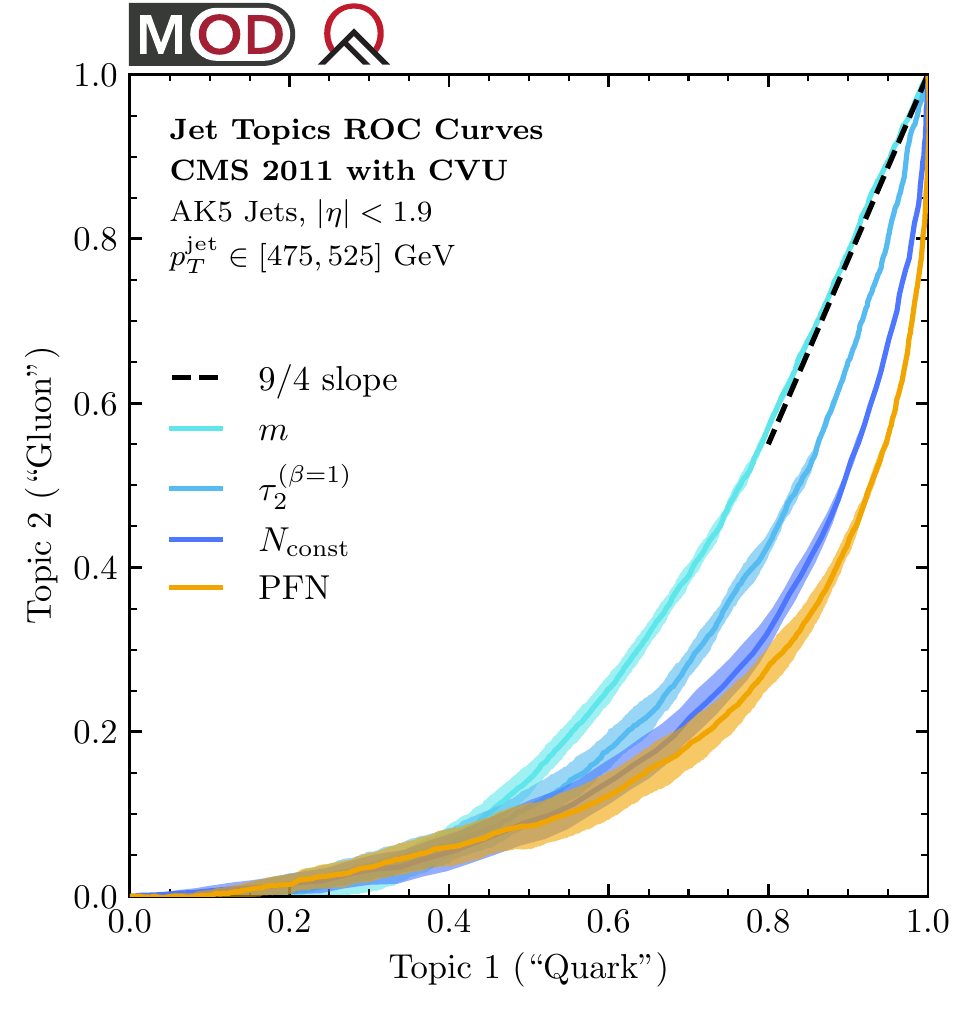}
\caption{
Extracted ROC curves from the CMS 2011 Open Data.
We consider four discriminants that span the range of quark/gluon tagging performance: jet mass, 2-subjettiness, constituent multiplicity, and PFN classification.
The fact that the ROC curves slopes are close to $0$ and $\infty$ at the endpoints imply that the jet topics are mutually irreducible, as desired.
For jet mass, the dashed black slope corresponds to the leading-logarithmic prediction in \Eq{eq:jetmassprediction}.
The error band corresponds to statistical uncertainties propagated through \Eqs{eq:p_q_cum}{eq:p_g_cum}.
}
\label{fig:ROC_extract}
\end{figure}

From the individual ``quark'' and ``gluon'' substructure distributions, we can present the tagging performance of each observable in the form of ROC curves.
Following \Eqs{eq:p_q}{eq:p_g}, the cumulative distributions are
\begin{align}
\label{eq:p_q_cum}
P_q(\figO) & =\frac{P_{M_1}(\figO)-\ka_{12}\, P_{M_2}(\figO)}{1-\ka_{12}},\\
\label{eq:p_g_cum}
P_g(\figO) &=\frac{P_{M_2}(\figO)-\ka_{21}\, P_{M_1}(\figO)}{1-\ka_{21}}.
\end{align}
When plotting uncertainties below, we ignore statistical correlations between the reducibility factors and the cumulative distributions.
The ROC curve for each observable is defined by the set of points 
\begin{equation}
\label{eq:pure_ROC}
\{P_{q}(\figO_{\rm cut}), P_{g}(\figO_{\rm cut})\},
\end{equation}
in analogy with \Eq{eq:mix_ROC}.
The statistical uncertainties in these ROC curves are dominated by those of the reducibility factors $\kappa_{12}$ and $ \kappa_{21}$.

In \Fig{fig:ROC_extract}, we show the ROC curves for four representative discriminants: jet mass, 2-subjettiness, constituent multiplicity, and PFN classification.
Recall from \Sec{subsection:roc} that the slopes of the ROC curve at its endpoints are related to the reducibility factors associated with the corresponding observable.
In the case of jet mass, the reducibility factors at leading-logarithmic order are \cite{Metodiev:2018ftz}:
\begin{equation}
\label{eq:jetmassprediction}
\kappa^{\rm mass}_{gq} = 0, \qquad  \kappa^{\rm mass}_{qg} = \frac{C_q}{C_g} = \frac{4}{9},
\end{equation}
where in the last step, we used $C_g = C_A = 3$ and $C_q = C_F = 4/9$.
This expected behavior is reflected in the $1/\kappa^{\rm mass}_{qg} \approx 9/4$ slope at the right endpoint of the jet mass ROC curve.

In the case of the PFN, we effectively have a self-calibrating classifier, since the PFN R-fit method was used to operationally define quarks and gluons.
As expected, the slopes of the PFN ROC curve are approximately $0$ and $\infty$ at the respective endpoints, since the samples are, by construction, mutually irreducible.

\subsection{Rapidity Spectrum}
\label{eq:rapidity_spectrum}

\begin{figure}[t]
	\includegraphics[width=0.45\textwidth]{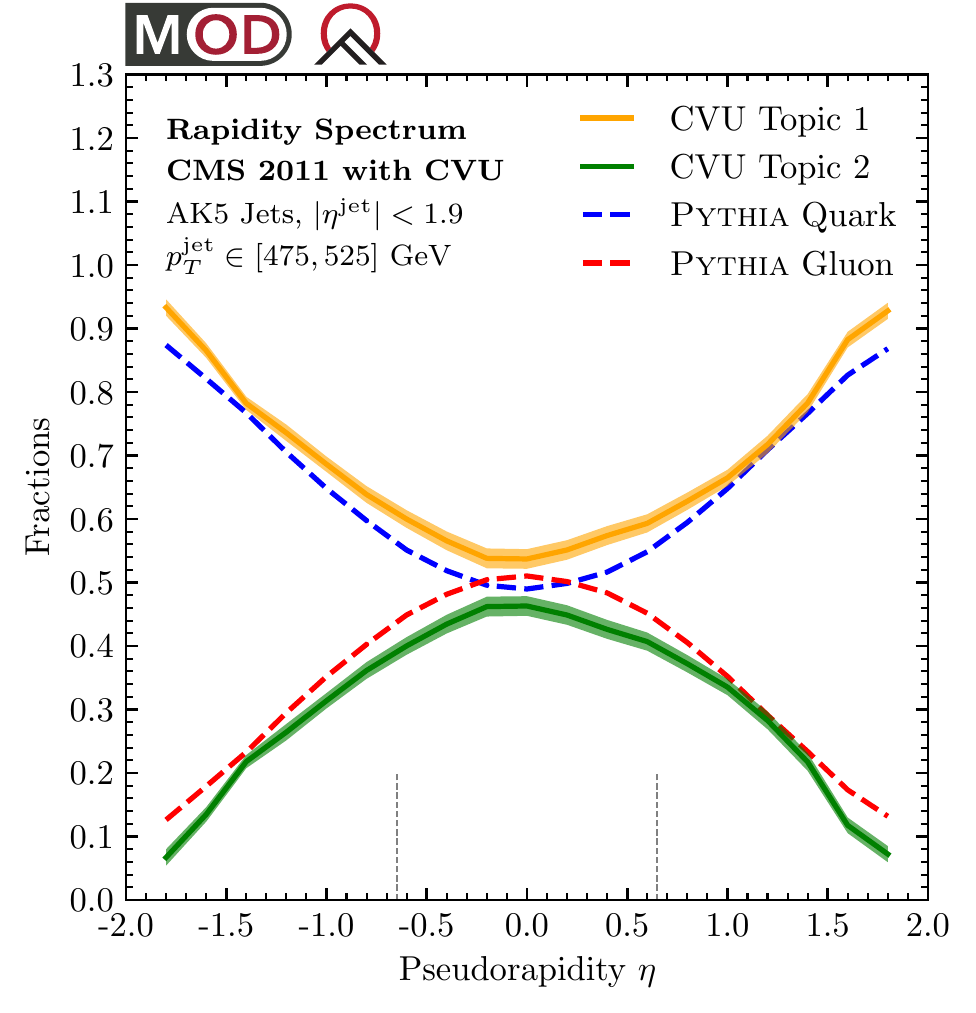}
	\caption{
		Extracted quark and gluon fractions as a function of pseudorapidity, comparing the topic extraction in the CMS 2011 Open Data to parton labels from \Pythia.
		The vertical gray lines correspond to the forward/central boundaries at $|\eta| = 0.65$.
		Compared to the \Pythia baseline, there is an overall increase in the ``quark'' content across the spectrum.
		The error band corresponds to statistical uncertainties in the fitted distribution.
	}
	\label{fig:rapidity_spectrum}
\end{figure}

The analysis thus far has focused on two wide rapidity regions, forward and central separated by the boundary $|\eta| = 0.65$.
With the quark and gluon topics in hand, we can revisit \Eq{eq:factorized_eta} and try to determine the rapidity spectrum for the quark and gluon jets separately.
The idea, adapted from \Ref{Metodiev:2018ftz}, is to bin the data into smaller rapidity slices, and fit the observable distribution in each slice to a linear combination of topics.

For this analysis, we use the PFN R-fit method to determine the jet topics and then fit linear combinations of the constituent multiplicity distribution.
The resulting quark and gluon fractions in rapidity slices of size $\Delta \eta = 0.2$ are shown in \Fig{fig:rapidity_spectrum}.
Here, the uncertainties are dominated by the those of the overall quark fractions.
The slight $\eta \to - \eta$ asymmetry is within the expectation from statistical uncertainties.
The kink in the rapidity spectrum around $|\eta| = 1.4$ mirrors that of \Fig{fig:unfolding_eta}, so it would be interesting to investigate potential detector mismodeling near that transition region.

\subsection{Correlation Dimension}
\label{sec:corrdim}

As our final application of the quark fractions in \Eq{eq:quark_fractions_PFN}, we compute separate correlation dimensions for the quark and gluon samples.
The correlation dimension is a scale-dependent measure of the number of effective degrees of freedom describing a dataset~\cite{Grassberger:1983zz,CAMASTRA20032945,NIPS2002_1177967c}, and it can be computed knowing just pairwise distances between data points, without needing explicit coordinates.
This analysis was presented for a mixed sample of quark and gluon jets in \Ref{Komiske:2019jim}, which contains more details about the analysis procedure.

As our pairwise distance measure between jets, we use the energy mover's distance (EMD), which quantifies the amount of ``work'' to rearrange one energy distribution into another~\cite{Komiske:2019fks}.
It is based on the well-known earth mover's distance from computer vision~\cite{DBLP:journals/pami/PelegWR89,Rubner:1998:MDA:938978.939133,Rubner:2000:EMD:365875.365881,DBLP:conf/eccv/PeleW08,DBLP:conf/gsi/PeleT13}, also known as the Wasserstein metric~\cite{wasserstein1969markov,dobrushin1970prescribing}.
The EMD has units of energy, and angular distances are normalized such that they are measured in units of the jet radius.
By computing all pairwise distances,
\begin{equation}
D_{k \ell} \equiv \text{EMD}(\mathcal{J}_k,\mathcal{J}_\ell),
\end{equation}
one can triangulate the ``space'' of jets and derive a variety of geometric quantities~\cite{Komiske:2020qhg}.
See \Refs{Cesarotti:2020hwb,CrispimRomao:2020ejk,Cai:2020vzx,Cai:2021hnn} for related EMD studies in particle physics.

One such geometric quantity is the correlation dimension~\cite{Grassberger:1983zz,CAMASTRA20032945,NIPS2002_1177967c}  defined as:
\begin{equation}
\label{eq:dim_def}
\text{dim}(Q) = \frac{\partial \ln N(Q)}{\partial \ln Q},
\end{equation}
where
\begin{equation}
\label{eq:NQ_def}
N(Q) = \sum_{k \ell} w_kw_\ell\,\Theta \big[ D_{k \ell} < Q \big]
\end{equation}
is the ``bi-cross section'' of the jet pairs within an EMD of $Q$ from each other, where we have accounted for jet weights $w_i$ in the case of weighted distributions.
For unweighted events (i.e.~$w_i = 1$), the bi-cross section simply reduces to the number of jet pairs.
For a uniform distribution in a compact domain of $\mathbb{R}^n$, we have $\text{dim}(Q) = n$ as long as $Q$ is small compared to the size of the domain.

To address the case of quark/gluon mixtures, we can modify \Eq{eq:NQ_def} to handle the case of multiple samples:
\begin{equation}
\label{eq:NQ_mod_def}
N_{AB}(Q) = \sum_{k \in A , \ell \in B}w_kw_\ell \,\Theta \big[ D_{k \ell} < Q \big].
\end{equation}
The correlation dimension derived from $N_{AB}(Q)$ has the interpretation of the dimensionality of the $B$ manifold at a scale $Q$ averaged over the $A$ manifold, and is symmetric between $A$ and $B$.

In practice, we choose a binning in $Q$ and work with the number of jet pairs falling into each EMD bin.
Concretely, let $\{Q_0,\,Q_1,\,\ldots,\,Q_S\}$ be the edges of $S$ bins in $Q$, and
\begin{equation}
\label{eq:ni_def}
n_{AB}(s)=\sum_{k\in A,\,\ell\in B}w_kw_\ell\,\Theta[Q_{s-1}\le D_{k\ell}<Q_s]
\end{equation}
be the number of jet pairs in bin $s=1,\ldots,S$.
Note that $n_{AB}(s)=n_{BA}(s)$ due to the symmetry of the EMD.
We can relate this expression to \Eq{eq:NQ_mod_def} via
\begin{equation}
\label{eq:Nn_relation}
N_{AB}(Q_s)=\sum_{t=1}^s n_{AB}(t).
\end{equation}

\begin{figure*}[t]
\centering
\subfloat[]{
\includegraphics[width=0.45\textwidth]{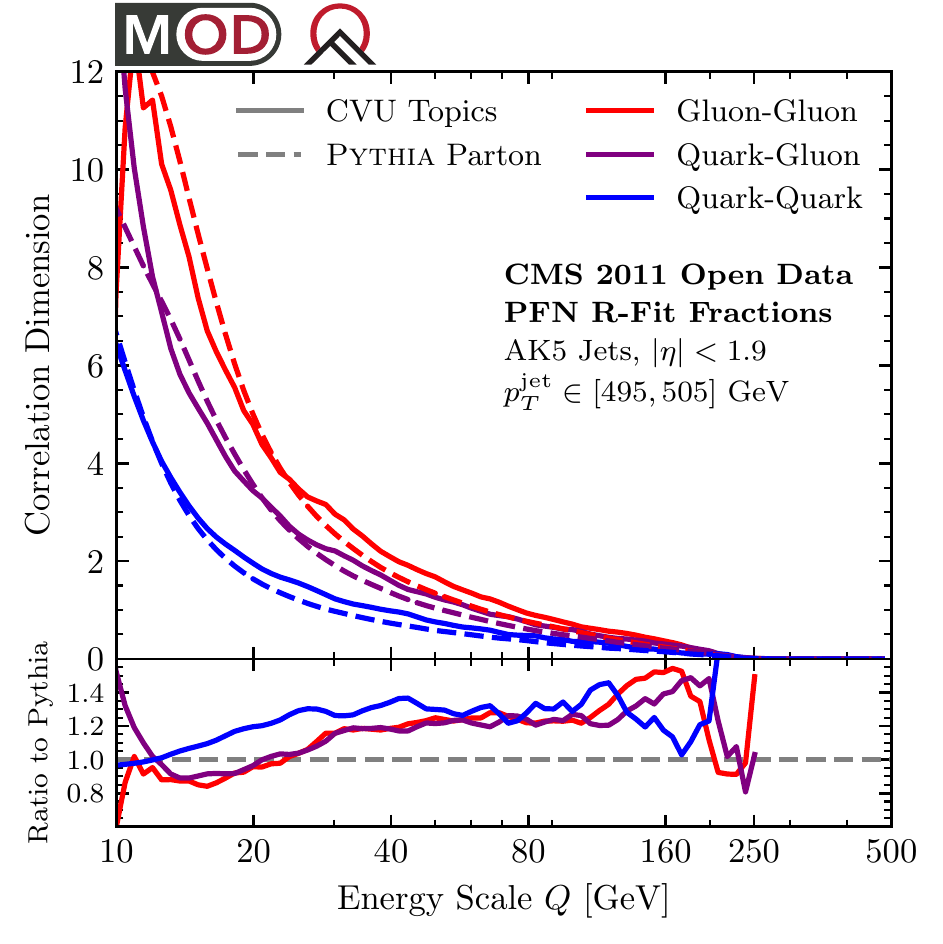}
\label{fig:correlation_dimension}
}
$\quad$
\subfloat[]{
\includegraphics[width=0.45\textwidth]{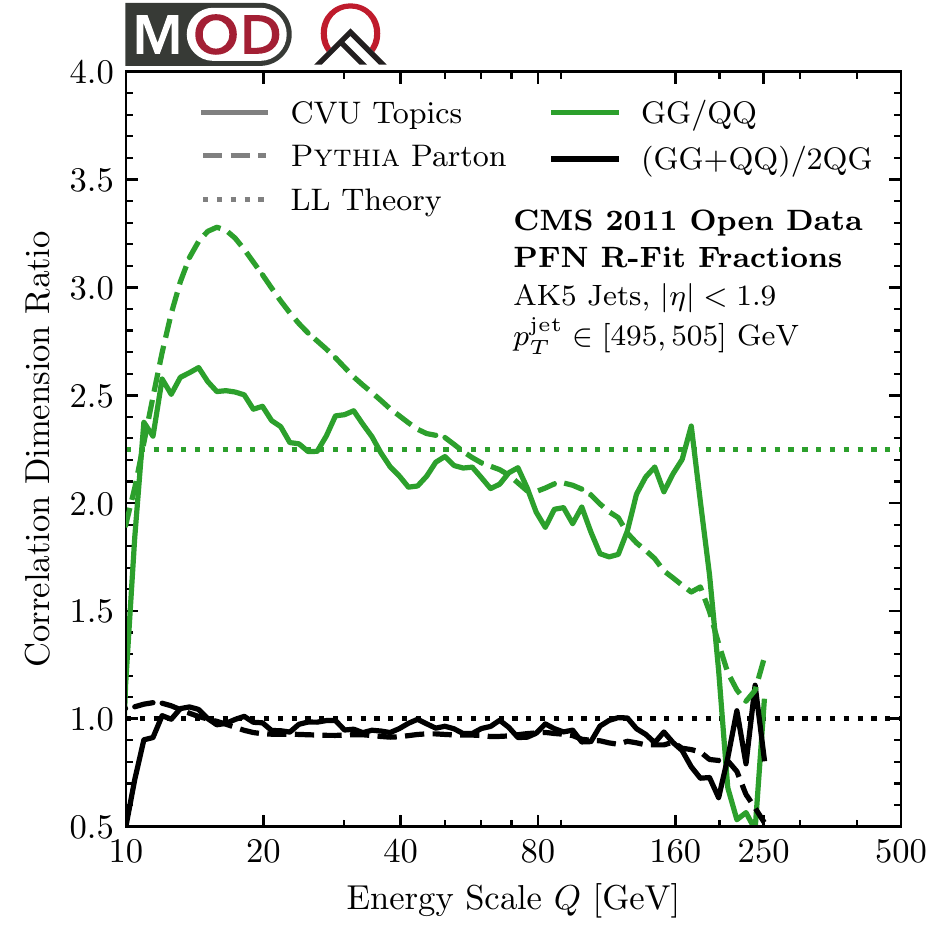}
\label{fig:correlation_dimension_ratio}
}
\caption{
(a) Correlation dimensions for three different types of jet pairs: gluon-gluon, quark-gluon, and quark-quark.
The \Pythia baseline uses parton-level labels.
For the CMS Open Data result with CVU, the quark fractions are given by \Eq{eq:quark_fractions_PFN}.
These fractions are used in in \Eq{eq:nMM_nQG_relations} to unmix the bi-cross sections of the central and forward jet combinations into quark and gluon combinations.
(b) Ratios of the correlation dimension from \Eqs{eq:dim_ratio}{eq:dim_sum}, compared to first-principles QCD calculations in the strongly-ordered limit.
}
\label{fig:correlation_dimension_all}
\end{figure*}

For two mixed samples $M_1$ and $M_2$ with known quark fractions $f_1$ and $f_2$, the three quantities:
\begin{equation}
\label{eq:nM1M2_quantities}
n_{M_1 M_1}(s),\, n_{M_1 M_2}(s),\, n_{M_2 M_2}(s),
\end{equation}
can be inverted bin by bin to determine the quark/gluon values:
\begin{equation}
\label{eq:nqg_quantities}
n_{qq}(s),\, n_{qg}(s),\, n_{gg}(s),
\end{equation}
and subsequently the correlation dimension for the quark-quark, quark-gluon, and gluon-gluon samples, respectively, via \Eqs{eq:Nn_relation}{eq:dim_def}.
The equations relating the quantities in \Eq{eq:nM1M2_quantities} to those in~\Eq{eq:nqg_quantities} can be derived by ensuring bi-cross sectional consistency with the known fractions $f_1$ and $f_2$ and total cross section of $M_1$ and $M_2$:
\begin{widetext}
\begin{equation}
\label{eq:nMM_nQG_relations}
\begin{bmatrix}
\vspace{0.1in}\displaystyle\frac{n_{M_1M_1}(s)}{\sigma_1^2}\\
\vspace{0.1in}\displaystyle\frac{n_{M_1M_2}(s)}{\sigma_1\sigma_2}\\
\displaystyle\frac{n_{M_2M_2}(s)}{\sigma_2^2}
\end{bmatrix}
=
\begin{pmatrix}
\vspace{0.1in} f_1^2 & 2f_1(1-f_1) & (1-f_1)^2 \\
\vspace{0.1in} f_1f_2 & f_1(1-f_2)+f_2(1-f_1) & (1-f_1)(1-f_2) \\
f_2^2 & 2f_2(1-f_2) & (1-f_2)^2
\end{pmatrix}
\begin{bmatrix}
\vspace{0.1in}\displaystyle\frac{n_{qq}(s)}{\big(f_1\sigma_1+f_2\sigma_2\big)^2}\\
\vspace{0.1in}\displaystyle\frac{n_{qg}(s)}{(f_1\sigma_1+f_2\sigma_2)\big((1-f_1)\sigma_1+(1-f_2)\sigma_2\big)}\\
\displaystyle\frac{n_{gg}(s)}{\big((1-f_1)\sigma_1+(1-f_2)\sigma_2\big)^2}
\end{bmatrix},
\end{equation}
\end{widetext}
where the total cross section of the mixed samples are:
\begin{equation}
\sigma_1=\sum_{i\in M_1}w_i,\quad \sigma_2=\sum_{i\in M_2}w_i.
\end{equation}
One can easily solve \Eq{eq:nMM_nQG_relations} for $n_{qq}(s)$, $n_{gg}(s)$, and $n_{qg}(s)$ for each $s$ given $n_{M_1M_1}(s)$, $n_{M_2M_2}(s)$, $n_{M_1M_2}(s)$, $\sigma_1$, $\sigma_2$, $f_1$, and $f_2$.

The results of this analysis are shown in \Fig{fig:correlation_dimension}, comparing the \OmniFold CVU results to those obtained from the parton-level labels in \Pythia.
Here, we have restricted our attention to a narrow jet range of $p_T \in [495,505]$ GeV, and we rescale the jet constituents such that the jet $p_T$ is always \SI{500}{GeV}. 
At the scale $Q = \SI{250}{GeV}$, the correlation dimensional is zero, since at this resolution scale ($Q \approx p_T/2$), a jet looks like a single parton.
As the $Q$ scale decreases, the dimensionality increases logarithmically as one resolves more features within the jet, with higher dimensionalities associated with the gluon samples compared to the quark ones.

We can gain additional insight into the correlation dimension by performing a first-principles calculation of this quantity in the strongly-ordered limit.
For simplicity of presentation, we ignore running of the strong coupling constant $\alpha_s$. 
Consider the EMD between two jets $i \in I$ and $j \in J$ with color factors $C_I$ and $C_J$.
In the strongly-ordered limit, we need only consider the case where one jet has emitted a soft-collinear gluon, whereas the other jet still looks like a single parton.
In that case, the EMD reduces to the $\beta = 1$ jet angularity~\cite{Ellis:2010rwa,Larkoski:2014uqa,Larkoski:2014pca} between one of the jets and the jet axis.
Considering an ensemble of such jet pairs, the cumulative distribution for the EMD is:
\begin{equation}
N_{IJ}(Q) = \exp \Big[- \frac{\alpha_s (C_I + C_J)}{\pi}\, \log^2 \frac{p_T/2}{Q}\Big].
\end{equation}
Here, the exponential is the Sudakov factor for \emph{neither} of the two jets to have an emission above the scale $Q$ (i.e.~the product of the Sudakov factors for the individual jets).
The factor of $p_T/ 2$ in the logarithm is because this is approximately the maximum EMD achievable between two anti-$k_t$ jets.

Taking the logarithmic derivative in \Eq{eq:dim_def}, the (mixed) correlation dimension is:
\begin{equation}
\label{eq:dim_QCD}
\text{dim}_{IJ}(Q) = \frac{2\alpha_s (C_I + C_J)}{\pi}\, \log \frac{p_T/2}{Q}.
\end{equation}
From this, we can predict that the ratio between the gluon-gluon and quark-quark correlation dimensions is:
\begin{equation}
\label{eq:dim_ratio}
\frac{\text{dim}_{gg}(Q)}{\text{dim}_{qq}(Q)} = \frac{C_g}{C_q} = \frac{9}{4},
\end{equation}
where we have used $C_g = C_A = 3$ and $C_q = C_F = 4/3$.
Intriguingly, this means that the correlation dimensions exhibit Casimir scaling.
We can also predict that the correlation dimensions satisfy the following relation:
\begin{equation}
\label{eq:dim_sum}
\frac{\text{dim}_{gg}(Q) + \text{dim}_{qq}(Q)}{2 \, \text{dim}_{qg}(Q)} = \frac{2 C_g + 2 C_q}{2 (C_g + C_q)} = 1.
\end{equation}

The two analytic predictions in \Eqs{eq:dim_ratio}{eq:dim_sum} are plotted in \Fig{fig:correlation_dimension_ratio}, along with the CMS Open Data results with CVU.
We see quite good agreement, especially given the relatively simplicity of the strongly-ordered analysis.
Interestingly, the tilt in these predictions, which are minimally expected from running coupling effects and subleading terms in the splitting function, are smaller in the CMS Open Data than in \Pythia.
We hope that this encourages future precision calculations of the correlation dimensions of quark, gluon, and other types of jets.

\section{Conclusions}
\label{sec:conc}

In this paper, we disentangled ``quark'' and ``gluon'' jet substructure distributions in the CMS 2011 Open Data using the operational approach advocated in \Ref{Komiske:2018vkc}.
We improved the statistical robustness of jet topics by introducing two new methods to extract reducibility factors:  the L-fit method based on the log-likelihood ratio and the R-fit method based on the ROC curve.
In order to mitigate sample dependence from inhomogeneities in the CMS detector response, we implemented central value unfolding with \OmniFold.
We found noticeable differences in both the quark/gluon fractions and the quark/gluon distributions compared to the \Pythia baseline, which would be interesting to investigate further in future work.

There are a number of directions to extend the proof-of-concept study presented here.
At minimum, one could apply our improved jet topics procedures to the unfolded ATLAS measurements from \Ref{ATLAS:2019rqw} and the unfolded CMS measurements from \Ref{CMS:2021iwu}.
Binned distributions from these studies are available at \textsc{HEPData}~\cite{hepdata.89321,hepdata.111308}, which is sufficient for the anchor bin method already implemented by ATLAS.
Note that \Ref{hepdata.111308} and v2 of \Ref{hepdata.89321} have covariance matrices, which is essential for this kind of analysis.
To make the most of the L-fit or R-fit methods, though, it would be beneficial to have results with finer binning (or unbinned~\cite{Arratia:2021otl}).
Similarly, it would be exciting to go beyond central value unfolding and account for systematic uncertainties in the context of \OmniFold.
This would require a more detailed understanding of the CMS detector response to make sure that the unfolding does not erroneously adjust the truth distributions to account for potential detector mismodeling.
Finally, one could apply jet topics to observables designed to isolate pure quark and gluon samples~\cite{Stewart:2022ari}.

Our analysis focused on a narrow jet $p_T$ range, but with more $p_T$ slices, one could study the evolution of the quark and gluon distributions as a function of energy scale.
This would be particularly interesting in the context of the correlation dimension analysis, to test the strongly-ordered prediction in \Eq{eq:dim_QCD} that the dimension depends on the ratio of $Q/p_T$.
Similarly, it would be interesting to augment this two-sample analysis with more jet samples, either to test the prediction of sample independence (see \Ref{Komiske:2018vkc} for a quantification strategy) or to try to identify additional jet categories (e.g.\ up-type versus down-type quarks).
For the latter, one would want to apply machine learning techniques sensitive to the charge and flavor of the jet constituents, such as the PFN-ID approach~\cite{Komiske:2018cqr}.
We look forward to future developments of the jet topics method, to improve the (operational) connection between long-distance measurements and short-distance QCD dynamics.

\begin{acknowledgments}
We thank Jasmine Brewer, Matt LeBlanc, Eric Metodiev, Benjamin Nachman, and Andrew Turner for helpful feedback.
We thank Radha Mastandrea, Eric Metodiev, and Preksha Naik for collaboration in the initial stages of this project.
We thank CERN, the CMS collaboration, and the CMS Data Preservation and Open Access (DPOA) team for making research-grade collider data available to the public.
We thank the letters $\Delta$ and $O$ for pointing out alternate meanings of $N_{95}$ in the public health literature.
This work was supported by the Office of Nuclear Physics of the U.S. Department of Energy (DOE) under Grant No.\ DE-SC0011090, by the DOE Office of High Energy Physics under Grant Nos.\ DE-SC0012567 and DE-SC0019128, and by the National Science Foundation under Cooperative Agreement PHY-2019786 (The NSF AI Institute for Artificial Intelligence and Fundamental Interactions, \url{http://iaifi.org/}).
\end{acknowledgments}

\appendix

\section{Comparison to Monte Carlo Results}
\label{sec:pythia_only_analysis}

\begin{figure*}[t]
	\subfloat[]{
	\includegraphics[width=0.45\textwidth]{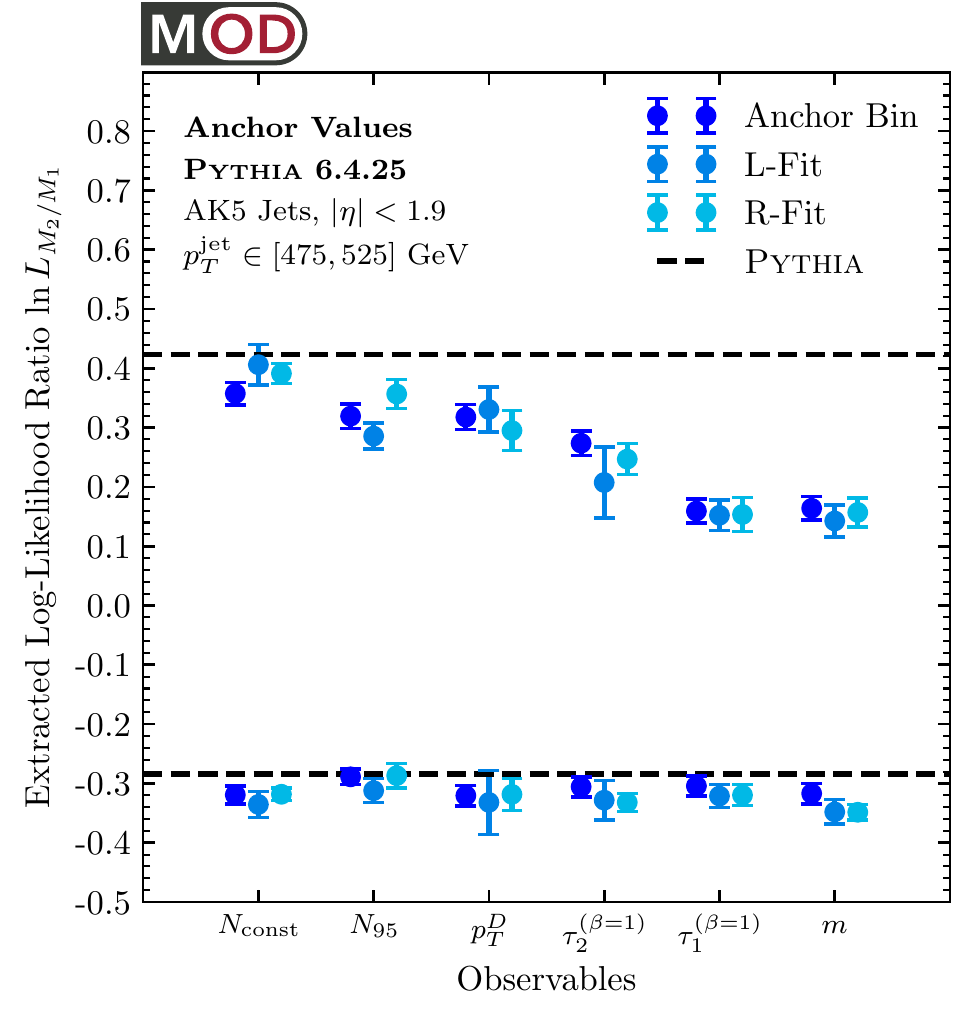}
	\label{fig:all_results_not_unf}
	}
	\subfloat[]{
		\includegraphics[width=0.45\textwidth]{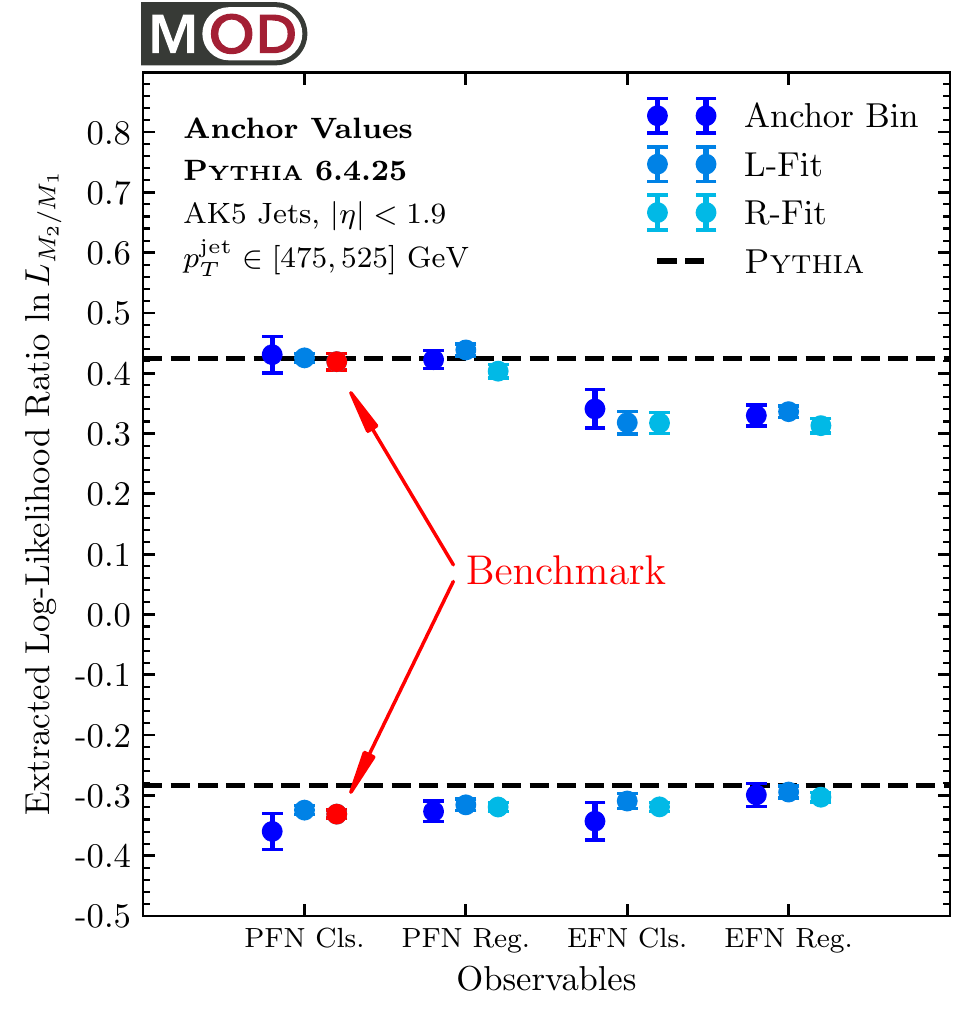}
	\label{fig:ml_results_not_unf}
	}
	\caption{
	Same as \Fig{fig:all_results} but applied to the \Pythia 6.4.25 dataset.  Interestingly, the extracted anchor values for the benchmark method in red does not match the horizontal dashed \Pythia parton-level expectation from \Eq{eq:frac_exp_pythia}.
	}
	\label{fig:pythia_red_fac_results}
\end{figure*}

In this appendix, we repeat the analyses from \Secs{section:extracting_k}{sec:quark_and_gluon_dist}, but applied to the \Pythia Monte Carlo samples.
This is a cross check of our new jet topics procedures on a sample without any unfolding complications.

\begin{figure}[t]
	\includegraphics[width=0.45\textwidth]{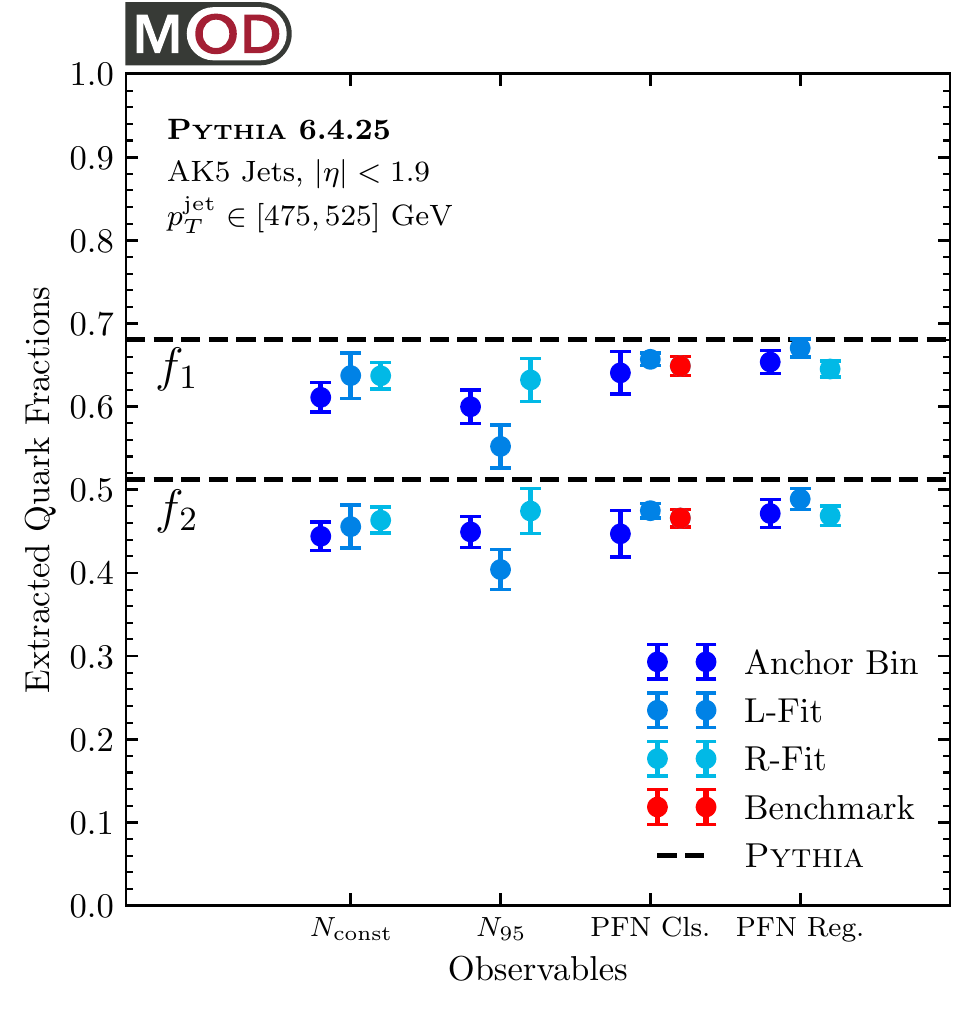}
	\caption{
	Same as \Fig{fig:all_fractions} but applied to the \Pythia 6.4.25 dataset.  Interestingly, the extracted jet topics are more gluon enriched compared to the expectation from \Eq{eq:frac_exp_pythia}. 
	}
	\label{fig:pythia_quark_frac_results}
\end{figure}

\begin{figure*}[t]
	\subfloat[]{
	\includegraphics[width=0.45\textwidth]{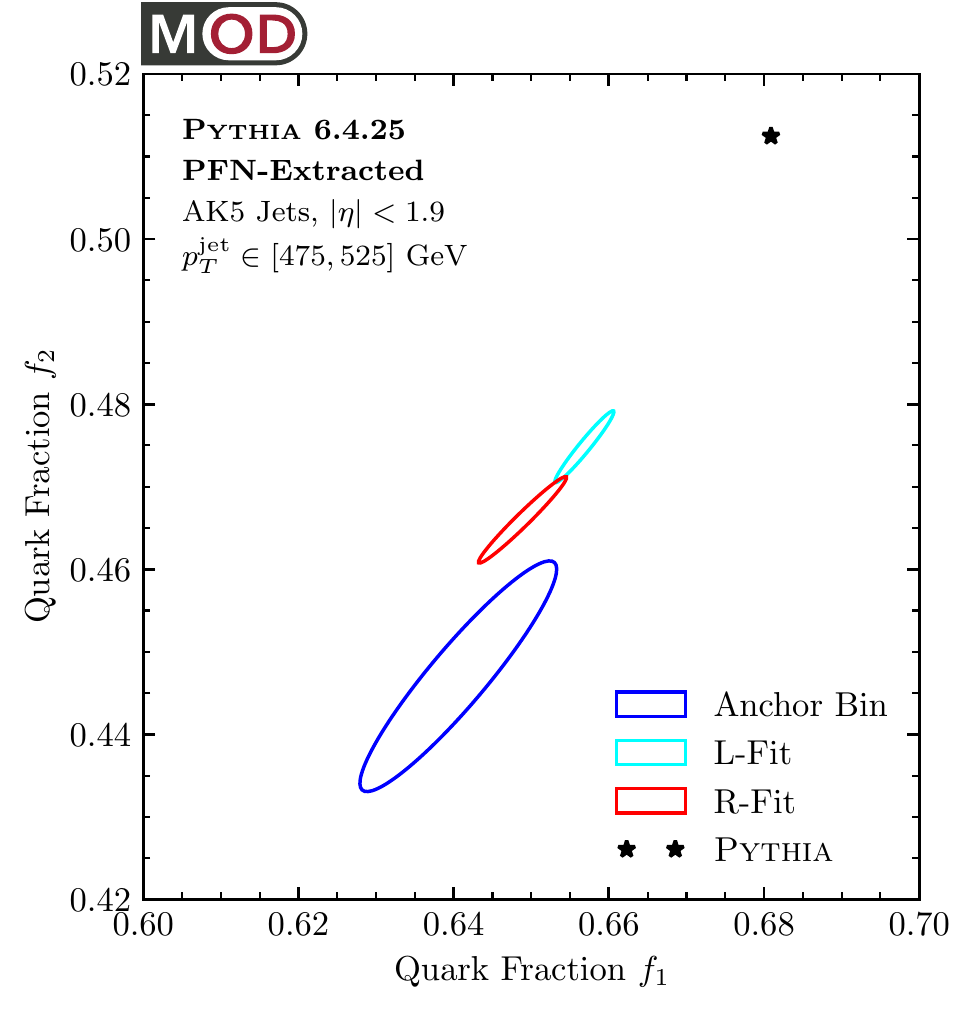}
		\label{fig:PFN_fractions_pythia}
	}
	\subfloat[]{
	\includegraphics[width=0.45\textwidth]{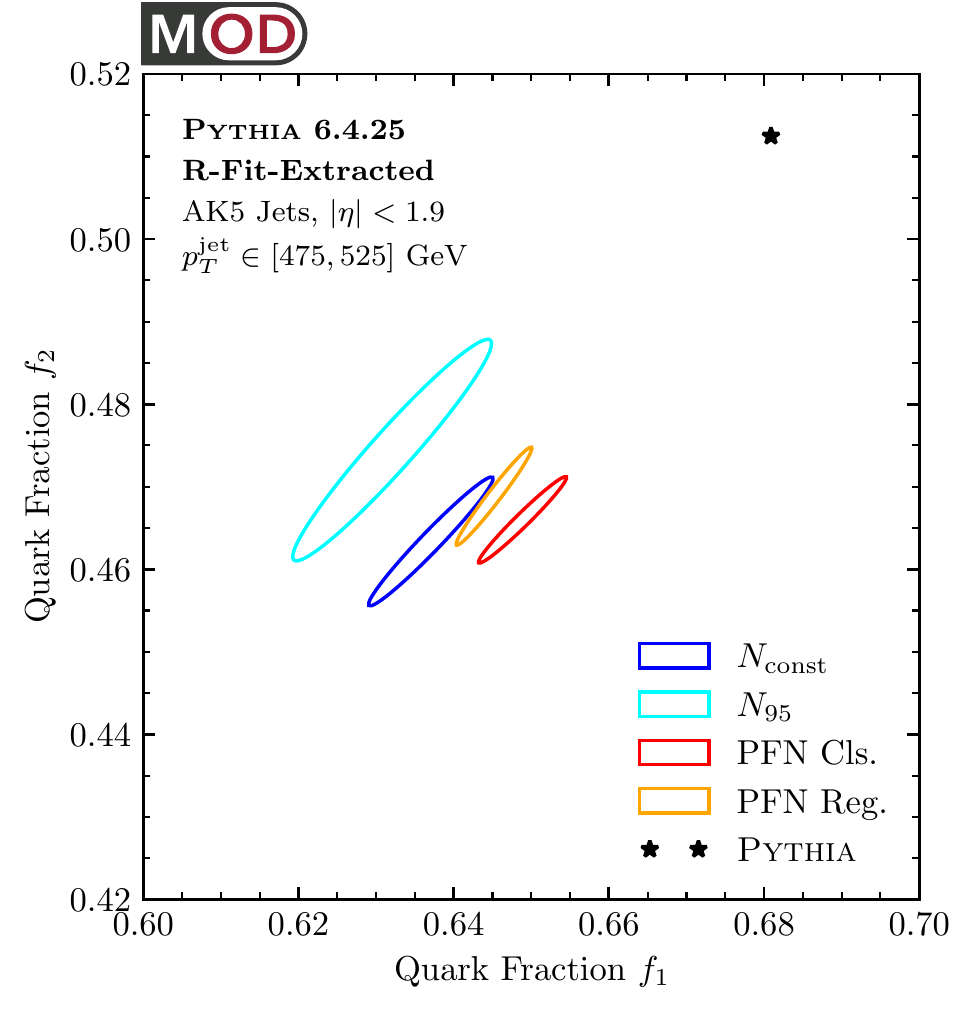}
	\label{fig:roc_fractions_pythia}
	}
	\caption{
	Same as \Fig{fig:ellipse_cvu} but applied to the \Pythia 6.4.25 dataset.  Interestingly, the ellipses are offset from, but diagonally aligned with, the parton label expectation from \Eq{eq:frac_exp_pythia}.
	}
	\label{fig:ellipses_pythia}
\end{figure*}

In \Fig{fig:pythia_red_fac_results}, we show results for the extraction of the anchor values.
Like in \Fig{fig:all_results}, we see that the better quark/gluon discriminants yield more separation in the anchor values.
Unlike in \Fig{fig:all_results}, the smaller anchor value is shifted systematically downward compared to the expectation from \Pythia parton labels in \Eq{eq:frac_exp_pythia}.

This downward shift is reflected in \Figs{fig:pythia_quark_frac_results}{fig:ellipses_pythia}.
Here, the quark fractions extracted from constituent multiplicity, image activity, PFN classification, and PFN regression are consistent with each other but smaller than the fraction of quark-parton-labeled jets in \Pythia.
Because the jet topics procedure acts at the level of distributions, we cannot identify which parton-level quark jets are being categorized as topics-level gluon jets.
One way this mismatch can happen is if a quark parton in \Pythia arises from a gluon splitting via $g \to q \bar{q}$.
Because the $g \to q \bar{q}$ splitting function is universal in the collinear limit, it would be ``correct'' to assign quarks from this splitting process to the gluon sample, though we were not able to find direct evidence that this is actually happening.
Interestingly, this mismatch did not show up in the CMS Open Data result in \Fig{fig:all_fractions}, so it is also possible that there is a more subtle issue with how \Pythia generates phase space configurations.

\begin{figure*}[t]
	\centering
	\subfloat[]{\includegraphics[width=0.33\textwidth]{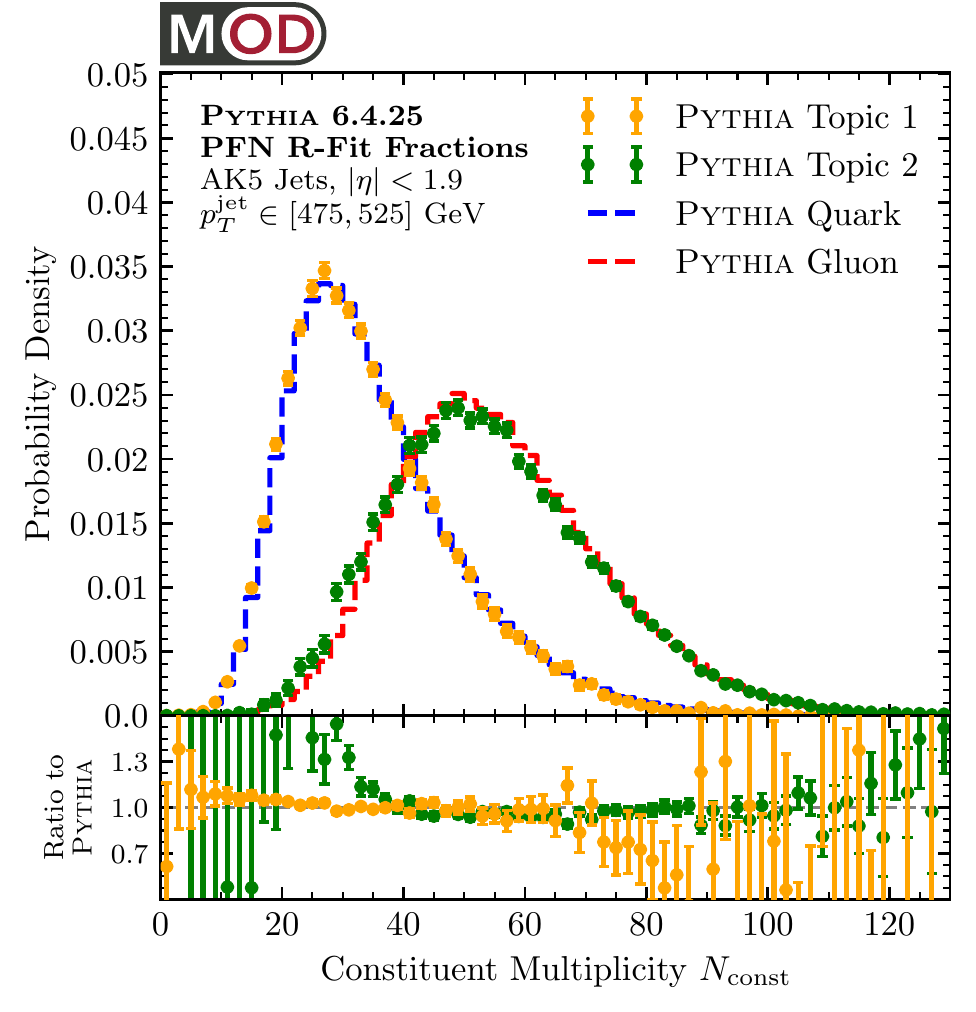}\label{fig:pyth_qg_mult}}
	\subfloat[]{\includegraphics[width=0.33\textwidth]{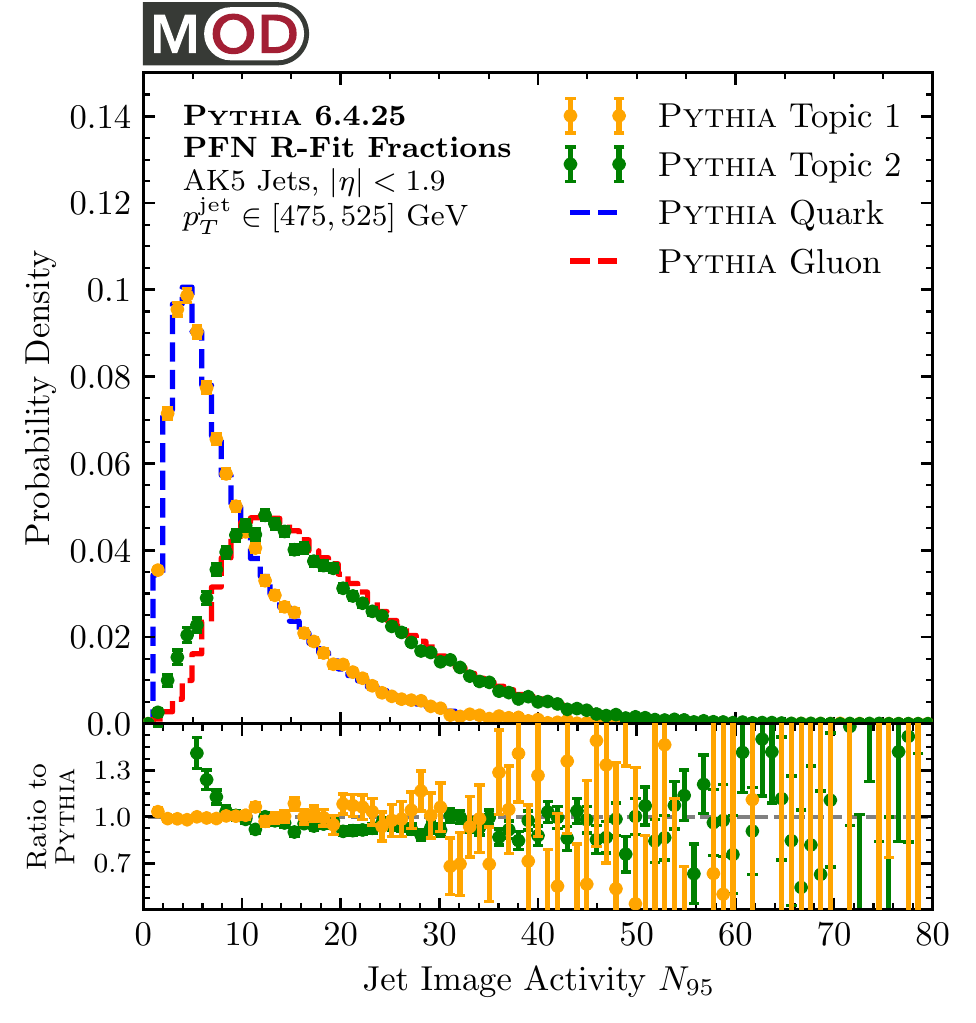}\label{fig:pyth_qg_n95}}
	\subfloat[]{\includegraphics[width=0.33\textwidth]{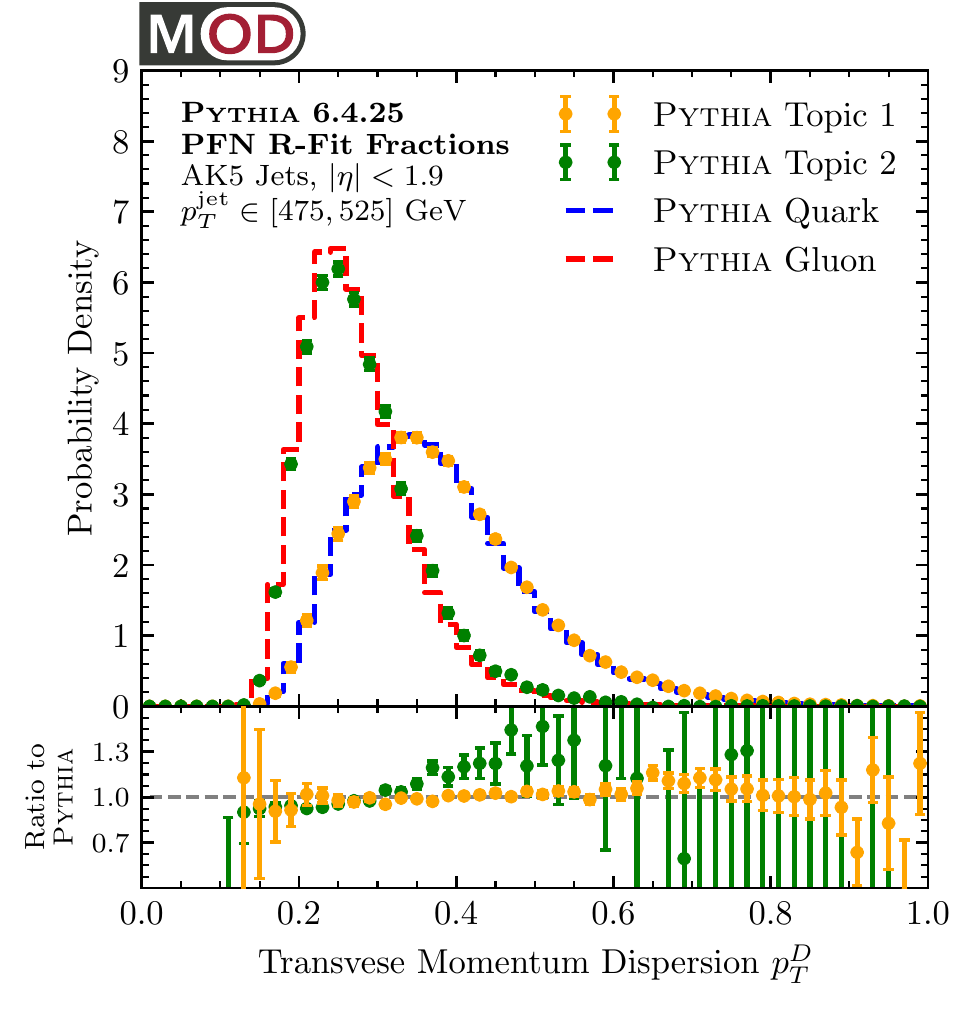}\label{fig:pyth_qg_ptd}}
	\\
	\subfloat[]{\includegraphics[width=0.33\textwidth]{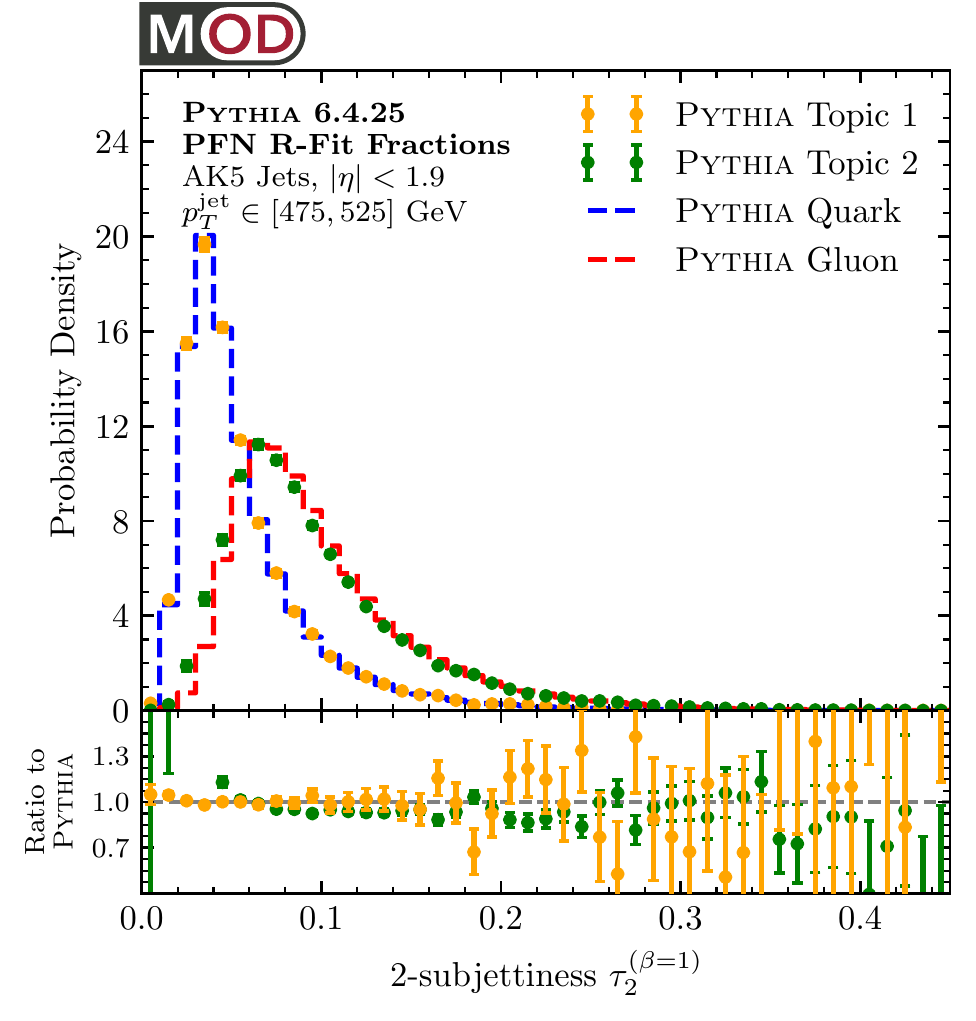}\label{fig:pyth_qg_nsub2}}
	\subfloat[]{\includegraphics[width=0.33\textwidth]{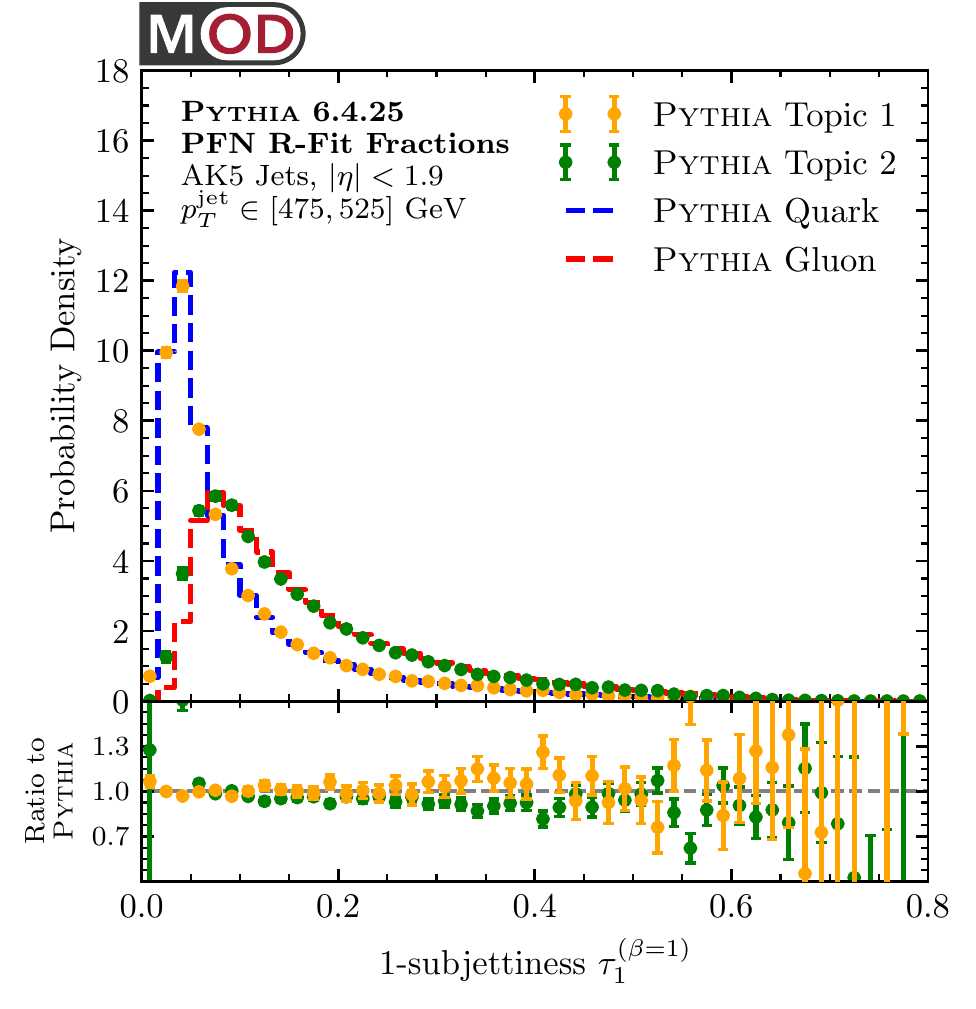}\label{fig:pyth_qg_nsub1}}
	\subfloat[]{\includegraphics[width=0.33\textwidth]{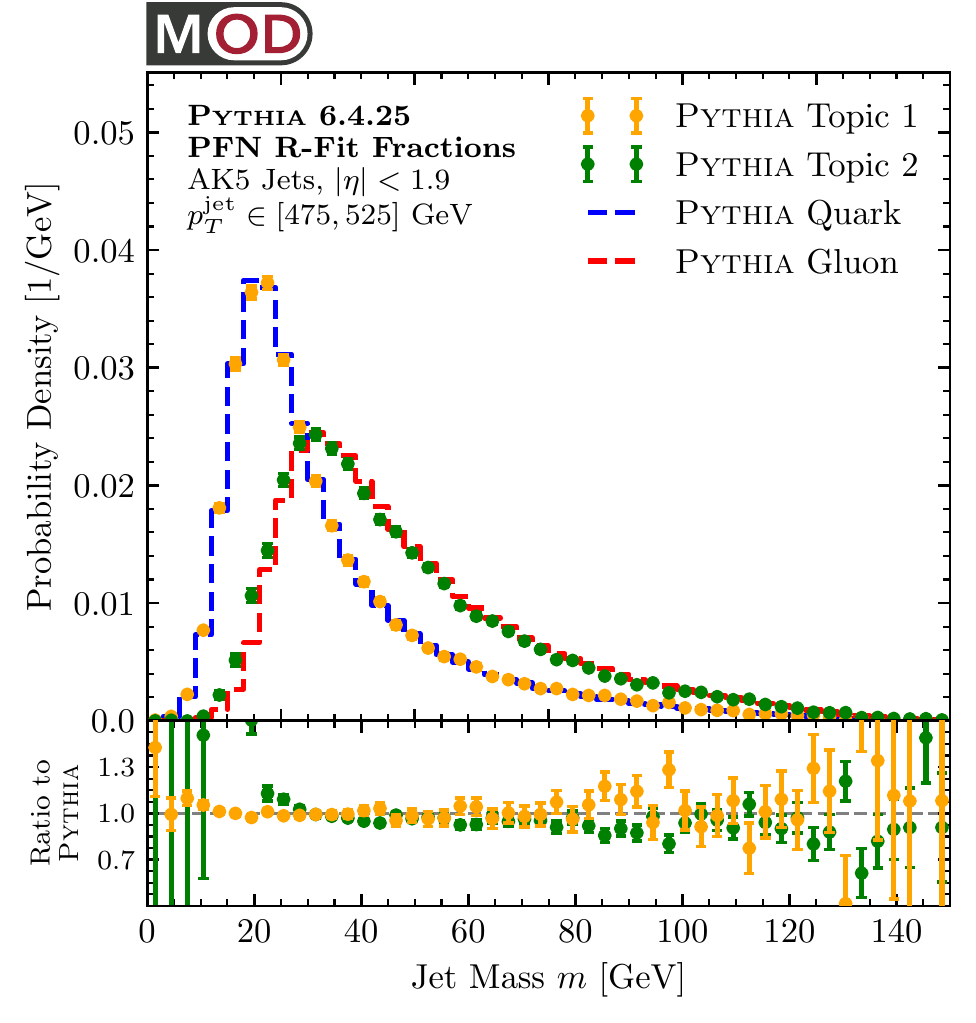}\label{fig:pyth_qg_mass}}
	\caption{
	Same as \Fig{fig:simqg} but applied to the \Pythia 6.4.25 dataset.
	Compared to the truth-parton-labeled samples, the extracted ``gluon'' distribution (green) is more quark like.
	}
	\label{fig:pyth_genqg}
\end{figure*}

Like in the main text, we use the R-fit method with PFN classification as our default strategy to extract quark fractions:
\begin{equation}
	\label{eq:quark_fractions_PFN_pythia}
	\text{\Pythia jet topics:} \quad
	\begin{aligned}
		f_1 &\simeq 0.649 \pm 0.012, \\
		f_2 &\simeq 0.466\pm 0.011.
	\end{aligned}
\end{equation}
In \Fig{fig:pyth_genqg}, we show the separate ``quark'' and ``gluon'' distributions using these fractions.
The difference between jet topics and parton-truth-labeled samples is most apparent in the IRC-unsafe observables, where the extracted gluon distributions are more quark-like than the partonic expectation.
Interestingly, the mismatch is much less evident, though still present, in the IRC-safe observables.
This motivates further studies with additional samples taken from different regions of phase space, to see if these quark/gluon trends persist.

\begin{figure}[t]
\centering
\includegraphics[width=0.45\textwidth]{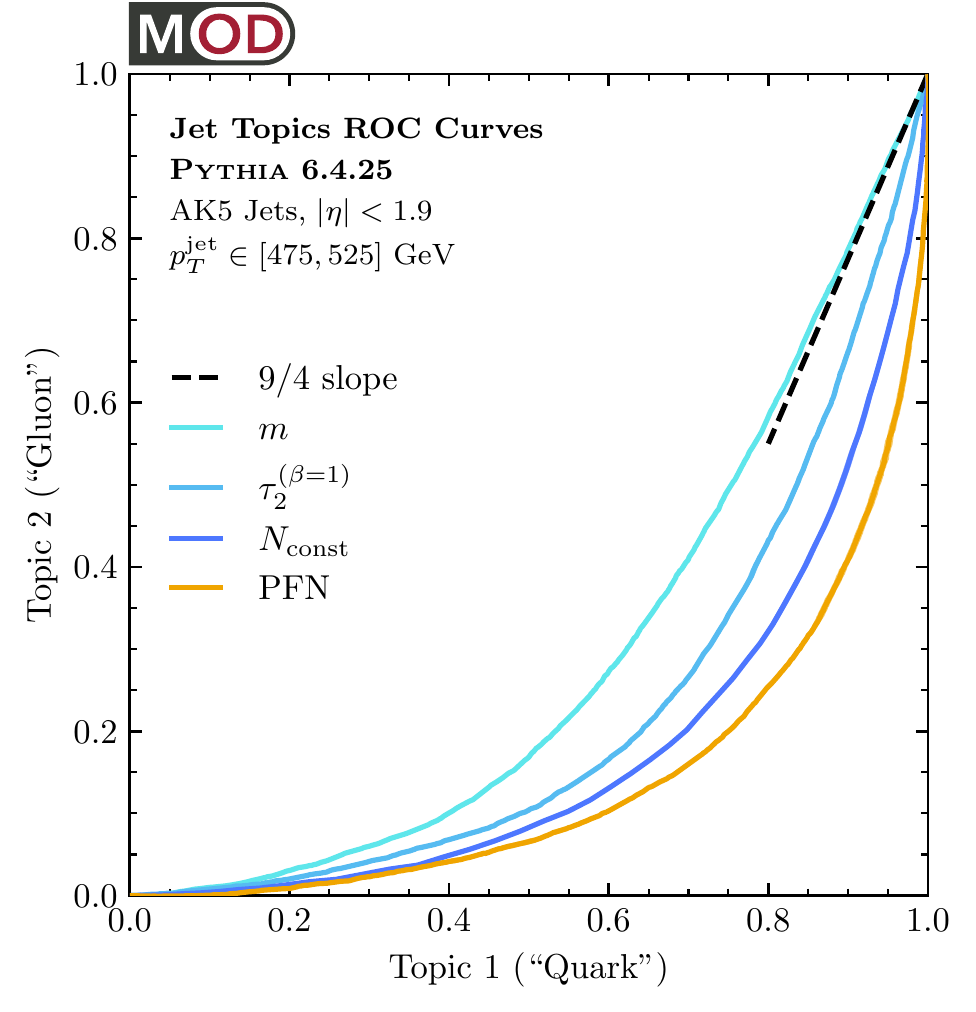}
\caption{
Same as \Fig{fig:ROC_extract} but applied to the \Pythia 6.4.25 dataset. 
}
\label{fig:ROC_extract_pythia}
\end{figure}

\begin{figure}[t]
	\includegraphics[width=0.45\textwidth]{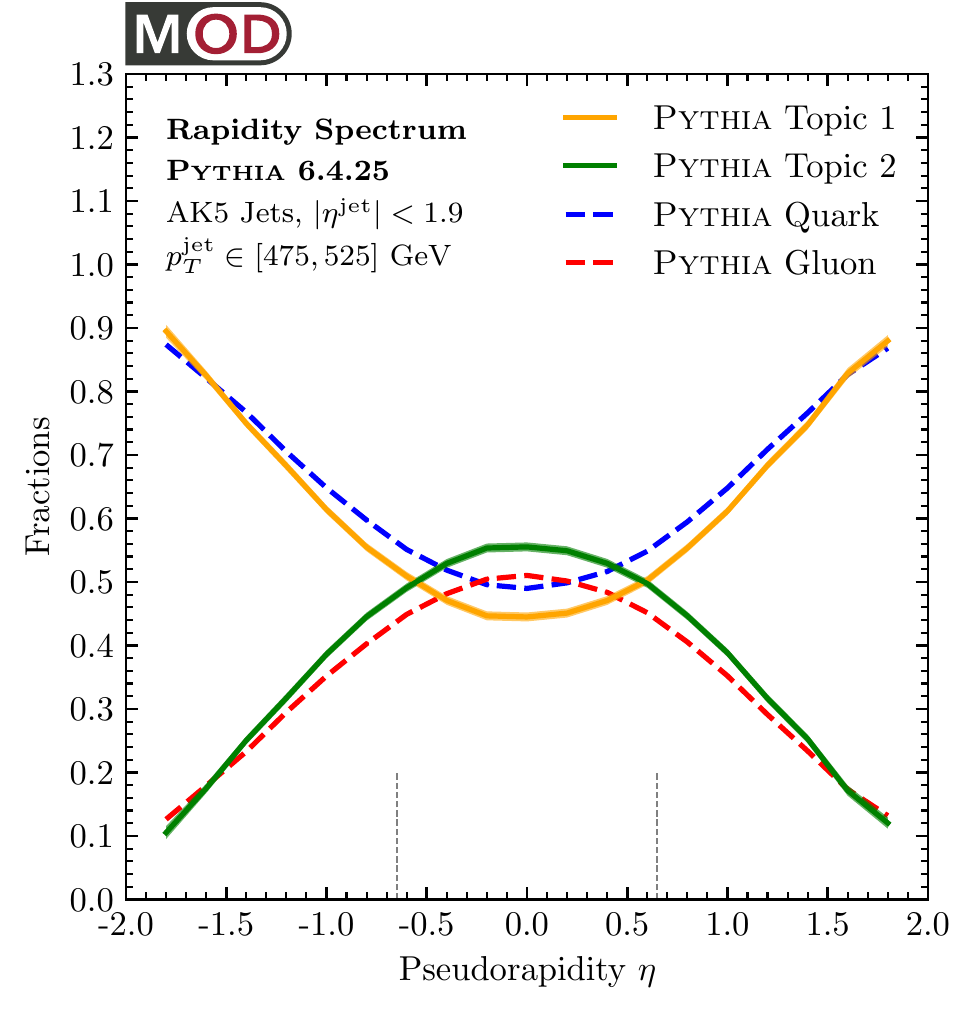}
	\caption{
		Same as \Fig{fig:rapidity_spectrum} but applied to the \Pythia 6.4.25 dataset. 
		Here, the topics extracted from \Pythia are more gluon-like compared to the parton labels.
	}
	\label{fig:rapidity_spectrum_pythia}
\end{figure}

We show plots of the extracted ROC curve in \Fig{fig:ROC_extract_pythia}, which agree qualitatively with the results from the CMS Open Data.
We show plots of the rapidity spectrum and correlation dimension in \Figs{fig:rapidity_spectrum_pythia}{eq:pythia_correlation_dimension_all}, respectively.
The difference between the topics-labeled curves and the parton-labeled curve highlights the ambiguity of quark and gluon labels beyond leading-logarithmic accuracy, motivating further studies using operational definitions that are well defined in data.

\begin{figure*}[t]
\centering
\subfloat[]{
\includegraphics[width=0.45\textwidth]{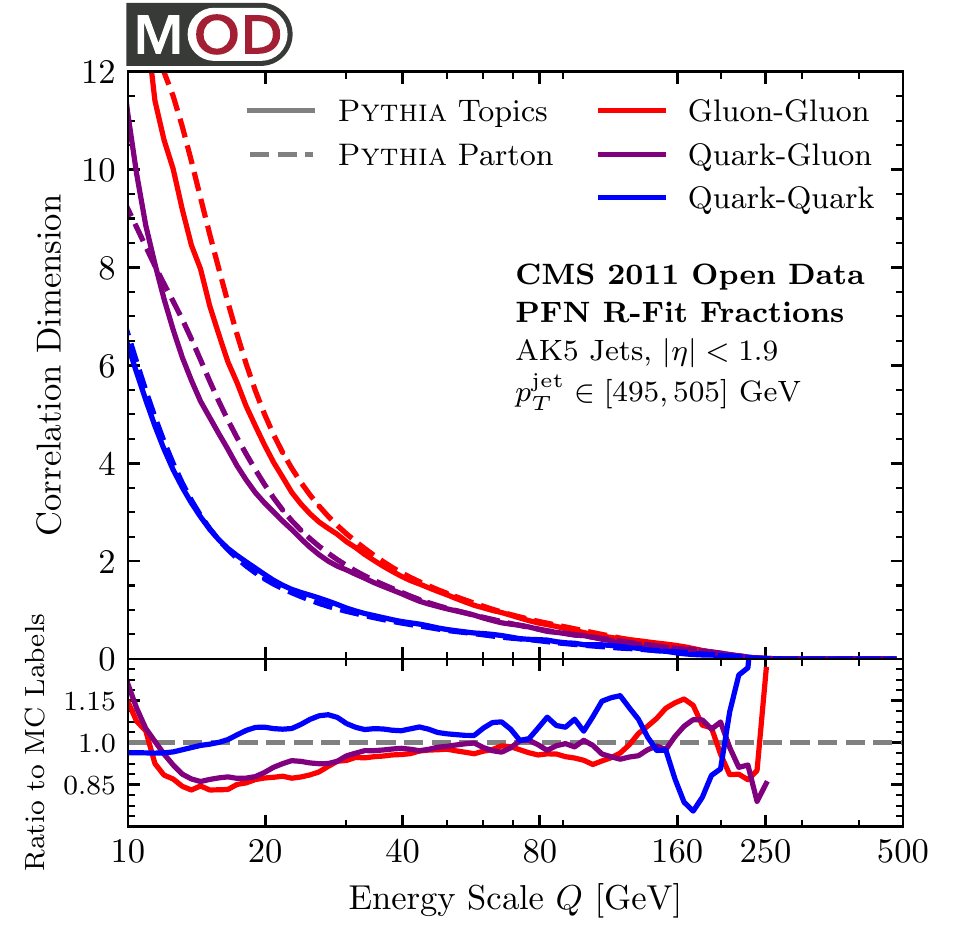}
\label{fig:pythia_correlation_dimension}
}
$\quad$
\subfloat[]{
\includegraphics[width=0.45\textwidth]{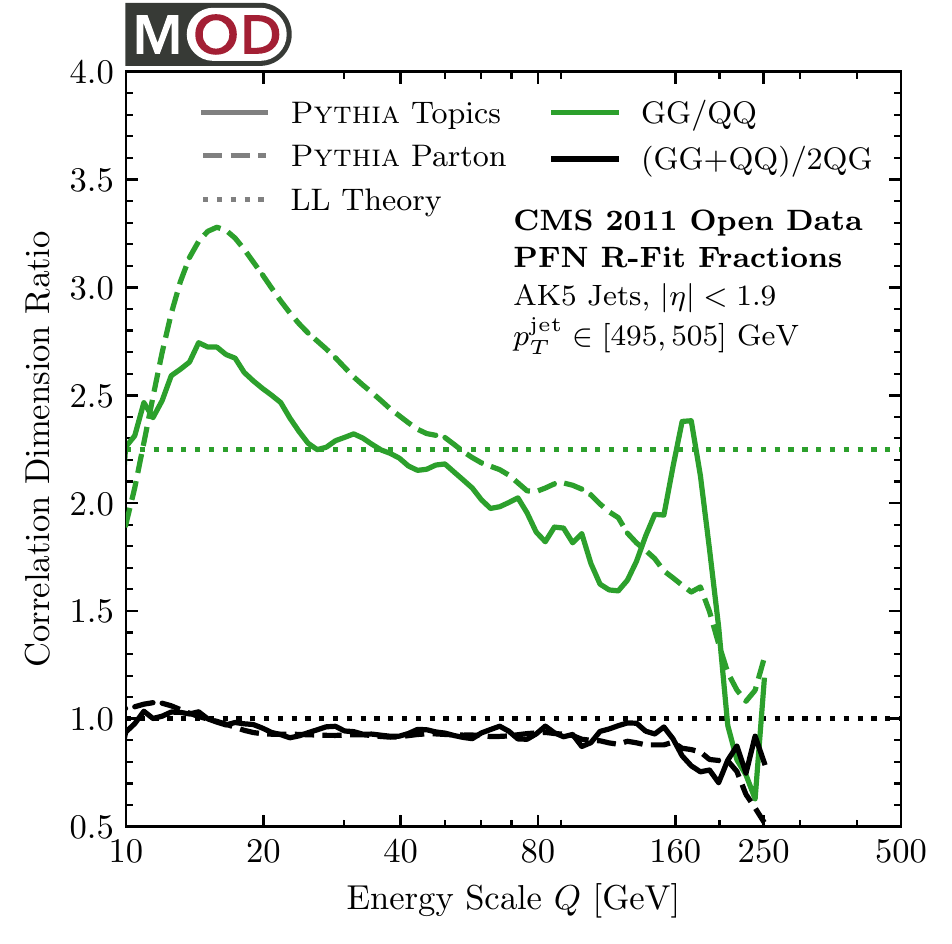}
\label{fig:pythia_correlation_dimension_ratio}
}
\caption{
Same as \Fig{fig:correlation_dimension_all} but applied to the \Pythia 6.4.25 dataset.}
\label{eq:pythia_correlation_dimension_all}
\end{figure*}

\bibliography{open_data_topics}

\begin{flushright}

\href{https://www.freepik.com/catalyststuff}{\includegraphics[width=0.13\textwidth]{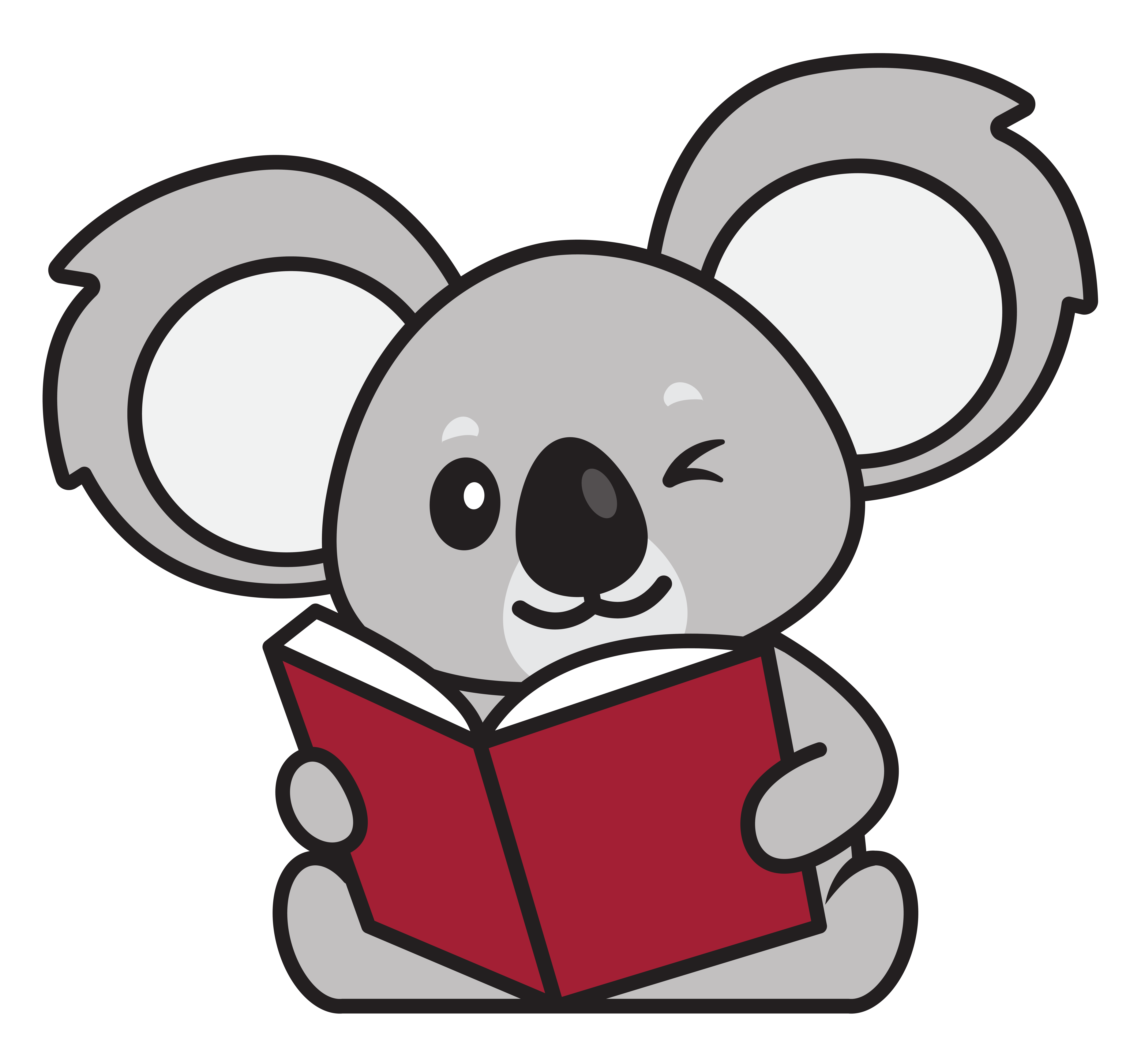}}
\end{flushright}

\end{document}